\documentclass[11pt]{article}
\usepackage[letterpaper,includefoot,top=1in,bottom=1in,left=1in,right=1in]{geometry}
\usepackage[centertags,sumlimits]{amsmath}
\usepackage{amsfonts}
\usepackage{amssymb,graphics,psfrag}
\usepackage{array,epsfig,multirow,stmaryrd,graphicx}
\usepackage[all,cmtip]{xy}
\usepackage{mathrsfs}
\usepackage{multirow}
\usepackage{verbatim}
\numberwithin{equation}{section}
\numberwithin{table}{section}

\renewcommand{\Im}{\operatorname{Im}}
\renewcommand{\Re}{\operatorname{Re}}

\newcommand{\beq}{\begin{equation}}
\newcommand{\eeq}{\end{equation}}
\newcommand{\be}{\begin{equation}}
\newcommand{\ee}{\end{equation}}
\newcommand{\bea}{\begin{eqnarray}}
\newcommand{\eea}{\end{eqnarray}}
\newcommand{\ben}{\begin{eqnarray*}}
\newcommand{\een}{\end{eqnarray*}}
\newcommand{\ba}{\begin{aligned}}
\newcommand{\ea}{\end{aligned}}
\newcommand{\bt}{\begin{tabular}}
\newcommand{\et}{\end{tabular}}
\newcommand{\bc}{\begin{center}}
\newcommand{\ec}{\end{center}}

\newcommand{\cO}{\mathcal{O}}

\newcommand{\cC}{\mathcal{C}}
\newcommand{\cD}{\mathcal{D}}
\newcommand{\cL}{\mathcal{L}}
\newcommand{\cS}{\mathcal{S}}
\newcommand{\cK}{\mathcal{K}}
\newcommand{\cN}{\mathcal{N}}
\newcommand{\cW}{\mathcal{W}}
\newcommand{\cG}{\mathcal{G}}
\newcommand{\cA}{\mathcal{A}}

\newcommand{\cF}{\mathcal{F}}
\newcommand{\cI}{\mathcal{I}}
\newcommand{\cJ}{\mathcal{J}}

\newcommand{\cV}{\mathcal{V}}

\newcommand{\cM}{\mathcal M}

\newcommand{\I}{\text{Im}}
\newcommand{\R}{\text{Re}}

\DeclareMathOperator{\vol}{vol}

\newcommand{\bbZ}{\mathbb{Z}}
\newcommand{\bbR}{\mathbb{R}}
\newcommand{\bbC}{\mathbb{C}}

\newcommand{\nn}{\nonumber}

\newcommand{\cref}{{\bf [check ref]}}

\newcommand{\Jc}{J_{\rm c}}

\entrymodifiers={+!!<0pt,\fontdimen22\textfont2>}

\begin{document}

\baselineskip=16pt
\setlength{\parskip}{6pt}

\begin{titlepage}
\begin{flushright}
\parbox[t]{1.4in}{
MPP-2011-42\\
BONN-TH-2011-07}
\end{flushright}

\begin{center}

\vspace*{1cm}

{\Large \bf The $\cN=1$ effective actions of D-branes \\[.1cm]
in Type IIA and IIB orientifolds 
}

\vskip 1cm

\begin{center}
        \normalsize \bf{Thomas W.~Grimm}$^a$ \footnote{\texttt{grimm@mppmu.mpg.de}}, \normalsize \bf{Daniel Vieira Lopes}$^b$ \footnote{\texttt{dlopes@th.physik.uni-bonn.de}}
\end{center}
\vskip 0.75cm

 \emph{$^{a}$ Max Planck Institute for Physics, \\ 
                       F\"ohringer Ring 6, 80805 Munich, Germany} 
\\[0.25cm]
\emph{$^b$ Bethe Center for Theoretical Physics, Universit\"at Bonn, \\[.1cm]
Nussallee 12, 53115 Bonn, Germany}
\\[0.15cm]
 \vspace*{0.75cm}

\end{center}

\vskip 0.0cm

\begin{center} {\bf ABSTRACT } \end{center}

We discuss the four-dimensional $\cN=1$ effective actions of single
space-time filling Dp-branes in general Type IIA and Type IIB Calabi-Yau
orientifold compactifications. The effective actions depend on
an infinite number of normal deformations and gauge connection modes.
For D6-branes the $\cN=1$ K\"ahler potential, the gauge-coupling function, the superpotential 
and the D-terms are determined 
as functions of these fields. They can be expressed as integrals 
over chains which end on the D-brane cycle 
and a reference cycle. The infinite deformation space will reduce to a 
finite-dimensional moduli space of special Lagrangian submanifolds
upon imposing F- and D-term supersymmetry conditions.
We show that the Type IIA moduli space geometry is
captured by three real functionals encoding the deformations
of special Lagrangian submanifolds, holomorphic
three-forms and K\"ahler two-forms of Calabi-Yau manifolds.
These elegantly combine in
the $\cN=1$ K\"ahler potential, which reduces after applying mirror
symmetry to the results previously determined for
space-time filling D3-, D5- and D7-branes. We also propose general 
chain integral expressions for the K\"ahler potentials of
Type IIB D-branes.

\end{titlepage}

\section{Introduction}

In recent years there has been vast progress in understanding the effective
supergravity theories arising in Type II string compactifications. From a phenomenological
perspective four-dimensional effective theories
which are $\cN=1$ supersymmetric and admit non-trivial gauge groups are of particular interest.
A prominent set-up admitting these features are
Calabi-Yau orientifold compactifications  with space-time
filling D-branes \cite{Blumenhagen:2005mu,Blumenhagen:2006ci,Douglas:2006es,Denef:2008wq,Cvetic:2011vz}. 
Such compactifications admit in addition to the bulk moduli also a universal class 
of deformation and Wilson line moduli associated to 
the D-branes. It will be the task of this work to 
study the four-dimensional $\cN=1$ characteristic data
encoding the dynamics of the combined open and closed sector 
moduli. We will first concentrate on Type IIA compactifications
with space-time filling D6-branes and later turn to Type IIB
compactifications with D3-, D5-, or D7-branes in the discussion 
of mirror symmetry.

Concentrating on Type IIA Calabi-Yau orientifolds 
the supersymmetric space-time filling objects are O6-planes and D6-branes
wrapped on a supersymmetric three-cycle in the internal
Calabi-Yau space. The orientifold planes are supersymmetric since
they wrap special Lagrangian cycles which arise as the fix-point locus
of an anti-holomorphic and isometric involution of the Calabi-Yau space.
On such cycles the K\"ahler form
and the imaginary part of the holomorphic three-form of the Calabi-Yau manifold
vanish. Similarly, in a supersymmetric background the D6-branes also 
have to wrap special Lagrangian
cycles which preserve the same supersymmetry as the 
O6-planes \cite{Becker:1995kb,Marino:1999af,Blumenhagen:2006ci}. 
We will mainly focus on a simple set-up and consider the dynamics of 
one space-time filling D6-brane and its non-intersecting orientifold image. 
Global tadpole cancellation conditions impose topological constraints on 
this configuration and generically imply that there will be 
additional D6-branes. Their dynamics can be included in the analysis, 
but will be neglected for simplicity.\footnote{For state of the art model building in 
Type IIA see, for examples, refs.~\cite{Palti:2009bt,Forste:2010gw,Blumenhagen:2002wn}, and references therein. 
Reviews can be found in \cite{Blumenhagen:2005mu,Blumenhagen:2006ci,Cvetic:2011vz}.}

In order to determine the $\cN=1$ effective theory one needs to include
the fluctuations of the fields around a given background. Therefore, the 
scalar fields in the effective theory will include the deformations both 
of the internal Calabi-Yau geometry as well as the
deformations of the D6-branes. As a first step the effective action 
including only the closed string zero modes in a Calabi-Yau orientifold 
background can be derived \cite{Grimm:2004ua,DeWolfe:2005uu}. The reduction  
considers a finite set of complex and K\"ahler structure deformations
which are compatible with the orientifold involution. This set 
of real deformations is complexified by the axion-like scalars arising 
as zero modes of the R-R and NS-NS form fields of Type IIA string theory. 
It was argued in refs.~\cite{Grimm:2004ua,Benmachiche:2006df} that the K\"ahler metric on 
the $\cN=1$ field space is captured by two real functionals that have 
been studied intensively by Hitchin in \cite{Hitchin:2000jd,Hitchin:2001rw}. Using 
mirror symmetry at large volume and large complex structure, the Type IIA 
K\"ahler potential can be exactly matched with its Type IIB counterparts \cite{Grimm:2004ua}. This 
reproduces the expressions found for Type IIB orientifolds with O3/O7 planes  
and O5 planes \cite{Giddings:2001yu,Grimm:2004uq}. In this work we
extend the computation of the $\cN=1$ characteristic 
data to the open string sector and complete the 
leading-order mirror identification including 
space-time filling D-branes.

In the first part of the present paper 
we will focus on the contribution from a D6-brane and its orientifold image. 
The degrees of freedom associated to a D6-brane wrapped on a supersymmetric 
three-cycle $L_0$ are easiest summarized when considering a fixed background Calabi-Yau 
geometry. In addition to the $U(1)$ gauge field the D6-brane can admit 
brane deformations and non-trivial Wilson lines. A massless deformation 
preserving the $\cN=1$ supersymmetry along a normal vector field is associated to 
a harmonic one-form on the special Lagrangian cycle $L_0$ \cite{McLean}. 
For a fixed background Calabi-Yau geometry there 
are $b^1(L_0) = \text{dim} H^1 (L_0,\bbR)$ real massless deformations, 
which combine with the Wilson line scalars into complex fields. We 
will derive the effective action for these massless modes which 
keep $L_0$ special Lagrangian. 
However, more interestingly, one can also include massive deformations around 
$L_0$ by extending the analysis to include non-harmonic one-forms on $L_0$.
These deformations either violate the Lagrangian condition or the condition 
that the three-cycle is `special', as we discuss in more detail in the main 
text. We will show that these deformations induce a scalar potential 
consisting of an F-term part, rendering non-Lagrangian deformations massive,
and a D-term part, rendering non-special deformations massive. Performing a
Kaluza-Klein reduction of the D6-brane action we derive 
the $\cN=1$ open string K\"ahler metric and gauge coupling function, 
and explicitly extract the D6-brane superpotential and D-terms. 
We argue that these functions take an elegant form when using 
chain integrals over a four-chain ending on the internal D6-brane
three-cycle and a reference cycle $L_0$.

In order to determine the K\"ahler potential it is crucial to include 
also Calabi-Yau deformations parameterizing the bulk  
degrees of freedom. In fact, we show that at the classical level 
the open string scalars only enter in the $\cN=1$ K\"ahler potential 
through a redefinition of the complex coordinates for the Calabi-Yau deformations. 
This is a generic feature which is already known from 
Type IIB Calabi-Yau orientifold compactifications with single D3-, D7- or 
D5-branes \cite{Grana:2003ek,Camara:2004jj,Jockers:2004yj,Grimm:2008dq}, as 
well as Type II orientifolded orbifolds \cite{Lust:2004cx,Blumenhagen:2006ci}. 
In the type IIA 
compactifications we find that the full $\cN=1$ K\"ahler 
potential has an elegant form in terms of the functionals for 
real two- and three-forms studied by Hitchin \cite{Hitchin:2000jd,Hitchin:2001rw}, and the 
K\"ahler potential arising in the study of the deformation space of 
special Lagrangian submanifolds \cite{Hitchin:1997ti,Hitchin:1999fh}. 
We also comment on a generalization 
of the $\cN=1$ data to an infinite set of D-brane deformations. As in a 
fixed background, this generalization will be crucial in the evaluation 
of the superpotential and D-terms. As in \cite{Grimm:2008dq} it will be crucial 
to keep the non-dynamical four-dimensional three-forms in the Kaluza-Klein Ansatz to 
derive the scalar potential. In addition, we 
will also be able to extract the kinetic mixing terms of the bulk and brane 
$U(1)$ vector fields in the effective action. Such mixing 
can have profound phenomenological applications \cite{Abel:2008ai}. 

In the last part of this paper we turn to the discussion of mirror symmetry 
at large volume and large complex structure. We match the known 
$\cN=1$ data for single D3-, D5- or D7-branes with the data found in the D6-brane 
reduction. This allows us to give chain integral expressions also for the Type IIB 
reductions which complete the results of 
\cite{Grana:2003ek,Camara:2004jj,Jockers:2004yj,Grimm:2008dq}. Our strategy 
to gain a better understanding of the structure of the 
brane couplings is to use the formulation of mirror symmetry proposed by Strominger, Yau and Zaslow (SYZ) \cite{Strominger:1996}. 
Neglecting singular fibers it allows to view the compactification Calabi-Yau space as a three-torus fibration over a three-sphere. 
This allows us to identify different brane wrappings in the Type IIA and Type IIB picture, and
yields a matching of the leading order $\cN=1$ data.

The present work is organized as follows. In section \ref{D6reduction} we calculate the IIA orientifold effective action
with a space-time filling D6-brane. We first recall the results for the bulk orientifold reduction, and 
summarize the conditions to include supersymmetric D6-branes in the orientifold background. We 
then derive the kinetic terms and the scalar potential for an infinite set of normal deformations,
Wilson line modes, and $U(1)$ vector modes on the D6-brane. These computations are performed using 
a Kaluza-Klein reduction of the Dirac-Born-Infeld and Chern-Simons actions around a supersymmetric 
configuration.
In section \ref{modulispace} we analyze the moduli space of our configuration. We focus 
on the finite set of massless bulk and brane modes.  As first steps we consider the moduli spaces for 
the bulk sector and the brane sector separately. We
introduce the necessary mathematical tools to describe these spaces as Lagrangian embeddings into 
vector space. In general, the geometry of the open-closed moduli space is more complicated. However,
we show that it is possible to encode the complete moduli space in a single elegant K\"ahler potential 
as a function of non-trivial local complex coordinates. In section \ref{modulispace} also the kinetic 
terms for the massless $U(1)$ gauge field on the D6-brane are discussed, and a kinetic mixing with the 
massless bulk $U(1)$'s is found. In section \ref{general} we discuss the infinite deformation 
space around a supersymmetric configuration. We give an explicit form of the 
K\"ahler potential using chain integrals. We also 
derive the non-vanishing D-terms and the superpotential for open deformations 
violating the special Lagrangian condition for supersymmetric D6-branes.
Our results should be mirror symmetric to Type IIB orientifold configurations. 
In section \ref{mirror} we argue that it is
possible to relate the moduli fields obtained in section \ref{modulispace} with the moduli space for 
IIB orientifold configurations with D3-, D5- and D7-branes. Relations between the homology 
of the cycles for the mirror configurations can be inferred using the SYZ construction of 
mirror symmetry. We find elegant expressions for the Type IIB K\"ahler potentials 
and $\cN=1$ complex coordinates.


\section{The dimensional reduction of the D6-brane action \label{D6reduction}}

We start our discussion by fixing the background geometry of our set-up.
In the following, we consider the direct product of a compact Calabi-Yau orientifold $Y/\mathcal{O}$ and
flat Minkowski space $\mathbb{R}^{1,3}$. We are interested in compactifications with
space-time filling D6-branes and O6-planes which preserve $\cN=1$ supersymmetry in
four space-time dimensions. This fixes the orientifold projection to be
of the form \cite{Brunner:2003zm,Grimm:2004ua}
\begin{equation} \label{OactionO6}
        \mathcal{O} = (-1)^{F_L} \Omega_p \sigma^*\ ,\qquad \quad \sigma^* J = - J \ ,\qquad \qquad \sigma^* \Omega = e^{2i\theta} \bar \Omega\ ,
\end{equation}
where $\theta$ is some real phase.
Here $\Omega_p$ is the world-sheet parity reversal, $F_L$ is the space-time fermion number in the left-moving
sector, and $\sigma$ is an
anti-holomorphic and isometric involution of the compact Calabi-Yau manifold
$Y$.
The four-dimensional spectrum consists of fields
arising as in the zero mode expansion of
the ten-dimensional closed string fields into harmonics of the internal
space as well as zero modes arising from massless open strings
ending on the D6-branes. In the following we will focus on the
chiral multiplets in both sectors.

\subsection{On the four-dimensional Kaluza-Klein spectrum \label{KKspectrum}}

\subsubsection*{Closed string sector}

Let us start with a brief discussion of the closed string sector
following ref.~\cite{Grimm:2004ua}. The four-dimensional scalars, vectors,
two- and three-forms will arise in the expansions of the ten-dimensional
fields into harmonic forms of $Y$ which have to transform in a specified
way under the orientifold parity to yield modes which remain in 
the orientifolded $\cN=1$ spectrum. More
specifically, the ten-dimensional metric and the dilaton are invariant under
the action of $\sigma$ while the NS-NS B-field transforms as $\sigma^* B_2 =
-B_2$. The R-R fields $C_1,C_3,C_5,C_7$
remain in the orientifold spectrum if they obey $\sigma^* C_p = (-1)^{(p+1)/2}
C_p$. Clearly, in type IIA string theory not all the odd-dimensional R-R
forms $C_p$ are independent. Denoting by $G_{p+1}$
the R-R field strengths are
\beq \label{def-fieldstrengths}
  G_2 = dC_1\ ,\qquad G_{p+1} = dC_p - H_3 \wedge C_{p-2}\ , \qquad H_3 = dB_2 \ .
\eeq
When considering all $G_{p+1},p=1...9$ 
a the duality constraint  
\beq \label{dual_constr}
    G_{p+1} = (-1)^{(p+1)/2} *_{10}G_{9-p}\ ,
\eeq 
has to be imposed to relate the lower and higher-dimensional forms.
This can be extended to include Romans mass $G_{0}$ which appears in the 
massive extension of Type IIA supergravity~\cite{Romans:1985tz}. Using all forms $G_{p+1}$ 
one can use a democratic formulation of Type II supergravity \cite{Bergshoeff:2001pv}.
The bosonic kinetic terms of the ten-dimensional action are then given by
\beq \label{democratic_action}
  S^{(10)}_{\rm dem} = -\int \tfrac12 R *_{10} 1 + \tfrac14 H_3 \wedge *_{10} H_3 + \sum_{p = 1}^9  \tfrac18 G_{p+1} \wedge *_{10} G_{p+1} \ .
\eeq
This is only an auxiliary action since the duality condition \eqref{dual_constr} has to be imposed by hand in addition to 
the equations of motion.
When coupling the bulk supergravity to a D-brane it turns out to be useful to also introduce another basis $\cA_q$ 
of $q$-forms with a redefined duality relation  
\beq \label{cA-def}
  \cA = \sum_q \cA_q = e^{-B_2} \wedge \sum_p C_p \ , \qquad d\cA_q = (-1)^{(q+1)/2} (*_B\, d\cA )_q\ ,
\eeq 
where the `B-twisted' Hodge star is given by $*_B = e^{-B_2} *_{10} e^{B_2} $. Clearly, the supergravity 
action \eqref{democratic_action} can be easily rewritten in terms of the $\cA_q$.

To perform the Kaluza-Klein expansion of the closed string fields
one notes that the anti-holomorphic involution does not preserve the $(p,q)$ split of
the Dolbeault cohomology groups, but rather maps a $(p,q)$- to a $(q,p)$-form.
One thus splits the real de Rham cohomologies into $\sigma^*$-eigenspaces
$H^{n}_\pm(Y)$. It was shown in ref.~\cite{Grimm:2004ua} that the $\cN=1$ coordinates on
the closed string field-space arise by expanding a complex
two-form $J_c$ and three-form $\Omega_c$ into a basis of $H^2_{-}(Y,\bbR)$ and
$H^{3}_+(Y,\bbR)$, respectively. More precisely, in accord with
\eqref{OactionO6}  we expand
\beq \label{expJB}
  \Jc = B_2 + i J = (b^a + iv^a)\, \omega_a = t^a \omega_a\ ,
\eeq
where $a = 1,\ldots, h^{(1,1)}_-$ labels a basis $\omega_a$ of
$H^2_-(Y,\bbR)$. We thus find the same complex
structure as in the underlying $N=2$ theory with the dimension of the K\"ahler moduli
space truncated from $h^{(1,1)}$ to $h^{(1,1)}_-$.

The complex three-form $\Omega_c$ contains the degrees of freedom arising from the
complex structure moduli, the dilaton as well as the scalars from the R-R
forms. We combine these as
\beq \label{def-Omegac}
  \Omega_c = 2\,\R(C \Omega) + i C_3^{\rm sc}\, =\, N'^k \, \alpha_k - T'_{\lambda} \,
  \beta^\lambda \ ,
\eeq
where $C \propto e^{-\phi + i \theta}$, as given in \eqref{defC}, contains the dilaton, and $k = 1,\ldots, n_-, \lambda = 1,\ldots, n_+$  label a basis
$(\alpha_k,\beta^\lambda)$ of $H^{3}_+(Y,\bbR)$. Here $C_3^{\rm sc}$
is the part R-R three-form which is also a three-form on the Calabi-Yau manifold $Y$
and hence descents to scalars in four dimensions.
We can use the
expansion of $\Omega$ into the full symplectic
basis $(\alpha_K,\beta^K)$ of
$H^3(Y,\bbR)$ as $\Omega = X^K \alpha_K -\cF_K \beta^K$.
Under $\sigma^*$ this basis splits into a
basis $(\alpha_k,\beta^\lambda)$ of $H^{3}_+(Y,\bbR)$ and a
dual basis $(\alpha_\lambda,\beta^k)$ of $H^{3}_-(Y,\bbR)$. We thus find the
 explicit expressions
\beq \label{def-N'T'}
  N'^k = 2\, \R (CX^k) + i\, \xi^k \ ,\qquad T'_\lambda =  2\, \R(C\cF_\lambda) +i\, \tilde
  \xi_\lambda \ .
\eeq
Note that the split of the $h^{(2,1)}+1$ basis elements of $H^{3}_+(Y,\bbR)$ into
$n_-$ elements $\alpha_k$ and $n_+$ elements $\beta^\lambda$ does depend on
the point in the complex structure moduli space on which one evaluates
$C\Omega$. In fact, at the large complex structure point the precise
split will determine whether this type IIA set-up is dual to an
orientifold with O3/O7 planes or O5/O9 planes as we will discuss in 
detail in section \ref{mirror}. It is important to point out,
that the complex coordinates $(N'^k,T'_\lambda)$ are the correct complex scalars
in the $\cN=1$ chiral multiplets in the absence of D6-branes, but
will receive corrections upon introducing dynamical D6-branes.

Before discussing the open string spectrum let us comment further on the
complex function $C$ appearing in \eqref{def-Omegac}. Since the orientifold
projection is an anti-holomorphic involution the complex structure
deformations will be real. In fact, $C$ has a phase factor $e^{-i\theta}$ and
is defined to compensate rescalings
of $\Omega$ such that $C\Omega$ has a fixed normalization
\beq \label{JJJ=OO}
 e^{2 \phi}  C \Omega \wedge \overline{C\Omega} = \tfrac{1}6 J \wedge J \wedge
 J\ .
\eeq
It turns out to be convenient
to introduce the four-dimensional dilaton $D$ by setting $e^{-2D} = e^{-2\phi} \cV$,
where $\cV = \frac16 \int_Y J \wedge J \wedge J$ is the string-frame volume of the
Calabi-Yau space. The compensator field is then given by
\beq \label{defC}
C = e^{-D-i\theta} e^{K^{\rm cs}/2}=e^{-\phi-i\theta} \cV^{1/2} e^{K^{\rm cs}/2},
\eeq
where  $K^{\rm cs} = -
\ln\big[-i\int\Omega\wedge \bar \Omega \big]$.

Let us note that the R-R three-form in general also leads to
$U(1)$ vectors in four space-time dimensions via the expansion
\beq \label{Cvec}
   C_3^{\rm vec} = A^\alpha \wedge \omega_\alpha\,
\eeq
where $\omega_\alpha$ is a basis of $H^{2}_+(Y,\bbR)$. Their holomorphic gauge coupling
function $f_{\alpha \beta}$ has also been determined in ref.~\cite{Grimm:2004ua}.
Denoting by $\cK_{\alpha \beta a} = \int_{Y} \omega_{\alpha} \wedge \omega_{\beta} \wedge  \omega_a$,
the intersection form of two elements of $H^{2}_-(Y,\bbR)$ with one element of $H^{2}_+(Y,\bbR)$
one finds that $f_{\alpha \beta} = i\cK_{\alpha \beta a} t^a$.

\subsubsection*{Open string sector: supersymmetric D6-branes \label{special_Lagr_sec}}

Let us next discuss the inclusion of
space-time filling D6-branes into our set-up. In the
background configuration these have to
be chosen such that they preserve the same supersymmetry
as the O6-planes which arise as the fix-point set of $\sigma$.
In fact, since $\sigma$ is an anti-holomorphic involution
the O6-planes wrap special Lagrangian cycles satisfying
\beq \label{sp_Lagr_O6}
   J|_{\text{O6-plane}} = 0 \ , \qquad \I (C\Omega)|_{\text{O6-plane}} = 0\ .
\eeq
Let us consider a single D6-brane wrapped on a three-cycle $L$ in
$Y$. We will consider the case where $L$ is mapped to a three-cycle
$L'=\sigma(L)$ which is in a different cohomology class and does
not intersect $L$.\footnote{Note that this is a non-generic situation for a three-cycle in 
a six-dimensional manifold. Generically D6-branes on three-cycles will intersect in points. At these 
intersections matter fields can be localized and have to be included in the reduction.} For this
situation the pair of the D6-brane and its image D6-brane is merely an auxiliary description of a single
smooth D6-brane wrapping a cycle in the orientifold $Y/\mathcal{O}$.
Note that the number of D6-branes is restricted by tadpole cancellation.
In cohomology one has to satisfy\footnote{Note that this condition
will be modified in the presence of NS-NS background flux $H_3$ and the Romans
mass parameter $m^0$ with an additional term proportional to $m^0 H_3$ (see, e.g.~, ref.~\cite{DeWolfe:2005uu}).}
\beq
   \sum_{\rm D6} \, [L+L'] = 4 [L_{\rm O6}]\ ,
\eeq
where the sum is over all D6-branes present in the compactification and $L_{\rm O6}$ is the
fix-point set of the involution indicating the location of the O6-plane.

In a supersymmetric orientifold background the D6-brane also has to wrap a
calibrated and hence supersymmetric cycle. These calibration conditions have
been determined in \cite{Becker:1995kb}. They imply that the
D6-brane wraps a special Lagrangian submanifold $L_0 \subset Y$ such
that
\beq \label{sp_Lagr}
  J|_{L_0} = 0 \ ,\qquad \I (C \Omega)|_{L_0} = 0 \ , \qquad 2\, \R (C\Omega)|_{L_0} = e^{-\phi} \text{vol}_{L_0}
\eeq
where $\text{vol}_{L_0}=\sqrt{\iota^* g_6} d^3 \xi$ is the induced volume form on $L$. Note
that the first condition in \eqref{sp_Lagr} implies that $L_0$ is Lagrangian, while
the second condition makes it special Lagrangian. We fixed the coefficient, in particular
the phase of $C\Omega$, such that the same supersymmetry is preserved as for the
orientifold planes \eqref{sp_Lagr_O6}. Finally, we note that it was also
shown in \cite{Marino:1999af} that in a  supersymmetric background
one has
\beq \label{F-B=0}
   F_{\rm D6} - B_2|_{L_0} = 0\ ,
\eeq
where $F_{\rm D6}$ is the field strength of the $U(1)$ gauge field $A$
living on the D6-brane. In the following we will always denote the
background special Lagrangian cycle wrapped by a supersymmetric D6-brane
by $L_0$.

For a \textit{fixed} background complex and K\"ahler structure we can discuss supersymmetric
deformations of the D6-branes.
In fact, the deformations of $L_0$ preserving the special Lagrangian conditions \eqref{sp_Lagr}
were studied by McLean \cite{McLean}. It was shown that a normal vector field $\eta$ to a
compact special Lagrangian cycle
is the deformation vector field to a normal deformation through special Lagrangian
submanifolds if and only if the corresponding 1-form $\theta_\eta = \eta \lrcorner J$
on $L_0$ is harmonic. This reduces the infinite dimensional space of maps $\iota$ to
a deformation space of dimension $b^1(L_0)=\dim H^{1}(L_0,\bbR)$.
Furthermore, there are no obstructions to extending a first order deformation to a finite deformation.
The tangent space to such deformations can be identified through the cohomology
class of the harmonic form with $H^1(L_0,\bbR)$.
We can thus write a basis of harmonic one-forms $\theta_i$ on $L_0$ as
\beq \label{def-theta}
  \theta_i = s_i \lrcorner J|_{L_0}\ ,\qquad *\theta_i = - 2 e^{\phi} s_i \lrcorner \I (C \Omega)|_{L_0} \ ,\qquad
  i=1,\ldots,b^1(L_0)\ ,
\eeq
where $s_i$ is a basis of the real special Lagrangian normal deformations.
Let us recall the derivation of the expression for $*\theta_i$~\cite{Hitchin:1999fh}.
We do this more generally, by determining the Hodge-dual of
a one form $\alpha=(X\lrcorner J)|_{L_0}$ for some $X \in TY|_{L_0}$. Note that
the vector dual to $\alpha$ by raising the index with the induced metric is $IX$
where $I$ is the complex structure on $Y$. Hence one checks
\beq
   *(X\lrcorner J)|_{L_0}  = (IX) \lrcorner \vol_{L_0}\ .
\eeq
However, on $L_0$ the volume form is identical to $2 e^{\phi}\R (C\Omega)$ by 
\eqref{sp_Lagr}. This implies
\beq \label{useful_id}
  *(X\lrcorner J)|_{L_0} = 2 e^{\phi} (IX\lrcorner \R (C\Omega))|_{L_0} = - 2 e^{\phi} (X\lrcorner \I (C\Omega))|_{L_0}
\eeq
where the minus sign arises from evaluating $I$ on the $(3,0)$-form $\Omega$,
$(IX)\lrcorner \Omega = i X \lrcorner \Omega$.

We have just introduced the general supersymmetric deformation
encoded by $b^1(L_0)$ scalars $\eta^i$ arising in the expansion
$\theta_\eta = \eta^i \theta_i$ of the harmonic form $\theta_\eta$. 
The $\eta^i(x)$ will be real scalar fields in the four-dimensional 
effective theory depending on the four space-time coordinates $x$.
Let us next discuss the degrees of freedom due to
$U(1)$ Wilson lines arising from non-trivial one-cycles on the D6-brane
world-volume. Later on we will show that these 
real scalars will complexify the $\eta^i$. The Wilson line scalars 
arise in the expansion of the $U(1)$ gauge boson $A_{\rm D6}$ on the D6-brane as
\begin{equation} \label{gauge-expansion}
        A_{\rm D6} = A +a^i\, \tilde \alpha_i \ ,
\end{equation}
where $A$ is a $U(1)$ gauge field 
and the $a^i(x)$ are $b^1(L_0)$ real scalars in four dimensions.
The forms $\tilde \alpha_i$ provide a basis of $H^1(L_0,\bbZ)$.
Note that in general the $U(1)$ field strength $F_{\rm D6}=dA_{\rm D6}$ can additionally
admit a background flux $\langle F_{\rm D6} \rangle=f_{\rm D6}$ in $H^2(L_0,\bbZ)$, which can be trivial 
or non-trivial in $H^2(Y,\bbR)$. Since we will focus on the kinetic terms in the
following we will set $f_{\rm D6}=0$ for most of the discussion.
Note that $F_{\rm D6}$ naturally combines with the NS--NS B-field into the combination
$ F_{\rm D6}- \iota^{\ast} B_2$.

To summarize, one finds as massless variations around a
supersymmetric vacuum $h^{(1,1)}_- + h^{(2,1)} + 1$ chiral multiplets from the
bulk and $b^{1}(L_0)$ chiral multiplets $(\eta^i,a^i)$ from the D6-brane.
The precise organization of these fields into $\cN=1$ complex coordinates
is postponed to section \ref{modulispace}.

\subsection{General deformations of D6-branes \label{gen-defD6}}

So far we have discussed the supersymmetric background D6-brane and its supersymmetric
deformations. However, in general $L_0$ admits an infinite set of
deformations which will render the D6-brane non-supersymmetric.
These deformations will be included in the following and
shown to be obstructed by a scalar potential.
In order to do that, one recalls
that the string-frame world-volume action for the D6-brane takes the form
\beq \label{eqn:effectiveaction}
        S^{\text{SF}}_{\text{D6}}=-\int_{\mathcal{W}_7}d^7\xi
        e^{-\phi}\sqrt{-\text{det}\left(\iota^{\ast}\left(g_{10}+B_2\right)- F_{\rm D6}\right)}
        +\int_{\mathcal{W}_7}\sum_{q\ \text{odd}} \iota^\ast(C_q) \wedge e^{ F_{\rm D6}-\iota^\ast(B_2)} \ .
\eeq
In this subsection we will derive the scalar potential arising from the first term 
in \eqref{eqn:effectiveaction}, the Dirac-Born-Infeld action.

\subsubsection*{Exponential map and normal coordinate expansion}

 A general fluctuation
of $L_0$ to a nearby three-cycle $L_\eta$ is described by
real sections $\eta$ of the normal bundle $N_Y L_0$.
Clearly, the space of such sections is infinite dimensional as is the
space of all $L_\eta$. To make the identification between $L_\eta$
and $\eta$ more explicit, one recalls that in a sufficiently small neighborhood of
$L_0$ the exponential map exp$_\eta$ is a diffeomorphism of $L_0$ onto $L_\eta$.
Roughly speaking, one has to consider geodesics through each point $p$ on $L_0$ with tangent
$\eta(p)$ and move this point along the geodesic for a distance of $||\eta||$ to obtain the nearby three-cycle $L_\eta$.

It is important to consider how the pull-backs of $J, \I( C\Omega)$ as well as other
two and three-forms of $Y$ behave when moving
from $L_0$ to $L_\eta$. To examine this change one introduces the
pull-back of the exponential map
\beq
  E_\eta(\gamma) =\exp_\eta^* (\gamma|_{L_\eta}) \ ,
\eeq
where $\eta \in NL_0$, and $\gamma \in \Omega^p(Y)$ are $p$-forms on $Y$.
Hence, $E_\eta$ pulls back $\gamma$ from $L_\eta$ to a $p$-form $E_\eta(\gamma) \in \Omega^p(L_0)$
on $L_0$.
Of particular interest are the evaluation of $E_\eta$ on $J$ and $\I( C\Omega)$.
In fact, one shows that \cite{McLean}\footnote{This can be deduced from the
fact that $J$ and $\I(C\Omega)$ are closed and one has in cohomology that $[E_\eta(\gamma)] = [\gamma|_{L_0}]$.}
\beq \label{def-muhat}
   E_\eta(J) = d \hat \mu_1\ ,\qquad E_\eta\big(\I (C\Omega) \big) = d\hat \mu_2\ ,
\eeq
which means that $E_\eta(J)$ and $E_{\eta}\big(C\Omega \big)$ are exact forms on $L_0$.
In order to study special Lagrangian deformations as in section \ref{special_Lagr_sec}
one thus has to consider the space of  deformations $\eta_{\rm sp}$
such that $E_{\eta_{\rm sp}}(J)=0$ and $E_{\eta_{\rm sp}}(\I(C\Omega))=0$ \cite{McLean}.

In the leading order effective action we will often  be interested in first
order deformation and the linearizations $E'_\eta(\gamma) := \partial_t E_{t\eta}(\gamma)|_{t=0}$ of $E_\eta$ will be of importance. A straightforward
computation shows that for $\gamma$ being a closed form on $Y$ one has
\beq \label{linearization}
  d\gamma=0:\qquad E'_\eta(\gamma) = \cL_\eta (\gamma)|_{L_0} = d(\eta \lrcorner \gamma)|_{L_0}\ .
\eeq
Here we have used the standard formula
for the Lie derivative on a form $\cL_\eta \gamma = d(\eta \lrcorner \gamma) + \eta \lrcorner d\gamma$.
Note that \eqref{linearization} immediately implies that
\beq \label{normal_exp_JOm}
  E'_\eta(J) = d\theta_\eta\ ,\qquad  E'_\eta(\I( C\Omega)) =  -2 e^\phi d * \theta_\eta\ .
\eeq
where $\theta_\eta = \eta \lrcorner J|_{L_0}$  and
we have again used the fact that $*\theta_\eta = 2 e^\phi \eta \lrcorner \I C\Omega|_{L_0}$ 
as in \eqref{def-theta}. One can proceed with the expansion of the
exponential map and determine the full normal coordinate expansion.
In particular, for a $p$-form one finds the small $t$ expansion
\bea \label{normal_C}
   E_{t \eta}(C_p) &=&  \tfrac{1}{p!}  \Big[ C_{i_1 \ldots i_p} + t \cdot \Big( \eta^n\partial_nC_{i_1 \ldots i_p}
                  - p \nabla_{i_1}\eta^nC_{ni_2 \ldots i_p} \Big)
              \\
 && \phantom{\tfrac{1}{p!}  \Big[ }+ \tfrac{1}{2} t^2 \cdot \Big( \eta^n\partial_n(\eta^m\partial_m C_{i_1 \ldots i_p}) p \nabla_{i_1}\eta^n\eta^m\partial_m C_{ni_2 \ldots i_p}
          -\tfrac{p(p-1)}{2}\nabla_{i_1}\eta^n\nabla_{i_2}\eta^m C_{nmi_3 \ldots i_p} \nn \\
    && \phantom{\tfrac{1}{p!}  \Big[ } +\tfrac{p-2}{2}R_{ni_1 m}^{j}\eta^n\eta^m C_{ji_2 \ldots i_p} \Big)
        +\, \cO(t^3)\ \big]\ d\xi^{i_1} \wedge \ldots \wedge d\xi^{i_p}.\nonumber
\eea
Such normal coordinate expansions have been used for D-branes of different dimensions, for example, in
refs.~\cite{Grana:2003ek,Jockers:2004yj,Grimm:2008dq}.

\subsubsection*{The scalar potential for Lagrangian deformations}

Using the exponential map and the corresponding normal coordinate expansion
one can study the geometric properties of $L_\eta$ when examined on $L_0$.
This in particular includes the variations of the volume functional
\beq \label{Volumefunc}
  V(L_\eta)= \int_{L_\eta} d^3\xi \, e^{-\phi}\,\sqrt{\det(\iota^* g)} =\int_{L_\eta}e^{-\phi} \vol_{L_\eta}\ ,
\eeq
where $\iota$ is the pull-back of the Calabi-Yau metric to $L_\eta$.
Despite the fact that the reference cycle $L_0$ is special Lagrangian
the analysis of $V(L_{\eta})$ is still rather involved for a general
deformation vector field $\eta$ \cite{McLean}. Note that the volume functional 
\eqref{Volumefunc} is obtained from the Dirac-Born-Infeld action in \eqref{eqn:effectiveaction} 
by temporarily ignoring the brane field strength $F_{\rm D6}$ and the NS-NS B-field $B_2$.

We will first restrict to the case that $L_\eta$ is Lagrangian. In this case
one can study the deformations of the volume functional employing a rather elegant computation.
Later on we include the additional terms obtained in the general linearized
analysis of \cite{McLean}.
Given a Lagrangian submanifold $L$ in $Y$ one notes that its induced
volume form is proportional to the pullback of the holomorphic three-form
$\Omega$ to $L$. The proportionality factor is in general a function depending
on the coordinates $\xi$ of $L$.
Thus, comparing $\Omega|_L$ with the volume form on $L$ one has
\beq \label{def-thetaD6}
  e^{-\phi} \text{vol}_L =  2 C_{\rm D6}\Omega|_L\ ,\qquad C_{\rm D6}(\xi) = |C| e^{-i\theta_{\rm D6}(\xi)}\ ,  
\eeq
where $|C|$ is introduced in \eqref{JJJ=OO}.
Recall that $C$ is constant on $Y$ since it only contains the constant phase of the O6-planes, in
accord with the fact that the O6-planes wrap special Lagrangian cycles \eqref{sp_Lagr_O6}.
In contrast, $\theta_{D6}(\xi)$ is a real map, generally depending on
the coordinates $\xi$ on $L$. As $\theta_{\rm D6}$ appears in \eqref{def-thetaD6}
with a $2\pi$ periodicity it is a map from $L$ to the circle, and induces a   
map $\theta_{\rm D6 *} : \pi(L) \rightarrow \pi(S^1)$. However, in order to
avoid anomalies, one demands that $\theta_{\rm D6}$ actually admits a lift
to a function with values on the full real axis. This implies that $\theta_{\rm D6*}(L)$
vanishes and translates to the condition that the class $[d\theta_{\rm D6}]$ vanishes.
These Lagrangian submanifolds are known as graded Lagrangians \cite{Seidel00}, and the lift of $\theta_{\rm D6}$
to a real valued function is the grading. 

Let us next consider a family of Lagrangian submanifolds $L(t)$ which
are obtained by deforming an initial Lagrangian $L(0)$
for a distance $t$ into the direction of the normal vector
field $\eta$. For this to be a Lagrangian deformation $\theta_\eta = \eta \lrcorner J|_L$
has to be closed. On each $L(t)$ we can introduce a coordinate dependent
phase $\theta_{\rm D6}(\xi,t)$.
We consider the $t$-derivative of $e^{i\theta_{\rm D6}(t)}\vol_{L(t)}$ by
evaluating
\beq
  \frac{d}{dt} (e^{-\phi+i\theta_{\rm D6}}\vol_{L}) = (\cL_{\eta} |C|\Omega)|_L = - e^{-\phi+i\theta_{\rm D6}} (d\theta_{\rm D6} \wedge \eta\lrcorner \vol_L + i(d^* \theta_\eta) \vol_L)\ ,
\eeq
where $d^* = * d*$ with $*$ being the $t$-dependent Hodge-star on $L(t)$.
Comparing real and imaginary parts one finds that
\beq \label{flow_gen}
  \frac{d}{dt} \theta_{\rm D6} = - d^* \theta_\eta\ ,\qquad \frac{d}{dt} \vol_L = - d\theta_{\rm D6} \wedge * \theta_\eta\ ,
\eeq

Note that a particularly interesting
case is when $\theta_\eta =d \theta_{\rm D6}$, since in this case the second equation
ensures that the volume of $L$ is decreasing along $\eta_{d\theta_{\rm D6}}$.
In fact, this normal vector precisely parameterizes the directions to $L$
in which its volume is most efficiently decreasing. This vector is
known as mean curvature vector. Such Lagrangian mean curvature flows 
have been discussed intensively in the mathematical literature (see, e.g., refs.~\cite{Thomas:2002,Wang}, and 
references therein).

We are now in the position to evaluate the $t$-derivatives of $\vol_L$
at the point $t=0$. We return to the case that $L(0)=L_0$ is
the background special Lagrangian. One then shows that
\beq \label{dt_of_vol}
  \frac{d}{dt} \vol_{L}|_{t=0} = 0\ , \qquad \frac{d^2}{dt^2} \vol_{L}|_{t=0} = (d d^* \theta_\eta) \wedge * \theta_\eta\ .
\eeq
In this computation it is crucial to use the fact that at $t=0$ one
has $\theta_{D6}(0)=\theta_{\rm O6}$ is constant on $L_0$. This immediately implies
the vanishing of the first derivative of $\vol_L$ using \eqref{flow_gen}. To evaluate
the second derivative both equations \eqref{flow_gen} have to be applied successively.
Finally, we can use \eqref{dt_of_vol} to evaluate the lowest order scalar potential
for a Lagrangian brane on $L(t)$ as
\beq
  \frac{d^2}{dt^2} V(L_{t\eta}) |_{t=0}= e^{-\phi} \int_{L_0} d * \theta_\eta \wedge * d* \theta_\eta =
  4  e^{\phi} \int_{L_0}  d(\eta \lrcorner \I C\Omega) \wedge * d(\eta \lrcorner \I C\Omega) \ ,
\eeq
where $V$ is the volume functional \eqref{Volumefunc}.
As we will show later on, this term provides a scalar potential which
corresponds to a D-term in the four-dimensional $\cN=1$ effective theory
for the D6-brane.

\subsubsection*{The scalar potential for general deformations}

Before turning to the details of the Kaluza-Klein reduction let us recall that
one can extend the analysis to deformations
$\eta$ for which $L(t)$ is no longer Lagrangian. In this case $d \eta \lrcorner J$ does
not necessarily vanish and \eqref{def-thetaD6} is not generally possible. However, one
can still evaluate the second derivative of the volume of $L(t)$ at the
point $t=0$ as \cite{McLean}
\beq \label{general_potential_metric}
  \frac{d^2}{dt^2} V(L_{t\eta}) |_{t=0} = e^{-\phi} \int_{L_0} d(\eta \lrcorner J) \wedge * d(\eta \lrcorner J)
+ 4 e^{\phi} \int_{L_0} d(\eta \lrcorner \I C\Omega) \wedge * d(\eta \lrcorner \I C\Omega) \ .
\eeq
The new term depending on $d(\eta \lrcorner J)$ is the obstruction for $L(t)$ to be Lagrangian.
In the  four-dimensional $\cN=1$ effective theory
for the D6-brane this term can be obtained as one of the F-term contributions
from a superpotential which we determine in section \ref{general}.

\subsubsection*{The scalar potential including the B-field}

So far we have discussed the scalar potential without the inclusion 
of the NS-NS B-field of Type IIA string theory and the brane field strength 
$F_{\rm D6}$. To compute the leading order potential including $F_{\rm D6}$
we note that only the part $\tilde F$ of $F_{\rm D6}$ contributes to the potential 
which has indices on the internal three-cycle wrapped by the brane. 
We perform a Taylor expansion of the Dirac-Born-Infeld action using 
\beq \label{det-exp}
  \sqrt{\det (\mathfrak{A} + \mathfrak{B})} =  \sqrt{\det (\mathfrak{A})}\Big[\mathbf{1} 
  + \tfrac12 \text{Tr}(\mathfrak{A}^{-1} \mathfrak{B}) +\tfrac{1}{8} \left( \big[\text{Tr}(\mathfrak{A}^{-1} \mathfrak{B})\big]^2  -2 \text{Tr}\big([\mathfrak{A}^{-1} \mathfrak{B}]^2\big) \right)+\ldots \Big]
\eeq  
for small fluctuations $\mathfrak{B}$ and invertible $\mathfrak{A}$. The matrix $\mathfrak{B}$
we want to identify with the normal coordinate expansion of $B_2 - \tilde F$ in \eqref{eqn:effectiveaction}, while $\mathfrak{A}$ is 
the background metric of the Calabi-Yau space restricted to $L_0$. Recall that the normal coordinate 
expansion $E_{t \eta }(B_2)$ was given in \eqref{normal_C}. 
One notes that the first term in the expansion \eqref{det-exp} is canceled by tadpole cancellation of the 
D6-branes with the O6-planes in the background. Moreover, the second and third term in \eqref{det-exp} do not 
contribute to the potential since $\mathfrak{A}$ is symmetric while $\mathfrak{B}$ is anti-symmetric.
Evaluating the remaining term $ \text{Tr}\big([\mathfrak{A}^{-1} \mathfrak{B}]^2\big)$ and adding the result 
\eqref{general_potential_metric} one finds 
\beq \label{full-potential}
  V_{\rm DBI}^{\rm SF} = e^{-\phi} \int_{L_0} \big[ d * \theta_\eta \wedge * d * \theta_\eta + d  \theta_\eta \wedge * d  \theta_\eta + (\tilde F -B_2 - d \theta^B_\eta ) \wedge * (\tilde F-B_2 - d \theta^B_\eta)\big]\ ,
\eeq
which is still expressed in the ten-dimensional string frame.
Here we have introduced the abbreviation 
\beq
  \theta_\eta^B = \eta \lrcorner B_2|_{L_0}\ ,
\eeq
which is the B-field analog of $\theta_\eta = \eta  \lrcorner J|_{L_0}$. 
This concludes the computation 
of the scalar potential from the Dirac-Born-Infeld action. In a next step we want to introduce 
a Kaluza-Klein basis and determine the complete leading order effective action including the
kinetic terms. 

\subsection{A Kaluza-Klein basis}

In performing a Kaluza-Klein reduction of the D6-brane action to
four space-time dimensions we like to include all massive
modes corresponding to arbitrary deformations of $L_0$ to $L_\eta$. This means
that we include sections $s_I$ of $NL_0$ which yield one-forms in the contraction
with $J$
\beq \label{def-gen_theta}
  \theta_I = s_I \lrcorner J|_{L_0} \quad \in\quad \Omega^1(L_0)\ .
\eeq
For a compact $L_0$ it is possible to label these one-forms by indices
$I=1,\ldots,\infty$ by considering the Kaluza-Klein
eigenmodes of the Laplacian $\Delta_{L_0}$. In this case the zero modes $\Delta_{L_0} \theta_i =0$
are precisely the harmonic forms $\theta_i$ introduced in \eqref{def-theta}.
However, the basis adopted to $\Delta_{L_0}$
is not always useful, since it explicitly depends on the metric inherited
form the ambient Calabi-Yau manifold. In the following we will therefore work with a
general countable basis of $\Omega^1(L_0)$, and later use the induced metric to interpret
the final expressions after performing the reduction. In general we will always
demand that the one-forms $\theta_I$ are finite in the $L^2$-metric
\beq \label{def-cG_L2}
  \cG(\tilde \alpha,\tilde \beta) = \int_{L_0} \tilde \alpha \wedge * \tilde \beta\ ,
\eeq
where $\tilde \alpha,\tilde \beta \in \Omega^1(L_0)$.

Let us now turn to the discussion of the $U(1)$ gauge field on the
D6-brane. It admits the general expansion
\beq \label{exp_AD6_gen}
  A_{\rm D6} = A^J\,  h_J + a^I\, \hat \alpha_I\ ,
\eeq
where $h_J \in C^\infty(L_0)$ is a basis of functions on $L_0$ and
$\hat \alpha_J \in \Omega^1(L_0)$ is a basis of one-forms
on $L_0$. Here again a countable basis can be chosen due to the compactness of
$L_0$. Note that the field-strength of $A_{\rm D6}$ is given by
\beq \label{exp_FD6_gen}
  F_{\rm D6} = F^J \,  h_J - A^J\wedge d h_J + da^I \wedge \hat \alpha_I + \tilde F\ ,\qquad
  \tilde F= a^I \, d\hat \alpha_I + f_{\rm D6}\ ,
\eeq
where $f_{\rm D6} \in H^2(L_0,\bbR)$ is a background flux of $F_{\rm D6}$ on $L_0$.
The terms $d h_J$ and $d\hat \alpha_I$ arise due to the fact that the functions
$h_J$ need not to be constant on $L_0$ and the one-forms $\hat \alpha_I$ 
need not to be closed.

We thus find that an infinite tower of scalars $a^I$ which are coefficients
of \textit{exact} forms are actually gauged by the gauge fields $A^J$ for which $dh_J \neq 0$.
Moreover, scalars $a^I$ arising in the expansion in \textit{non-closed} forms
appear without four-dimensional derivative in the expansion \eqref{exp_FD6_gen}.
To see this, we introduce a special basis adopted to the metric induced
on $L_0$. More precisely, via the Hodge decomposition each
one-form $\hat \alpha_I$ can be uniquely decomposed into a harmonic form, an
exact form $d\hat h_I$ and an co-exact form $d^* \hat  \gamma_I$ on $L_0$ as
\beq
   \hat \alpha_I =  \mu_{I}^i \tilde \alpha_i + d\hat h_I + d^* \hat \gamma_I\ ,
\eeq
where $\tilde \alpha_i$ are the
$b^1(L_0)$ harmonic forms introduced in \eqref{gauge-expansion}.
We thus pick a basis of the space of exact forms $\Omega^1_{\rm ex}(L_0)$ denoted by $dh_I$
and a basis $d^* \gamma_I$ of the space $\Omega^1_{\text{co-ex}}(L_0)$ which are exact with respect to $d^*$.
By appropriate redefinition we can introduce scalars
$\hat a^J$ parameterizing the expansion in  $dh_J$.
Denoting the coefficients of the non-closed forms $d^*\gamma_I$ by $\tilde a^I$, and the
coefficients of the harmonic forms by $a^j$ the
expansion \eqref{exp_FD6_gen} reads
\bea \label{exp_FD6_spbasis}
   F_{\rm D6} &=& F^I \,  h_I +da^j \wedge \tilde \alpha_j + \cD \hat a^I\wedge d h_I +d \tilde a^J \wedge d^* \gamma_I + \tilde F\ ,\\
  \cD \hat a^I &=& d \hat a^I - A^I\ , \qquad \tilde F = \tilde a^I dd^* \gamma_I + f_{\rm D6}\ . \nn
\eea
From this we conclude that precisely the scalars $\hat a^I$
are gauged by $A^I$.
Since the four-dimensional effective theory is an $\cN=1$ supersymmetric
theory one infers that there will be D-terms induced due to these gaugings $\cD \hat a^I$,
while F-terms are induced due to $\tilde F$.
We will determine the D-term in section \ref{general}, and check that it matches the moment map
analysis of ref.~\cite{Thomas:2001ve}.

\subsection{The four-dimensional effective action \label{kin_action}}

We can now determine the kinetic terms for the chiral multiplets of the
D6-brane coupled to the bulk supergravity.
Since the bulk action has been Kaluza-Klein reduced
on the orientifold background in ref.~\cite{Grimm:2004ua}
we will focus on the reduction of the D6-brane action
\eqref{eqn:effectiveaction}. The contributions entirely due
to bulk fields are later included in the determination of the
$\cN=1$ characteristic data.

\subsubsection*{Dirac-Born-Infeld action}

Let us start by considering the Kaluza-Klein reduction of the first term
in \eqref{eqn:effectiveaction}, i.e.~the Dirac-Born-Infeld action. We expand the
determinant in \eqref{eqn:effectiveaction} to quadratic order in
the fluctuations around the supersymmetric background.
These are precisely the fluctuations of the embedding $\iota$ of $L$
parameterized by the fields $\eta^i$ of
\eqref{def-theta} and the Wilson line scalars $a^i$ introduced in
\eqref{gauge-expansion}.
The normal coordinate expansions of the ten-dimensional metric
on the D6-brane world-volume is given to leading order by
\beq\label{g_exp}
        \iota^{\ast}g_{10} = \big( e^{2D}\eta_{\mu
          \nu} + g(\partial_{\mu}\eta,\partial_{\nu} {\eta})\big)\, dx^{\mu}\cdot dx^{\nu}
          +(\iota^\ast g+\delta (\iota^\ast g))_{mn} d\xi^m\cdot d \xi^n\ ,
\eeq
where $g_{mn}$ is the induced metric on $L$, and $\delta (\iota^\ast g)_{mn}$
is the metric variation induced by the variation of the background K\"ahler
and complex structure.
Note that the four-dimensional metric $\eta_{\mu \nu}$ is rescaled to the four-dimensional
Einstein frame.\footnote{Recall
  that the four-dimensional metric in the Einstein frame $\eta$ is related to
  the string frame metric $\eta^{\rm SF}$ via $\eta = e^{-2D}\, \eta^{\rm
    SF}$.}
One first performs the Taylor expansion of the
determinant while using \eqref{g_exp}. Inserting the result
together with $F_{D6}$ given in \eqref{exp_FD6_spbasis}
into the first part of (\ref{eqn:effectiveaction}) we obtain the
four-dimensional action
\bea \label{eqn:DBI}
         S^{(4)}_{\text{DBI}}
         &=&-\int \tfrac{1}{2} \R f_{\text{r}\, IJ} \, F^I\wedge * F^J
         + e^{2D} \mathcal{G}_{ij}\, d a^i \wedge * d{a}^{j}
         + e^{2D} \tilde \cG_{IJ}\, d \tilde a^I \wedge * d \tilde a^J  \nn  \\
         &&\qquad \quad + e^{2D} \cG_{IJ}\, \cD \hat a^I \wedge * \cD \hat a^J+ e^{2D} {\widehat \cG}_{IJ}\, d\eta^I \wedge \ast d \eta^{J}
         +  V_{\rm DBI} *1  \ ,
\eea
in the four-dimensional Einstein frame. The covariant derivative $\cD \hat a^I$
was introduced in \eqref{exp_FD6_spbasis} and indicates the gauging of the infinite tower of scalars
$\hat a^I$. The potential term $V_{\rm DBI}$
depends on the deformations $\delta (\iota^\ast g)_{mn}$ of the calibration conditions \eqref{sp_Lagr} induced
by the variation of the induced metric on $L_\eta$ which we computed in \eqref{general_potential_metric}. Moreover,
one obtains an additional term depending on the modes violating the background condition $F_{\rm D6}-B_2|_{L_0}=0$
as in \eqref{full-potential}.
Explicitly we find
\beq \label{DBI_potential}
 V_{\rm DBI} =  \frac{e^{3 \phi}}{\cV^2}\int_{L_0} d^*\theta_\eta \wedge * d^*\theta_\eta
               + \frac{e^{3 \phi}}{\cV^2}\int_{L_0}\Big( d\theta_\eta \wedge * d\theta_\eta  + (\tilde F-B_2 - d \theta_\eta^B)\wedge * (\tilde F-B_2- d \theta_\eta^B) \Big)\ ,
\eeq
where $\tilde F$ is defined in \eqref{exp_FD6_spbasis}.
In the following we will discuss the metric functions appearing in the kinetic terms
of \eqref{eqn:DBI}.

The first term in (\ref{eqn:DBI}) is the kinetic term for the $U(1)$ gauge bosons
$A^I$. The gauge coupling function is thus given to leading order by
\beq \label{gauge_coupling}
 f_{{\rm r}\, IJ}= \int_{L_0} \big(2\, \R(C\Omega) + i C_3) h_I h_J ,
\eeq
where the volume of $L_0$ has been replaced using \eqref{sp_Lagr}. Note that $\R f_{{\rm r}\, IJ}$
admits a simple geometrical interpretation as $L^2$-metric on the space of functions on $L_0$.
More generally, without introducing a specific basis and restricting to
a special Lagrangian one writes for two functions $h,\tilde h$ on $L$
\beq
  \R f_{\rm r}(h ,\tilde h)|_L = e^{-\phi}\int_{L} h \wedge * \tilde h\ ,
\eeq
which readily reduces to \eqref{gauge_coupling} on $L=L_0$ using $*1=\vol_L$ and  \eqref{sp_Lagr}.

The second, third and fourth term in \eqref{eqn:DBI} are the kinetic terms for the Wilson line moduli $a^i,\tilde a^I, \hat a^I$, where the
later appear with the covariant derivative $\cD \hat a^I = d\hat a^I + A^I$ as introduced in \eqref{exp_FD6_spbasis}.
The appearing metrics take the form
\beq \label{metric_a}
   \cG_{ij}= \tfrac12 e^{-\phi}\cG(\tilde \alpha_i,\tilde \alpha_j)\ , \qquad
   \tilde \cG_{IJ} = \tfrac12 e^{-\phi}\cG(d^* \gamma_I, d^* \gamma_J)\ ,\qquad
   \cG_{IJ}= \tfrac12 e^{-\phi}\cG(dh_I, dh_J)\
    ,
\eeq
where $\cG$ is the $L^2$-metric defined in \eqref{def-cG_L2}, and $\tilde \alpha_i$, $dh^I$ and $d^* \gamma_I$ 
are the one-form basis introduced in \eqref{exp_FD6_spbasis}.
The fifth term in \eqref{eqn:DBI} contains the field space metric for the
deformations $\eta^I$ and is of the form
\beq  \label{eqn:metrics}
 \widehat \cG_{IJ} = \int_{L_0} g(s_I,s_J) \R (C \Omega) = \tfrac12 e^{-\phi} \cG(\theta_I, \theta_J)\ .
\eeq
where $\theta_I$ are the one-forms on $L_0$ introduced in \eqref{def-gen_theta}. Let us comment
on the derivation of the second identity in \eqref{eqn:metrics}. Here we
first have to use the fact that $g(s_i,s_j)=J(s_i,I s_j)=(Is_j) \lrcorner \theta_i$, where $J$ is the
K\"ahler form and $I$ is the complex structure on $Y$. Next we deduce from $J \wedge \R(C\Omega)=0$ that
we can move the $I s_j$ to obtain $\theta_i \wedge (Is_j) \lrcorner \R
C\Omega$. However, since $C\Omega$ is a $(3,0)$-form one deduces using
\beq \label{useful_id_2}
  2(Is_j) \lrcorner \R C\Omega = -2 s_j \lrcorner \I( C\Omega) = e^{-\phi} * \theta_j\ ,
\eeq
and the identity \eqref{def-theta} the second equality in \eqref{eqn:metrics}.

This completes our reduction of the Dirac-Born-Infeld action. Let us stress that 
the reduction so far only included the leading order terms. In order to fully 
extract the $\cN=1$ characteristic data, however, we will need to match also 
higher order terms. It turns out that an efficient strategy to proceed is to include 
these by using supersymmetry and a careful study of the  
the Chern-Simons action. We will turn to the Kaluza-Klein reduction of this 
part of the D-brane action in the following.

\subsubsection*{Chern-Simons action}

Let us now turn to the dimensional reduction of the Chern-Simons part of the D6-brane action.
In the reduction one can again perform a normal coordinate expansion of the form-fields
appearing in the action. However, we will take here a somewhat different route and parameterize
the normal variations by introducing a four-chain $\cC_4$ which
contains the three-cycle $L_\eta$ in its boundary
\beq \label{def-cC4}
  \partial \cC_4 = L_\eta - L_0 \ ,
\eeq
where $L_0$ is the reference three-cycle, the supersymmetric background cycle.

We consider the Chern-Simons action containing the R-R forms $C_3,C_5$ and $C_7$ given by
\beq \label{csaction}
 S_{\rm CS}= \int_{\cW_7^{(0)}}  e^{F-B_2} \wedge (C_3+C_5+C_7) + S_{\rm CS}^{\cC_4}.
\eeq
Here $\cW_7^{(0)}=\cM^{3,1} \times L_0$,
\beq \label{ext_CS}
  S_{\rm CS}^{\cC_4} = \int_{\cW_8} d \big[ e^{F-B_2} \wedge (C_3+C_5+C_7) \big]\ ,
\eeq
and $\cW_8 = \cM^{3,1}\times \cC_4$ such that $\cW_7 \subset \partial \cW_8$.
This is in a similar spirit as the constructions in \cite{Witten:1996hc}.
To perform the Kaluza-Klein reduction of \eqref{ext_CS} we consider the expansion of 
$\cA$, the wedge product between the R-R forms and the B-field introduced in \eqref{cA-def}, as
\bea \label{KK_RR}
   \sum_{p=3,5,7,9} e^{-B_2} \wedge C_p  &=& 
    (\xi^k \alpha_k - \tilde \xi_\lambda \beta^\lambda) 
   + (A^\alpha \wedge \omega_\alpha + A_\alpha \wedge \tilde \omega^\alpha) \\
   && 
    + (C_2^{\lambda} \wedge \alpha_\lambda - \tilde C^2_k \wedge \beta^k) 
    + (C^0_3 + C_3^a \wedge \omega_a+ C^3_a \wedge \tilde \omega^a + C^3_0 \wedge \vol_Y)\ .\nn
\eea
In \eqref{KK_RR}, $(\alpha_\lambda,\beta^k)$ is a basis of $H^{3}_-(Y,\bbR)$,  $\omega_a,\omega_\alpha$ are
basis of $H^2_-(Y,\bbR),H^2_+(Y,\bbR)$, and $\tilde \omega^a,\tilde \omega^\alpha$ are a basis of $H^4_+(Y,\bbR),H^4_-(Y,\bbR)$.
Here we introduced the four-dimensional two-forms $(C_2^{\lambda},\tilde C^2_k)$ which are
dual to the scalars $(\xi^k,\tilde \xi_\lambda)$, already introduced in \eqref{def-N'T'}.
The vectors $A^\alpha$ have been already introduced in \eqref{Cvec}, and $A_\alpha$ are their four-dimensional
duals.
Moreover, the Kaluza-Klein expansion \eqref{KK_RR} also contains the four-dimensional three-forms
$(C^0_3, C_3^a,C^3_a,C_0^3)$ which are non-dynamical, but will crucially contribute to the scalar potential
as in ref.~\cite{Grimm:2008dq}.

Note also that the fields defined in \eqref{KK_RR} are not the expansions 
from the R-R forms alone, but in general combine with the NS-NS two-form $B_2$. 
Denoting by a hat $\hat {}$ the fields which arise in the expansion 
of the R-R forms alone, one finds, for example, that
\beq \label{B-corrected_vectors}
 B_2\text{-corrected:}\quad \left\{ \begin{array}{ccl} \text{vectors:} &\quad & A^\alpha = \hat A^\alpha \ , \qquad A_\alpha=\hat A_{\alpha} - \hat A^\beta b^a \cK_{\beta a \alpha} \ ,\\[.1cm]
 \text{3-forms:} && C^0_3 = \hat C^0_3 \ , \qquad C_3^a = \hat C_3^a + \hat C_3^0\, b^a \ , \qquad \text{etc.} 
 \end{array} \right. \qquad \quad
\eeq
where $\hat A^\alpha$, $\hat C_3^0$ and $\hat A_\alpha$, $\hat C_3^a$ denote the space-time vector bosons and 
three-forms coming from the expansion
of $C_3$ and $C_5$, respectively. In contrast, the scalars and two-forms in \eqref{KK_RR} have no mixing with the B-field 
such that 
\beq
   \text{no}\ B_2\text{-correction:} \qquad \quad\text{scalars:} \quad (\xi^k,\tilde \xi_\lambda) 
     \qquad \text{2-forms:} \quad (C_2^\lambda,\tilde C_k^2)\ . 
\eeq
As discussed in more detail in section \ref{mirror} the situation is precisely reversed 
under mirror symmetry. In fact, using the results on the side without $B_2$ corrections 
mirror symmetry can be used to compute the corrected couplings.

The Chern Simons action is dimensionally reduced by
inserting \eqref{KK_RR} into \eqref{ext_CS}. Focusing on the
couplings of $A^\alpha$ and $(C_2^{\lambda},\tilde C^2_k)$ in
favor over their duals, one finds \footnote{One could also include
the couplings to $A_\alpha$ and $(\xi^k,\tilde \xi_\lambda)$. In this case one
has to analyze also the bulk action keeping all forms and their duals as in ref.~\cite{Bergshoeff:2001pv}.}
\bea \label{eqn:CS}
        S^{(4)}_{\text{CS}} &=& \int \tfrac{1}{2} \I f_{\text{r}\, IJ}\, F^I\wedge F^J
       - (\delta_{I \lambda} d C_{2}^\lambda - \delta^k_I d \tilde C^{2}_k) \wedge A^I \\
       && \qquad \qquad \qquad \qquad \quad - (\cI_{I \lambda}\, d C_{2}^\lambda - \cI_{I}^k\, d\tilde C^2_k) \wedge d a^I
       + \cL_{\rm mix} + \cL_{3} \ . \nn 
\eea
Here $\cL_{\rm mix}$ corresponds to the mixing of the brane and bulk gauge bosons
\beq \label{def-Lmix}
      \cL_{\rm mix} = \big(a^J \Delta_{(I) J \alpha} + \Gamma_{(I) \alpha} \big)\, dA^\alpha \wedge F^I 
+ \tilde \cJ_{(I)}^\alpha dA_\alpha \wedge F^I\ ,
\eeq
and $\cL_3$ is the term which depends on the three-form field strengths as
\beq \label{def-L3}
\cL_3 = d C_3^0 \,
          \big(
              \tfrac{1}{2} a^I a^J \Delta_{I J}
            + a^J \tilde \Gamma_{J} \big)
            + d C_3^a\, \big(a^J \Delta_{J a}
            + \Gamma_{a}\big) + d C^3_a \, \tilde \cJ^a \ . 
\eeq

In order to display the couplings appearing in this action we first define
the integral $\cI(\tilde \alpha,\alpha)$ between a one-form $\tilde \alpha$ on $L_\eta$ and a three-form $\alpha$
on $Y$, as well as the integral $\cJ(\tilde \beta,\omega)$ between a two-form $\tilde \beta$ on $L_\eta$ and a
two-form $\omega$ on $Y$. To do that we again extend the forms defined on $L_0$ to the chain $\cC_4$ such that they are
constant along the normal directions of $L_\eta$ in $Y$. We define
\beq \label{def-cI_gen}
  \cI (\tilde \alpha,\alpha) = \int_{\cC_4} \tilde \alpha \wedge \alpha\ ,\qquad
  \cJ(\tilde \beta,\omega) = \int_{\cC_4} \tilde \beta \wedge \omega\ .
\eeq
Furthermore, we will also need a pairing $\delta$ between a 
a function $h$ on $L_0$ and three-form $\alpha$ on $Y$, as well 
as a pairing $\Delta$ between a one-form $\gamma$ on $L_0$ and 
a two-form on $Y$. Hence, we set
\beq \label{def-delta 2}
   \delta(h,\alpha)=\int_{L_0} h \, \alpha + \cI(d h,\alpha), \qquad
   \Delta(\gamma ,\beta)=\int_{L_0} \gamma \wedge \beta + \cJ(d\gamma,\beta)\ .
\eeq
Note that these latter definitions 
include terms supported on $L_0$ which are non-vanishing 
even in the limit of vanishing normal displacement 
$\eta$. This redefinition is necessary since $\cI$ and $\cJ$ vanish 
for a vanishing normal displacement. 
In fact, we can expand \eqref{def-cI_gen} to first order in $\eta$
for small normal displacement in $\cC_4=L_\eta - L_0$ and obtain
\begin{equation} \label{def-cI_exact}
   \cI(\tilde \alpha,\alpha) = \int_{L_0} \tilde \alpha \wedge \eta\lrcorner \alpha\ + ... \ \ ,
   \qquad   \cJ(\tilde \beta,\omega) =\int_{L_0} \tilde \beta \wedge \eta\lrcorner \omega\ + ... \ \ ,
\end{equation}
which has a leading term linear in $\eta$.

Having introduced the pairings we can display the couplings in \eqref{eqn:CS}, \eqref{def-Lmix} 
and \eqref{def-L3}.
Let us start with the couplings in \eqref{eqn:CS} obtained 
as 
\beq \label{def-cI}
  \cI_{I \lambda} = \cI(\hat \alpha_I, \alpha_\lambda) \ , \qquad
  \cI_{I}^k = \cI(\hat \alpha_I,\beta^k)\ , \qquad 
  \delta_{I \lambda} = \delta(h_I,\alpha_\lambda) \ ,\qquad
  \delta_I^k = \delta(h_I,\beta^k)\ .
\eeq
Furthermore, in the mixed term $\cL_{\rm mix}$, given in \eqref{def-Lmix}, for 
the gauge bosons one finds
\beq
   \Delta_{(I) J \alpha} = \Delta(h_I \hat \alpha_J , \omega_\alpha)\ ,\qquad \quad 
   \Gamma_{(I) \alpha} = \cJ(h_I f_{\rm D6}, \omega_\alpha)\ , \qquad \quad 
   \tilde \cJ_{(I)}^\alpha = \int_{\cC_4} h_I \tilde \omega^\alpha\ .
\eeq
Finally, we introduce the coefficients in \eqref{def-L3} as
\beq
  \Delta_{J a} = \Delta(\hat \alpha_J , \omega_a) \ ,\qquad \Gamma_a = \cJ(f_{\rm D6}, \omega_a)\ ,
\eeq
for couplings between the ambient space two-forms $\omega_a$ and forms $\hat \alpha_J$ and $f_{\rm D6}$
on the D6-brane. The remaining couplings are
\beq \label{def-tildecJ}
  \Delta_{IJ}  = \int_{L_0} \hat \alpha_I \wedge d \hat \alpha_J\ , \qquad  \tilde \Gamma_{J} = \int_{L_0} \hat \alpha_J \wedge f_{\rm D6} \ , \qquad    \tilde \cJ^{a} = \int_{\cC_4} \tilde \omega^a \ .
\eeq

It is not hard to interpret the different terms appearing in the
action (\ref{eqn:CS}). The first term corresponds to the theta-angle
term of the gauge theory on the D6-brane and thus contains the imaginary
part of the gauge kinetic function.
The second is a Green-Schwarz term
which indicates that the scalar fields $(\xi^k,\tilde \xi_\lambda)$
dual to the two-forms $(C^2_k,\tilde C_2^\lambda)$ are gauged by the D6-brane vector fields $A^I$.
In fact, upon elimination of $(C^2_k,\tilde C_2^\lambda)$ one finds the covariant derivative
\beq \label{gauge_xii}
  D \xi^k = d \xi^k + \delta_I^k A^I\ ,\qquad \quad D \tilde \xi_\lambda = d \tilde \xi_\lambda + \delta_{I\lambda} A^I\ ,
\eeq
We will show that the corresponding D-term appears in $V_{\rm DBI}$ as expected from
supersymmetry in section \ref{general}. The third term in \eqref{eqn:CS} will be of importance
for the derivation of the K\"ahler potential and complex coordinates on the $\cN=1$
field space. Upon elimination of $(C^2_k,\tilde C_2^\lambda)$ it induces a mixing
of the kinetic terms of $a^I=(a^i,\hat a^I)$ and $(\xi^k,\tilde \xi_\lambda)$. More precisely,
one finds the modified four-dimensional kinetic terms
\beq
  \cL^{\rm kin}_{C_3} = G_{kl}\nabla \xi^k \wedge * \nabla \xi^l  +
  G^{\lambda \kappa} \nabla \tilde \xi_\lambda \wedge * \nabla  \tilde \xi_\kappa
   + 2 G_{k}^{\ \lambda} \nabla \xi^k \wedge * \nabla \tilde \xi_\lambda
\eeq
where the modified derivatives $\nabla$ are defined by
\beq
\nabla \xi^k \equiv D \xi^k + \cI_I^k da^I
\quad \quad \textrm{and} \quad \quad
\nabla \tilde \xi_\lambda \equiv
D \tilde \xi_\lambda + \cI_{I\lambda} da^I,
\eeq
with the metric $G$ given as in the closed string case,
\beq \label{hodge-G}
  G_{kl} = \tfrac12\, e^{2D} \int_Y \alpha_k \wedge * \alpha_l\ ,\qquad
  G^{\lambda \kappa} =  \tfrac12\, e^{2D} \int_Y \beta^\lambda \wedge * \beta^\kappa\ ,\qquad
  G_k^{\ \lambda} = -\tfrac12\, e^{2D} \int_Y \alpha_l \wedge * \beta^\lambda\ .
\eeq
Note that the form of the metric $G$ for $\nabla \xi^k$ and $\nabla \tilde \xi_\lambda$
closely resembles the form of the metric $\cG_{ij}$
for the scalars $a^i$ as seen from \eqref{eqn:DBI} and \eqref{metric_a}.
We will exploit this observation in the detailed study of the 
moduli space geometry later on. This similarity
only occurs in the $\cN=1$ orientifold for which the field space metric
is K\"ahler. In the underlying $\cN=2$ set-ups the moduli space containing
the R-R scalars is a quaternionic manifold. 

The $\cL_{\rm mix}$  is a kinetic mixing
term between the $U(1)$ from the brane with the vector field from the $C_3$ expansion. This term will be important
in the derivation of the gauge coupling function in section \ref{modulispace}.

The term $\cL_3$ given by \eqref{def-L3} contains the four-dimensional three-forms which
arise in the expansion of $C_3,C_5,C_7$. Very similar to the analysis in ref.~\cite{Grimm:2008dq} they
will be crucial to complete the
scalar potential contributions in $V_{\rm DBI}$ to supersymmetric F-terms which
can be obtained from a superpotential. To find the scalar potential from the
three-form potential one has to eliminate the forms $dC_3^0$, $dC_a^3$ and $dC^a_3$
from the complete four-dimensional effective action. In particular, in addition to $\cL_3$
one also has to include the reduction of the ten-dimensional kinetic term
in \eqref{democratic_action}. 
The resulting action for the three-forms will be given in terms of the 
matrix $\cN_{\hat A \hat B}$ defined as 
\beq \label{def-cN}
  \cN_{\hat A \hat B} = \left( \begin{array}{cc} -\tfrac13 \cK_{abc} b^a b^b b^c  &  \tfrac{1}{2} \cK_{B a b} b^a b^b \\
                                                  \tfrac{1}{2} \cK_{A a b} b^a b^b & - \cK_{AB a} b^a  \end{array} \right) -  i \cV \left( \begin{array}{cc} 1 + 4 G_{ab} b^a b^b   &  - 4 G_{Ba} b^a \\
                                              -4 G_{A a} b^a & 4 G_{AB} \end{array} \right)\ ,
\eeq 
where $\hat A = \{0, a,\alpha \}$, and one has to use $\cK_{a b \alpha }= \cK_{\alpha \beta \gamma} =0$.
Using these definitions we find after rescaling to the Einstein that
\beq
  S_{\text{3-form}} =  \int \tfrac14 e^{-4D}  (\I \cN)^{-1\, \hat a\hat b}(dC^3_{\hat a}  - \cN_{\hat a\hat c}\, dC_3^{\hat c}) \wedge * (dC^3_{\hat b } - \bar \cN_{\hat b\hat d}\, dC^{\hat d}_3) +  \cL_3\ ,
\eeq
where $C^{\hat a}_3= (C_3^0,C_3^a)$ and $C_{\hat a}^3=(C_0^3,C^3_a)$, and
$\cL_3$ is the D-brane coupling defined in \eqref{def-L3}. As in ref.~\cite{Grimm:2008dq} we next dualize
$dC_3^0,dC_3^a$ and $dC^3_a,dC^3_0$ into flux scalars $e_0,e_a,m^a,m^0$. In ref.~\cite{Beasley:2002db} the interpretation
of these scalars as quantized fluxes has been provided. They also arise as background values of the
field strengths $F_2=m^a \omega_a, F_4=e_a \tilde \omega^a$ and $F_6=e_0 \vol_Y$ as there expansions
into harmonic forms on $Y$. In addition there is Romans mass parameter $F_0 = G_0 = m^0$.
After dualization of the three-forms on finds the scalar potential
\beq \label{Vcs}
  V_{\rm flux+CS} = \tfrac14 e^{-4D}  (\I \cN)^{-1\, \hat a\hat b}(\tilde e_{\hat a}  - \cN_{\hat a\hat c}\, \tilde m^{\hat c}) \wedge * (\tilde e_{\hat b } - \bar \cN_{\hat b\hat d}\, \tilde m^{\hat d}) \ ,
\eeq
where
\bea \label{tildeflux}
  \tilde e_0 &=& e_0 + \tfrac12 \int_{\cC_4} \tilde F \wedge \tilde F + \tfrac12 \int_{L_0} \tilde F\wedge a^I \hat \alpha_I , \\
  \tilde e_a &=& e_a + \int_{\cC_4} \tilde F \wedge \omega_a \ + \int_{L_0} a^I \hat \alpha_I \wedge \omega_a , \nn \\
  \tilde m^a &=& m^a + \int_{\cC_4} \tilde \omega^a\ , \qquad \tilde m^0 = m^0\ . \nn
\eea
The additional terms in the definitions \eqref{tildeflux} arise precisely because of the
term $\cL_3$ form the D6-brane. Luckily, apart from these shifts, the closed string
moduli dependence of the potential \eqref{Vcs}
agrees with the analog expression found in ref.~\cite{Grimm:2004ua}, and we will thus be able to integrate
it into a superpotential without much effort.

\subsubsection*{Restriction of the brane action to harmonic modes}

To conclude our reduction of the D6-brane action let us also give the
result which is obtained by restricting to harmonic forms. This 
corresponds to a truncation of the Kaluza-Klein tower of the brane fields to include only 
the lightest states. The resulting action will be useful in the next section 
when analyzing the moduli space.
The Kaluza-Klein Ansatz for the D6-brane field strength, eqn.~\eqref{exp_FD6_spbasis}, simplifies to
\beq \label{FD6-light}
F_{\rm D6}=F + da^i \wedge \tilde \alpha_i + f_{\rm D6}\ .
\eeq
This implies that the DBI action reduces to 
\beq \label{eqn:DBIlight}
         S^{(4)}_{\text{DBI}}
         =-\int \tfrac{1}{2} \R f_{\text{r}} \, F\wedge * F
         + e^{2D} \mathcal{G}_{ij}\, d a^i \wedge * d{a}^{j}
         + e^{2D} {\widehat \cG}_{ij}\, d\eta^i \wedge \ast d \eta^{j}
           \ ,
\eeq
with the metric $\cG_{ij}$ being the same as in 
\eqref{metric_a}, and  $\widehat \cG_{ij}$ the restriction of 
\eqref{eqn:metrics} to supersymmetric deformations (i.e., harmonic one-forms $\theta_i$).
The gauge coupling function \eqref{gauge_coupling} 
simplifies to 
\beq \label{gauge_couplinglight}
 \Re f_{{\rm r}}= \int_{L_0} 2\, \R(C\Omega) \ ,
\eeq
as we restrict $h_I$ to the only harmonic function, the constant function which we normalized to $1$.
We did not include the scalar potential $V_{\textrm{DBI}}$ 
since it vanishes when restricting to the harmonic subset of forms, as we 
will show in section \ref{general}.

The truncation of the Chern-Simons action to the harmonic modes is
\bea \label{eqn:CSlight}
 S^{(4)}_{\text{CS}} &=& \int \tfrac{1}{2} \I f_{\text{r}\, }\, F\wedge F
       - (\delta_{\lambda} d C_{2}^\lambda - \delta^k d \tilde C^{2}_k) \wedge A
       - (\cI_{i \lambda}\, d C_{2}^\lambda - \cI_{i}^k\, d\tilde C^2_k) \wedge d a^i \\
&&     + \big(a^j \Delta_{j \alpha} + \Gamma_{\alpha} \big) dA^\alpha \wedge F + \tilde \cJ^\alpha dA_\alpha \wedge F
       + dC_3^a \big(a^j \Delta_{j a} + \Gamma_{a}\big)+ dC^3_a\, \tilde \cJ^a\nn
       + dC_3^0\, \big( a^j \tilde \Gamma_{j} \big) \nn \ ,
\eea
with couplings
\bea
   \delta_{\lambda} &=& \int_{L_0} \alpha_\lambda, \qquad 
   \delta^k = \int_{L_0} \beta^k,\qquad
   \cI_{i \lambda} = \int_{\cC_4} \tilde \alpha_i \wedge \alpha_\lambda, \qquad
    \cI_{i}^k = \int_{\cC_4} \tilde \alpha_i \wedge \beta^k\ , \\
   \Delta_{i A} &=& \int_{L_0} \tilde \alpha_i \wedge \omega_A,\qquad 
   \Gamma_{A} = \int_{\cC_4} f_{\rm D6} \wedge \omega_A, \quad A=\{a,\alpha\}, \qquad 
    \tilde \Gamma_{i} = \int_{L_0} \tilde \alpha_i \wedge f_{\rm D6} \ , \nn 
\eea
and $\tilde \cJ^{A} = \int_{\cC_4} \tilde \omega^A$ as defined in \eqref{def-tildecJ}.
One realizes that the couplings $(\delta_{\lambda},\delta^k)$ and $\Delta_{i A},\tilde \Gamma_{i} $
are constants, while the couplings $(\cI_{i \lambda},\cI_{i}^k)$ and $\Gamma_A$ depend on the brane deformations through the 
chain $\cC_4$.

Let us take a closer look at the three-form couplings $\cL_3$. We can expand the $\cC_4$ chain around the $L_0$ cycle to
see the explicit dependence on the brane deformations. Just like \eqref{def-cI_exact}, we obtain, up to first order in the open fields,
\beq  \label{cL3restr}
  \cL_3 = dC_3^a  \int_{L_0} \big( a^j \tilde \alpha_j \wedge \omega_a
            + \eta^j  s_j \lrcorner \omega_a \wedge f_{\rm D6} \big) 
     + dC^3_a\,  \int_{L_0} \eta^j s_j \lrcorner \tilde \omega^a 
       + dC_3^0\,  \int_{L_0} a^j \tilde \alpha_j \wedge f_{\rm D6}. \quad
\eeq
Note that this implies that $\cL_3$ is non-vanishing also in the case we 
restrict to harmonic forms only. However, note that \eqref{cL3restr}
describes a coupling between the open and closed sector. In fact,
the scalar potential \eqref{Vcs} arising 
from \eqref{cL3restr} is obtained as an F-term potential when varying the
superpotential with respect to the closed string fields $t^a$.

\section{The open-closed moduli space and the Hitchin functionals \label{modulispace}}

In this section we discuss the geometry of the
moduli space of the bulk sector and brane sector
in more detail. In the first part, section \ref{closed_orientifold_space}, we assume that
the open moduli are frozen and discuss the geometry
of the moduli space $\cM^Q$ of the dilaton and
the real complex structure deformations following \cite{Grimm:2004ua}.
In section \ref{special_Lagr_mod} we discuss the moduli space of
special Lagrangian deformations $\eta^i$ following the work of Hitchin
\cite{Hitchin:1997ti,Hitchin:1999fh}. This description will be slightly 
extended by including the NS-NS B-field.
The open moduli space has finite dimension $b^1(L_0)$
and can be encoded by the variation of harmonic one- or two-forms on $L_0$.

In the complete set-up, with varying open and closed modes, 
the definition of being special
Lagrangian crucially depends on both the K\"ahler as well as the complex
structure moduli of $Y$. In fact, the normal vectors $s_i$ used in order to
define the one-forms $\theta_i=s_i \lrcorner J$ need to be chosen
such that $\theta_i$ is harmonic.
This notion changes when varying
the complex and K\"ahler structure of $Y$. 
Nevertheless, if such
a change does not alter the topology of $Y$ and $L_0$, one
expects to find a new embedding map $\iota'$ which makes $L_\eta$
supersymmetric in $Y$ and posses also $b^1(L_0)$ special Lagrangian
deformations. This suggest to view the full moduli space as
fibration of the open string moduli space $\cM_o^\bbC$ over the
closed string moduli space $\cM^K_{\bbC} \times \cM^Q_{\bbC}$, where $\cM^K_{\bbC}$
is the space spanned by the complexified K\"ahler deformations. In
section \ref{subsec:Kpot} we will explore the local geometry of this full
moduli space in more detail. Note that we are still dealing with only 
a finite set of deformations. In the absence of background fluxes these 
remain massless due to the vanishing of the scalar potential.

In section \ref{gauge-coupling-functions} we also analyze the gauge coupling function 
and the kinetic mixing for the brane and bulk $U(1)$ gauge fields. 
In particular, we comment on its holomorphicity properties.

\subsection{The orientifold moduli space \label{closed_orientifold_space}}

Let us first discuss the moduli space $\cM^Q$ for the closed
string modes $e^D$ and the $h^{(2,1)}$ real complex structure deformations denoted
by $q^K$. Its metric takes the form
\beq \label{def-G}
  \tfrac12 G = dD\cdot dD + K^{\rm cs}_{KL}\, dq^K \cdot dq^L \ ,
\eeq
where $K^{\rm cs}_{KL}$ is the Weil-Petersson metric restricted
to the slice of real complex structure deformations preserving
the orientifold constraint \eqref{OactionO6}. As suggested already
in \eqref{def-Omegac} and \eqref{def-N'T'} one describes the geometry of this space by considering the
the three-form $2\,\R (C\Omega) \in H^{3}_+(Y,\bbR)$ with periods
$U^k=2\, \R(CX^k)$ and $U_\lambda =2\, \R(C\cF_\lambda)$.
In these new coordinates $U^K = (U^k,U_\lambda)$ the metric $G$ in
\eqref{def-G} is obtained as a second
derivative of the real function \cite{Grimm:2004ua}
\beq \label{def-Kc}
  K^Q (V) = -2\ln\Big[i \int_Y C\Omega \wedge \overline{C\Omega} \Big] = -2 \log \big[e^{-2D}\big]\ .
\eeq
Then the first derivatives of $K_c$ are given by
\beq \label{K-der}
   \frac{1}2 \frac{\partial K^Q}{\partial U^k}= 2\, e^{2D} \I (C\cF_k)\equiv V_k \ , \qquad
   \frac{1}2 \frac{\partial K^Q}{\partial U_\lambda} =-2\, e^{2D} \I (CX^\lambda)\equiv  V^\lambda\ .
\eeq
The second derivatives of $K^Q$ can be evaluated explicitly as
well
\beq \label{G_as_der}
  G = \frac{\partial^2 K^Q}{\partial U^K \partial U^L}\ dU^K \cdot dU^L = G_{kl}\ dU^k \cdot dU^l + G^{\lambda \kappa}\ dU_\lambda \cdot dU_\kappa        + 2 G_{k}^{\ \lambda}\ dU^k \cdot dU_\lambda\ .
\eeq
Here one checks that the components of $G$ in these coordinates are precisely
as defined in \eqref{hodge-G} by either using a truncated version of
$\cN=2$ special geometry as in ref.~\cite{Grimm:2004ua} or by applying the techniques developed
by Hitchin in ref.~\cite{Hitchin:2000jd} as done in \cite{Benmachiche:2006df}.\footnote{$H(\R(C\Omega))=i \int_Y C\Omega \wedge \overline{C\Omega}$ is also known as entropy or Hitchin functional of the real three-form $\R(C\Omega)$.}
The fact that \eqref{hodge-G} is the
metric for the scalars $(\xi^k,\tilde \xi_\lambda)$ allows us to identify
local complex coordinates $N'^k=U^k + i \xi^k$ and $T'_\kappa = U_\lambda + i \tilde \xi_\lambda$
on a K\"ahler manifold $\cM^{Q}_\bbC$ which is
locally of the form $\cM^Q \times H^3_+ (Y,\bbR/\bbZ)$. The K\"ahler potential
for the metric $G$ on $\cM^Q_\bbC$ is precisely $K^Q(N+\bar N, T+\bar T)$ given in \eqref{def-Kc}.

Note that originally
$\cM^Q_\bbC$ was found as the $\cN=1$ field-space obtained by truncating
the underling quaternionic geometry spanned by the $\cN=2$
hypermultiplets. Each hypermultiplet has been truncated to a single $\cN=1$
chiral multiplet such that $\cM^Q$ has half the real dimension of
the quaternionic space. However, in order to prepare for the discussion
of the moduli space of special Lagrangian submanifolds, one notes that
$\cM^Q$ can also be viewed as a Lagrangian submanifold of a vector space.
The map embedding $\cM^Q$ in this special way is of the form
\beq \label{def-F_c1}
  F^Q:\quad \cM^Q \hookrightarrow V \times V^*\ , \qquad V =  H_+^3(Y,\bbR)\ ,
\eeq
where $V^*\cong H^{3}_-(Y,\bbR)$ is the dual vector space of $V$.
$\cM^Q$ is Lagrangian
with respect to the natural symplectic structure $\mathfrak{w}$ on $V\times V^*$,
i.e.~$F^{Q*} \mathfrak{w}=0$, and its metric is induced by the natural metric
$\mathfrak{g}$ on $V\times V^*$, i.e.~$F^{Q*} \mathfrak{g} = G$. Explicitly, $\mathfrak{w}$ and
$\mathfrak{g}$ are given by
\beq \label{def-wg}
  \mathfrak{w}((a,a'),(b,b')) = a'(b) - b'(a)\ , \qquad \mathfrak{g}((a,a'),(b,b')) = a'(b) + b'(a)\ ,
\eeq
where $a'(a)$ is the application of $a' \in V^*$ to $a \in V$. In the case at hand, $\mathfrak{w}$ and
$\mathfrak{g}$ can be evaluated using the wedge product $a'(b) = \int_Y a' \wedge b$, and
the map $F^Q$ is given by
\beq \label{def-F_c2}
   F^Q :\quad (D,q^K) \ \mapsto \ (U^k \alpha_k - U_\lambda \beta^\lambda, V^\lambda \alpha_\lambda+ V_k \beta^k) =
                             (2\R(C\Omega),-2e^{2D} \I(C\Omega))\ .
\eeq
Since $\cM^Q$ is a Lagrangian subspace of $V\times V^*$ with induced metric $G$,
it can be obtained as a graph of a function $K^Q$ which is the potential introduced
in \eqref{def-Kc}. Note that the embedding of $\cM^Q$ satisfies another special property,
since
\beq \label{no-scale}
  \mathfrak{g}(F^Q(p),F^Q(p))= 2(V_k U^k + V^\lambda U_\lambda) = 4\ ,
\eeq
for every point $p$ on $\cM^Q$.
This additional condition corresponds to the fact that, upon complexification with the
R-R scalars, the K\"ahler metric satisfies the no-scale type condition
\beq
  K^Q_{M^{\prime K}} K^{Q \, M^{\prime K} \bar M^{\prime L}} K^Q_{\bar M^{\prime L}} = 4\ ,
\eeq
where $M^{\prime K} = (N'^k,T'_\lambda)$
are the complex coordinates introduced in \eqref{def-N'T'}.

It worthwhile to mention that a similar logic can also be applied to
the K\"ahler sector of the orientifold theory. In this case the functional
$K^K$ is simply given by the logarithm of the Calabi-Yau volume. One finds
that for the coefficients $v^a$ of $J = v^a \omega_a$ that
\beq
  \frac{\partial^2 K^K}{\partial v^a \partial v^b} = \frac{1}{4\cV}\int_Y \omega_a \wedge * \omega_b\ ,\qquad
    K^K = -\ln\Big[ \tfrac43 \int_Y J \wedge J \wedge J \Big]\ ,
\eeq
which is the analog of \eqref{def-Kc} and \eqref{G_as_der}.
Here also one finds a natural Lagrangian embedding $F^K$ of the moduli space $\cM^K$
into a vector space, which is now of the
form $V\times V^* = H^{2}_-(Y,\bbR) \times H^4_+(Y,\bbR)$. Here the
special non-scale property of $K^K$  translates to $\mathfrak{g}(F^{\rm ks}(p),F^K(p))=3$
for each $p$ in $\cM^K$. The complexification of $\cM^K$ is via the $B_2$ scalars as in \eqref{expJB}
and we locally have $\cM^K_{\bbC} = \cM^K \times H^{2}_-(Y,\bbR/\bbZ)$.

Let us conclude the discussion of the moduli space $\cM^Q \times \cM^K$ by presenting
yet another way to motivate its geometrical structures. In an orientifold compactification
it is well-known that the orientifold planes, located on the fix-points of the involution $\sigma$, are
not dynamical and hence do not posses moduli at weak string coupling. Hence, all deformations
in $\cM^Q \times \cM^K$ need to preserve the embedding of the fix-planes and thus the conditions
\eqref{sp_Lagr_O6}.
Clearly, this is indeed the case for the scaling of $e^{D}$.
Also the real complex structure and K\"ahler structure deformations chosen such
that $\I(C\Omega)$ and $J$ remain
elements of $H^3_-(Y,\bbR)$ and $H^{2}_-(Y,\bbR)$ ensure that these forms
vanishes on the fix-point locus of $\sigma$.
In the discussion of the D6-brane moduli space we will turn the story around and consider
the variations of the D-brane embedding maps $\iota$ which preserve the conditions \eqref{sp_Lagr}
for fixed closed string fields.

\subsection{The moduli space of D6-branes on special Lagrangian submanifolds  \label{special_Lagr_mod}}

In the following we will discuss the moduli space of a supersymmetric
D6-brane wrapped on a special Lagrangian cycle on a Calabi-Yau manifold $Y$
with fixed complex and K\"ahler structure following \cite{Hitchin:1997ti,Hitchin:1999fh}.
At the end of this subsection we propose a simple modification to include the B-field.

\subsubsection*{The geometry of the moduli space of  special Lagrangian submanifolds}

To begin with, recall that the space of maps from a three-dimensional
manifold $L$ into $Y$ is infinite dimensional if no further restrictions are imposed.
However, we have derived the potential \eqref{DBI_potential} for modes violating the supersymmetry
constrains \eqref{sp_Lagr} rendering these fields massive.
Reducing the general deformation problem to
embeddings $\iota$ which preserve \eqref{sp_Lagr} reduces the
problem to the study of a finite dimensional deformation space $\cM_o$.
In section \ref{KKspectrum} we already stated McLeans result that this
moduli space $\cM_o$ is $b^1(L_0)$-dimensional. At linear order
its geometry can be studied by considering
the variations of harmonic one-forms $\eta^i \theta_i =\eta^i s_i \lrcorner J$ on $L_0$.
Here $s_i$ is a normal vector parameterizing a deformation through special
Lagrangian submanifolds, and $J$ is a fixed background K\"ahler form which
vanishes on $L_0$. The Hodge dual to $\theta_i$ on $L_0$ can be obtained
as contraction of $\I(C\Omega)$ with $s_i$ as given in \eqref{useful_id_2}.
The variations of the $\theta_i$ and $*\theta_i$ are analyzed by expanding these
forms in an integral basis $\tilde \alpha_i$ of $H^{1}(L_0,\bbZ)$ and $\tilde \beta^i$ of
$ H^{2}(L_0,\bbZ)$ respectively,
\beq \label{theta-periods}
  \theta_i = \lambda_{i}^j\, \tilde \alpha_j\ ,\qquad  \qquad \tfrac12 e^{-\phi} *\theta_i = \mu_{ji}\, \tilde \beta^j\ ,
\eeq
where $\lambda_{i}^j(\eta)$ and $\mu_{ij}(\eta)$ define the periods of $\theta_i$ and $e^{-\phi} *\theta_i$.
Explicitly they are given by 
\beq
  \lambda_{i}^j = \int_{L_0} s_i \lrcorner J \wedge \tilde \beta^j\ , \qquad \quad \mu_{ij} = -\int_{L_0} s_j \lrcorner \I (C\Omega) \wedge \tilde \alpha_i\ .
\eeq 
Note
that we have introduced an additional factor of the dilaton, which is constant for a fixed background,
but will later allow us to make contact to the metrics found in section \ref{D6reduction}. 
By using the closedness of $J$ and $\I(C\Omega)$ one shows that there exist functions $(u^i,v_i)$
such that \cite{Hitchin:1997ti}
\beq \label{def-uv}
  \frac{\partial u^i}{\partial \eta^j} = \lambda_j^i\ ,\qquad \qquad \frac{\partial v_i}{\partial \eta^j} = \mu_{ij}\ .
\eeq
In fact, $(u^i,v_i)$ are the analogs of $(U^K,V_K)$ for the orientifold moduli space \eqref{K-der}.

Let us point out that the harmonic one-forms $\theta_i^\eta$ can be constructed on each 
$L_\eta$ obtained by a supersymmetric deformation of $L_0$ \cite{Hitchin:1997ti}. 
Generalizing \eqref{theta-periods} we can pull back $\theta_i^\eta$ from 
$L_\eta$ to $L_0$ using the exponential map $E$ introduced in section \ref{gen-defD6}.
Following the 
strategy of section \ref{kin_action} we can then use the chain $\cC_4$ to write
\beq \label{lambda_mu_chain}
  \lambda_{i}^j  = \partial_{\eta^i} \int_{\cC_4} J \wedge \tilde  \beta^j \ , \qquad  \quad \mu_{ji}  = -\partial_{\eta^i} \int_{\cC_4} \I C\Omega \wedge \tilde \alpha_j\ .
\eeq
which at linear order reproduces \eqref{theta-periods} on $L_0$.
Inserting \eqref{lambda_mu_chain} into \eqref{def-uv} this provides us with a chain integral expression for the coordinates $(u^i,v_i)$.

To obtain the differential geometrical structure on $\cM_o$
one follows a similar logic as in \eqref{def-F_c1} and \eqref{def-F_c2},
and defines the map
\bea
   F_o : & \cM_o & \hookrightarrow\ V \times V^*= H^{1}(L,\bbR) \times H^2(L,\bbR)\ ,\\         
         & \eta^i & \mapsto\ (u^i \tilde \alpha_i, v_i \tilde \beta^i) \nn \ .
\eea
Using this map $\cM_o$ is embedded as a Lagrangian submanifold with respect
to the natural symplectic form $\mathfrak{w}$ in \eqref{def-wg} on $V\times V^*$,
where now $a'(b)=\int_L a'\wedge b$ \cite{Hitchin:1997ti}.
Moreover, the induced metric obtained from $\mathfrak{g}$, defined in \eqref{def-wg}, is evaluated
to be
\beq
  F_o^* \mathfrak{g} = \cG_{ij}\ du^i \cdot du^j = \widehat \cG_{ij}\ d\eta^i \cdot d \eta^j\ ,
\eeq
where $\cG_{ij}$ is explicitly given in \eqref{metric_a} and $\widehat \cG_{ij}$ can be found in \eqref{eqn:metrics}.
It is straightforward to evaluate the metrics in terms of the periods $\lambda_{i}^j$ and $\mu_{ij}$
using \eqref{theta-periods} and \eqref{def-uv} as
\beq \label{cG_lambda_2}
   \widehat \cG_{ij} = \mu_{ki} \, \lambda_{j}^k \ ,\qquad \cG_{ij} = \mu_{ik}\, (\lambda^{-1})_j^k\ .
\eeq
From the fact that $\cM_o$ is a Lagrangian submanifold
one finds that it can be locally represented by a single function
$K_o$ with $v_i = \partial K_o / \partial u^i$. This is the
direct analog of \eqref{K-der}. Moreover, using the fact that $F_o^* \mathfrak{g} = du_i \cdot dv^i$
the metric on $\cM_o$ is the Hessian of $K_o$ with respect to $u^i$, i.e.~$\cG_{ij} = \partial^2 K_o/\partial u^i \partial u^j$.

As in the case of the orientifold moduli space, we next have
to define a complexification of $\cM_o$ to
obtain the space $\cM_o^\bbC$. Let us first consider
the case of vanishing B-field. Since the metric $\cG_{ij}$ in the
coordinates $u^i$ agrees with the metric for the Wilson line moduli $a^i$, found in
\eqref{eqn:DBI}, one defines complex coordinates $\zeta^i$ on
$\cM_o^\bbC$ as
\beq \label{def-zeta_noB}
   \text{no B-field:} \qquad \zeta^i = u^i + i a^i\ ,
\eeq
and identifies $K_o(\zeta+\bar \zeta)$ as a K\"ahler potential
such that
\beq \label{open_metric}
   \cG_{ij} = \frac{\partial^2 K_o}{ \partial u^i \partial u^j}=
   4 \frac{\partial^2 K_o}{ \partial \zeta^i \partial \bar \zeta^j}\ .
\eeq
The metric $\cG_{ij}$ on $\cM_o^\bbC$ satisfies an important additional property. In
fact, it turns out that $\cM_o^\bbC$ is actually a non-compact Calabi-Yau manifold with
non-vanishing holomorphic $b^1(L_0)$-form $\widehat \Omega = d\zeta^1 \wedge \ldots \wedge d\zeta^{b^1}$
with constant length with respect to the K\"ahler form on $\cM_o^\bbC$ \cite{Hitchin:1997ti}.
However, it is important to note that $K_o$ cannot be simply extended to a
K\"ahler potential on a compact Calabi-Yau manifold due to its apparent
shift symmetry $\zeta \rightarrow \zeta + i c$, for constants $c$. As well-known
these shift symmetries will however be broken by non-perturbative effects coupling
with instanton factors $e^{-\zeta^i}$. There has been much progress in understanding
such corrections for in the holomorphic superpotential by explicitly 
computing the Type IIB chain integrals. Recent works in this direction include~\cite{Sum1,Jockers,Alim:2009rf,Grimm:2009ef}, and references 
therein. The study of corrections to the K\"ahler potential is significantly more involved, since 
it is not protected by holomorphicity. 

\subsubsection*{Open coordinates with B-field}

So far we have analyzed in this subsection the open moduli space for vanishing $B_2$ and 
$f_{\rm D6}$. We want to generalize this in the following.
To include the B-field we note from \eqref{lambda_mu_chain} and \eqref{def-uv} that $u^i$ can be written by using the 
four-chain in \eqref{def-cC4} as 
\beq \label{u-chain}
  u^i = \int_{\cC_4} J \wedge \tilde \beta^i = \int_{L_0} \eta \lrcorner J \wedge \tilde \beta^i + \ldots ,
\eeq
where we have also given the $\eta$ expansion for small fluctuations around $L_0$. 
One can now replace $J$ in \eqref{u-chain} by 
$-i J_c = J -i B_2$ as used for the closed coordinates in \eqref{expJB}. 
This leads us to modify \eqref{def-zeta_noB} as 
\beq  \label{def-zeta_withB}
  \zeta^i = u^i_c + i a^i \ , \qquad \quad u^i_c = - i \int_{\cC_4} J_c \wedge \tilde \beta^i\ .
\eeq
Note that $u^i_c$ is the complexification of $u^i$ with a  
B-field correction which can be absorbed by a shift of $a^i$. This implies that 
\eqref{open_metric} remains to be valid.

In the definition \eqref{def-zeta_withB} we have used the chain $\cC_4$ with boundaries 
$L_0$ and $L_\eta$. It is desirable to introduce a similar extension which allows to include 
the gauge field. To do that we introduce an extension $\cF_{\rm D6} = d \cA_{\rm D6}$ 
of the gauge connection $A_{D6}$ to the chain $\cC_4$ such that 
\beq \label{cA-boundary}
  \cA_{\rm D6}|_{L_0} =  A_{\rm D6}^0\ ,\qquad \quad \cA_{\rm D6}|_{L_\eta} = A_{\rm D6}^0 - a^I \hat \alpha_I\ ,
\eeq
where $\hat \alpha_I$ and $A_{\rm D6}^0$ have been transported trivially from $L_0$ to $L_\eta$ along the geodesic given by $\eta$.
Here $A_{\rm D6}^0$ is a background gauge bundle on $L_0$ which for fixed 
$B_2$ allows to satisfy the supersymmetry conditions on $L_0$. In other words, for a constant $B_2$ along the chain, $\cF_{\rm D6}$ 
might satisfy the supersymmetry 
conditions on $L_0$ but violate the supersymmetry conditions on $L_{\eta}$ due to non-trivial Wilson line scalars $a^I$. 
Importantly this prescription can also be used for $\eta \rightarrow 0$. In this case, one does not 
deform $L_0$ but changes the gauge connection by non-trivial scalars $a^I$. The imaginary part of the $\cN=1$ 
coordinates arising from the gauge connection $A_{\rm D6}$ can now be also written as a chain integral
$\int_{\cC_4} \cF_{\rm D6} \wedge \tilde \beta^i$. Thus, we find that the $\zeta^i$ are given by the elegant expression 
\beq \label{def-zetai_simple}
   \zeta^i = -i \int_{\cC_4} (J_c - \cF_{\rm D6}) \wedge \tilde \beta^i\ .
\eeq
At leading order in the $\eta$-expansion the complex coordinates $\zeta^i$ are 
encoded by a one-form $\cA_c$ on $L_0$ with expansion 
\beq
  \cA_c =  - i \eta \lrcorner J_c + i A_{\rm D6} = \zeta^i \tilde \alpha_i \ ,
\eeq 
into a basis $\tilde \alpha_i$ of $H^{1}(L_0,\bbZ)$. Let us close by noting that 
\eqref{def-zetai_simple} naturally includes a possible D6-brane flux. It would be 
interesting to evaluate all expressions found below including this flux. However, 
we will keep $f_{\rm D6} =0$ in most of the computations.

\subsection{The open-closed K\"ahler potential and $\cN=1$ coordinates \label{subsec:Kpot}}

In the following we determine the $\cN=1$ data for the
kinetic terms of the four-dimensional effective action by
specifying the $\cN=1$ complex coordinates, the K\"ahler potential and
the gauge coupling function for the U(1) gauge theory on the
D6-brane. We will do this by only including a finite set of deformations
specified in the last two subsections. Note that these
deformations will be obstructed by a scalar potential, since
one always needs to impose the supersymmetry conditions \eqref{sp_Lagr}
for the deformed D6-brane which depend on both the open as
well as closed moduli. One thus expects that only a space
of complex dimension smaller than $\frac12 b^3(Y)+h^{1,1}_-(Y)+b^1(L_0)$
can be studied as a true open-closed moduli space
which is classically un-obstructed by a scalar potential in the
absence of background fluxes.
This can be also understood by noting that Type IIA compactifications
with D6-branes will admit an M-theory embedding as
a compactification on a $G_2$-manifold \cite{Harvey:1999as,Gukov:1999gr,Beasley:2002db}. The
finite number of massless deformations of this manifold
will incorporate the subset of the closed and open deformations of section \ref{closed_orientifold_space}
and \ref{special_Lagr_mod} which are flat directions of the supersymmetry conditions
\eqref{sp_Lagr}.

Let us start by noting that the D6-brane degrees of freedom are
still encoded by the complex coordinates
$\zeta^i$ which have been introduced in \eqref{def-zeta_noB} and \eqref{def-zeta_withB}.
From the closed string sector we find the complexified K\"ahler structure
deformations $t^a$ introduced in \eqref{expJB}. As we will check later on,
the definition of the remaining
closed string complex coordinates is corrected by a functional
depending on the open coordinates
$\zeta^i$. More precisely, they arrange very elegantly as
\beq \label{N=1coords}
   N^k = U^k - 2\, \partial_{V_k} (e^{2D} K_o) + i \xi^k\ ,\qquad
   T_\lambda = U_\lambda - 2\, \partial_{V^{\lambda}} (e^{2D} K_o) + i \tilde \xi_\lambda ,
\eeq
where the real scalars $(\xi^k,\tilde \xi_\lambda)$
arise in the expansion \eqref{def-Omegac}, and we recall that
$U^k =2\R( CX^k)$, $U_\lambda=2\R(C\cF_\lambda)$ as well as $V_{k}=2 e^{2D}\I(C \cF_k)$, $V^\lambda=-2 e^{2D}\I(CX^\lambda)$
are periods of $C\Omega$. In summary, we can simply write
\beq \label{N=1coords1}
  \zeta^i = u^i_c + i a^i\ ,\qquad M^K = U^K - 2\, \partial_{V_K} (e^{2D} K_o) + i \xi^K\ ,
\eeq
where $\xi^K=(\xi^k,\tilde \xi_\lambda)$ and the abbreviations $U^K=(U^k,U_\lambda)$ and $V_K=(V_k,V^\lambda)$
are as in \eqref{K-der}. The real function $K_o$ is now dependent on both $u^i$ as
well as $U^{K}$ (or rather $V_K$). To see this, note that $e^\phi *\theta_i = 2 s_i \lrcorner \I (C\Omega)$   
as introduced in \eqref{def-theta}, clearly depends on $\I(C\Omega)$.
Performing the $\eta$-expansion of $K_o$ around $\eta=0$ one finds
\bea \label{coord_expansion}
   - 2\, \partial_{V_k} (e^{2D} K_o) &=& -\, \partial_{V_k} (e^{2D} \cG_{ij})|_{\eta = 0} u^i u^j  + \ldots \ , \\[.2cm]
      &=& - \frac12 \int_{L_0} \tilde \alpha_i \wedge s_l \lrcorner \beta^k 
		\Big(\int_{L_0} \tilde \beta^j \wedge s_l \lrcorner J \Big)^{-1}  u^i u^j  + \ldots  \ ,
 \nn
\eea
as we derive in detail in appendix \ref{derivation_K}. Together with a similar expression for $\partial_{V^\lambda} (e^{2D} K_o)$, replacing $\beta^k \rightarrow \alpha_\lambda$, one can use \eqref{coord_expansion} to derive the 
leading order effective action. In order to do that, we also need to specify the 
K\"ahler potential, to which we will turn next. Realize that
as a trivial check of \eqref{N=1coords1} one recovers
the bulk $\cN=1$ coordinates $(N'^k,T'_k)$ given in \eqref{def-N'T'} if $K_o=0$.

To encode the leading order D6-brane effective action found in
\eqref{eqn:DBI} and \eqref{eqn:CS}, we finally need to specify the
K\"ahler potential. It is given by
\beq \label{eqn:kaehler-pot}
    K=K^{\rm ks}+ K^{\rm Q}= -\ln\Big[ \tfrac43 \int_Y J \wedge J \wedge J \Big] -2\ln\Big[i \int_Y C\Omega \wedge \overline{C\Omega} \Big]\ , \qquad
    e^{K} =\frac{1}{8} e^{4D} \cV^{-1}.
\eeq
Note that $K$ has to be evaluated in terms of the $\cN=1$
coordinates \eqref{N=1coords} and thus only depends on
$\zeta^i+\bar \zeta^i$, $M^K +\bar M^K$ and $t^a-\bar t^a$. This can be done explicitly for
the first term $K^{\rm ks}$ since
\beq
  K^{\rm ks}(t,\bar t) = - \ln \big[\tfrac i6 \cK_{abc} (t-\bar t)^a (t-\bar t)^b (t - \bar t)^c \big]\ ,
\eeq
where $\cK_{abc}=\int_Y \omega_a \wedge \omega_b \wedge \omega_c$ are the triple intersection numbers.
It corresponds to the volume of the Calabi-Yau manifold $Y$ and will be corrected
by perturbative and non-perturbative string worldsheet contributions. For the second
term $K^{\rm Q}$ it is in general hard to find an explicit expression in terms of the
$\cN=1$ coordinates. However, we are nevertheless able to check
that the general kinetic terms determined by the derivatives of $K^{\rm Q}$
match the leading order terms found by dimensional reduction.

Let us summarize the derivatives of the K\"ahler potential $K^Q$.
We note that the derivatives with respect to the closed
string moduli $N^k,T_\lambda$ take the same form as in \eqref{K-der},
$\partial_{N^k} K = V_k$, $\partial_{T_\lambda}K = V^\lambda$.
However, $(V_k,V^\lambda)$ now depend implicitly on the open string coordinates $\zeta^i$
through the evaluation of the closed string expressions in terms of the
$\cN=1$ coordinates \eqref{N=1coords}, i.e.~one has to view $V_K(u^i,U^K)$. The derivatives
with respect to $\zeta^i$ will be postponed to section \ref{general}.
In summary one finds that
\beq\label{first-derK}
  {K_i}  =  e^{2D} v_i \ ,\qquad \quad K_k  = 2\, e^{2D} \I (C\cF_k) \ ,
  \qquad \quad K^\lambda= -2\, e^{2D} \I (C X^\lambda)\ .
\eeq
where $K_i =\partial K / \partial {\zeta^i}$, $K_k={\partial K}/{ \partial {N^k}}$
and $K_\lambda= {\partial K}/{\partial {T_\lambda}}$.
Also the K\"ahler metric can be evaluated explicitly.
One finds for the derivatives with respect to $(N^k, T_\lambda,\zeta^i)$
that
\bea \label{Kaehler_metric}
  K_{k \bar l} &=& G_{kl}\ , \hspace{3.5cm} K_{\lambda \bar \kappa}\ =\ G^{\lambda \kappa}\ ,\hspace{1.2cm}  K_{k \bar \lambda} \
  =\ G_{k}^\lambda\ ,\\
  K_{i \bar \jmath} &=& e^{2D} \cG_{ij} +  \cI_{i}^K G_{KL}  \cI_j^{L}\ , \qquad
   K_{i \bar k} \ =\  \cI_{i}^L G_{L k}\ ,\qquad K_{i \bar \lambda}\ =\ \cI_i^L G_{L }^\lambda\ , \nn
\eea
where $G_{KL}=(G_{kl},G^{\lambda \kappa},G_{k}^\lambda)$ was given in \eqref{hodge-G}, and
$\cI_i^K = (\cI_{i}^k,\cI_{i\lambda})$ are the derivatives
\beq
  \cI_i^k = \frac{\partial^2 K_o}{\partial V_k \partial \zeta^i}\ ,\qquad
  \cI_{i\lambda} = \frac{\partial^2 K_o}{\partial V^\lambda \partial \zeta^i} \ .
\eeq
In appendix \ref{derivation_K} we will check these expressions by an explicit computation, and
match these data with the leading order effective action obtained in section \ref{D6reduction}.

Let us comment on the special form of the K\"ahler metric \eqref{Kaehler_metric}. It can
be directly inferred by making use
of the invariance of the kinetic terms under the shift symmetries
\beq \label{shifts}
  N^k \ \rightarrow \ N^k + i \Lambda^k \ ,\qquad T_\lambda \ \rightarrow \ T_\lambda + i \Lambda_\lambda\ ,
\eeq
for arbitrary constants $(\Lambda^k,\Lambda_\lambda)$. If such shift symmetries
exist in the full four-dimensional effective action one can replace
the chiral multiplets $N^k$ and $T_\lambda)$ by linear multiplets
$(V_k,C^2_k)$ and $(V^\lambda,C_2^\lambda)$, as described in more details in appendix \ref{linm}.
Here $V_K=(V_k,V^\lambda)$ are the scalars dual to
$(\R N^k, \R T_\lambda)$ given in \eqref{first-derK} and $(C^2_k,C_2^\lambda)$
are two-forms dual to the scalars from $C_3$. The chiral multiplets and linear multiplets are connected
by a Legendre transform, and the new real function encoding
the kinetic terms of the multiplets is given by
\bea \label{correctedK}
  \tilde K(V,\zeta + \bar \zeta) &=& K(V) - V_k (N^k + \bar N^k) - V^\lambda (T_{\lambda} + \bar T_\lambda)\\
    &=& K(V) + 4\frac{\partial(e^{2D} K_o)}{\partial V_K} V_K - 4\ , \nn
\eea
where we have inserted \eqref{N=1coords1} and used \eqref{no-scale} to obtain the constant term $-4$.
The key point to notice is that in this dual picture all quantities are functions
of $V_K,\zeta^i$. In particular, this implies that now $K(V)=-2 \ln (e^{-2D}) =-2\ln (i \int C\Omega \wedge \overline{C\Omega})$
is independent of $\zeta^i$, and all equalities found
for the moduli space of special Lagrangian cycles of section \ref{special_Lagr_mod}
can be directly applied.
Since the linear multiplet picture is just an equivalent dual description
one can equally express the kinetic terms in the chiral multiplet picture
in terms of the derivatives of $\tilde K$.
Let us denote by $\tilde K^{KL} =
\partial_{V_K}\partial_{V_L} \tilde K$, and by $\tilde K_{KL}$ its inverse. Similarly,
we denote by $\tilde K_{\zeta^i}^K$ and $\tilde K_{\zeta^i \zeta^j}$ the remaining
second derivatives with respect to $\zeta^i$ and $V_K$.
The expression for the kinetic terms then has the form
\bea \label{general_lin}
 \cL^{\rm kin} &=& - (\tilde K_{\zeta^i \bar \zeta^j} + \tilde K_{\zeta^i}^K \tilde K_{KL} \tilde K_{\bar \zeta^i}^L)\ d\zeta^i \wedge * d\bar \zeta^j
    + \tilde K_{K L}\left( d\R M^I \wedge * \R M^J + d\xi^K \wedge * d\xi^J \right) \nn \\
&&  - 2 \, \tilde K_{K L} \tilde K^L_{ \zeta^i} \left(d\R M^I \wedge * du^j +  d\xi^I \wedge * d a^j\right)
\eea
This is precisely the form of the K\"ahler metric \eqref{Kaehler_metric} and it remains
to check that indeed $\tilde K_{KL}=G_{KL}$, $\tilde K_{\zeta^i \bar \zeta^j} = e^{2D} \cG_{ij}$
and $\tilde K_{\zeta^i}^K = \cI_i^K$. For the leading order actions found in section \ref{D6reduction} 
this is done in appendix \ref{derivation_K}. Note that the form of the metric \eqref{general_lin} is
also inherited if only a potential term breaks the
shift-symmetries \eqref{shifts}.

Let us make a brief comment on the appearance of the term $d\R M^I \wedge * du^j$.
 This term corresponds to a kinetic mixing between complex structure and brane deformations, 
and would be expected to appear in higher order expansions of the Dirac-Born-Infeld action. 
In this section however it was obtained by simply analyzing the $\cN=1$ characteristic data 
and the moduli space.

\subsection{Gauge coupling functions and kinetic mixing for finite deformations} \label{gauge-coupling-functions}

Having discussed the kinetic terms for the scalars in the $\cN=1$ effective 
theory we will now turn to an analysis of the kinetic terms for the $U(1)$
vectors fields. We have shown in section \ref{D6reduction} in the case one 
focuses on harmonic modes in the reduction that the spectrum contains 
a D6-brane $U(1)$ vector $A$ as well as $h^{(1,1)}_+$ bulk $U(1)$ vectors $A^\alpha$.
The leading gauge coupling function for the brane $U(1)$ was 
derived in section \ref{kin_action} and given by
\beq \label{red_fr}
 f_{{\rm r}}= \int_{L_0} \big(2\, \R(C\Omega) + i C_3)  = \delta_k N'^k - \delta^\lambda T'_\lambda \ ,
\eeq
where $\delta_k=\int_{L_0} \alpha_k$ and $\delta^\lambda=\int_{L_0} \beta^\lambda$. 
However, as we have discussed in section \ref{subsec:Kpot}, the inclusion of the 
open moduli forces us to introduce the modified complex coordinates $N^k,T_\lambda$ 
given in \eqref{N=1coords}. In order to obtain a holomorphic 
gauge coupling function it is expected that \eqref{red_fr} is modified to
\beq \label{U1couplingfull}
 f= \delta_k N^k - \delta^\lambda T_\lambda\ .
\eeq
The modifications in \eqref{U1couplingfull} did not appear in 
our leading order dimensional reduction, but are expected to arise 
a higher order in the brane deformations.
As we will see shortly open moduli corrections to $f_{{\rm r}}$
are also obtained after a careful treatment of the two dual bulk 
gauge fields $A^\alpha,\, A_\alpha$ introduced in \eqref{B-corrected_vectors}. 
Recall that the gauge coupling function for the bulk R-R $U(1)$ vectors 
$A^\alpha$ is simply given by  \cite{Grimm:2004ua}
\beq \label{RRcoupling}
 f_{\alpha \beta}= i \int_{Y} \omega_{\alpha} \wedge \omega_{\beta} \wedge \omega_{a} \,
t^a = i \cK_{\alpha \beta a} t^a = -i \bar \cN_{\alpha \beta} \ .
\eeq
where $\cN_{\alpha \beta}$ is the complex matrix already introduced in \eqref{def-cN}. 
Clearly, $f_{\alpha \beta}$ is holomorphic in the complex fields $t^a$. Since 
the $t^a$ are not corrected by the open moduli one expects the result \eqref{RRcoupling}
to remain valid also in the leading order reduction with a D6-brane. We will show in the following that this is indeed
the case. More interestingly, we find that there are further corrections 
depending on the open moduli and D6-brane fluxes which induce a kinetic mixing of 
the brane and bulk $U(1)$ gauge fields.

Let us now turn to a more careful analysis of the gauge coupling functions including the 
brane moduli. In order to do that we summarize the action for all vector fields 
including the dual $A_\lambda$ introduced in \eqref{B-corrected_vectors}. 
The mixing terms proportional to $dA^\alpha \wedge F$ and $dA_\alpha \wedge F$ 
have appeared in in the reduction of the Chern-Simons action in \eqref{eqn:CSlight}. 
The brane couplings have to be taken into account when 
eliminating $A_\alpha$ in favor of $A^\alpha$ by using vector-vector duality in 
four dimensions as enforced by \eqref{dual_constr}.
A detailed calculation can be found in appendix \ref{ap_mix} which uses a procedure 
similar to the one of ref.~\cite{Jockers:2004yj}. Here we just present the results. 
The action obtained after a careful elimination of $A_\lambda$ is
\bea
S^{(4)}_{\text{vec}} 
&=-& \int \tfrac12 \R f_\alpha dA^\alpha \wedge * F \nn
	+ \tfrac12 \I f_\alpha dA^\alpha \wedge F 
	+\tfrac12 \I \cN_{\alpha \beta} dA^\alpha \wedge * dA^\beta  \nn
	\\
&&	+\tfrac12 \R \cN_{\alpha \beta} dA^\alpha \wedge dA^\beta
	+ \tfrac{1}{2} \R f_{\text{cor}\, } 	 F \wedge * F
	+ \tfrac12 \I f_{\text{cor}} F \wedge F \nn
\eea
where the gauge coupling function $f_\alpha$ 
encoding the kinetic mixing between bulk and brane $U(1)$'s is given by
\beq \label{couplingmix}
f_\alpha 
	=- 4 (i \bar \cN_{\alpha \beta} \tilde \cJ^{\beta} 
	+ i a^j \Delta_{j \alpha} 
	+ i \Gamma_\alpha)  \ ,
\eeq
and the corrected gauge coupling function $f_{\text{cor}}$ for the brane $U(1)$ is
\beq \label{fcor}
	f_{\text{cor}}
=	f_{\text{r}} + 4 (i \bar \cN_{\alpha \beta} \tilde \cJ^\alpha 
	+ i a^j \Delta_{j \beta} + i \Gamma_\alpha) \tilde \cJ^\beta \ .
\eeq
The coefficient functions are given by $\tilde \cJ^\alpha = \int_{\cC_4} \tilde \omega^\alpha,\ \Delta_{j \alpha}= \int_{L_0} \tilde \alpha_j \wedge \omega_\alpha$ 
and $\Gamma_\alpha = \int_{\cC_4} \omega_\alpha \wedge f_{\rm D6}$ as introduced in section \ref{D6reduction}.
Recall that $\Delta_{j \alpha}$ is independent of the moduli, while $\tilde \cJ^\beta, \Gamma_\alpha $
depend on the brane deformations through the chain $\cC_4$.

To study the holomorphicity properties of the gauge couplings we discuss $f_\alpha$ and 
$f_{\text{cor}}$ in turn.
One notes that the first term in \eqref{couplingmix} 
can be rewritten as
\beq  \label{cj_expansion}
	i \bar \cN_{\alpha \beta} \tilde \cJ^\alpha
	= \int_{\cC_4}i \bar \cN_{\alpha \beta} \tilde \omega^\alpha 
	= \int_{\cC_4} (J- i B) \wedge \omega_\beta
	= u_c^j  \, \Delta_{j \beta}\ ,
\eeq
where we have used \eqref{def-zeta_withB} to obtain the factor $u_c^j$.
Using this expression it is straightforward to rewrite the 
gauge coupling $f_\alpha$ in the absence of brane fluxes as
\beq
	f_\alpha
	= - 4 \zeta^j \Delta_{j \beta} \ ,
\eeq
which is clearly holomorphic on the open moduli 
$\zeta^i = u_c^i + i a^i$. It would be interesting to extend 
these arguments to include the D6-brane flux $f_{\rm D6}$.

Let us now turn to the analysis of 
the corrected gauge coupling function $f_{\rm cor}$ of the brane $U(1)$. 
Using \eqref{fcor} and \eqref{couplingmix} one sees that it can be written as
\beq \label{fcor_rewrite}
	f_{\text{cor}} =f_{\text{r}}-f_\alpha \tilde \cJ^\alpha \ ,
\eeq
the additional term is at least of second order in the open moduli. 
One notes that the real part of $f_{\text{cor}}$ is given by
\beq \label{Rf_simple}
	\R f_{\text{cor}} =  \R f_{\text{r}} + 4 \I \cN_{\alpha \beta} \tilde \cJ^\alpha \tilde \cJ^\beta
	= \R f_{\text{r}} + \R f_\alpha \R f^{\alpha \beta} \R f_\beta \ ,
\eeq
which can be inferred from \eqref{couplingmix} and \eqref{fcor_rewrite}.
This result generalizes to the space of infinite deformations by replacing $f_{\rm r}$ with $f_{\text{r}\, IJ}$, and $f_\alpha$ with
$f_{\alpha I}$. The expressions for these are straightforward generalizations of \eqref{couplingmix}-\eqref{Rf_simple} with 
the abbreviations introduced in section \ref{kin_action}. Hence, the real part of the gauge coupling function takes the form 
\beq
  \R\, \mathbf{f} = \left( \begin{array}{cc}  \R f_{\text{r}\, IJ} + \R f_{\gamma I } \R f^{\gamma \delta} \R f_{\delta J}  \qquad  &   \R f_{I \alpha}\\
                      \R f_{J \beta} & \R f_{\alpha \beta} \end{array} \right), 
\eeq
and can be easily inverted.
This result will be important in section \ref{general}, 
when we compute the scalar potential coming from D-terms since it involved the inverse $ (\R\, \mathbf{f})^{-1} $.

Let us close this section by making some general remarks 
about the holomorphicity of the gauge coupling 
function $f_{\rm cor}$ in \eqref{fcor_rewrite}. 
In order to do that, one has express it in terms of the 
$\cN=1$ coordinates $N^k,T_\lambda, t^a$ and $\zeta^i$. 
However, recall from \eqref{N=1coords} that also the $N^k$ and $T_\lambda$ \
receive corrections by the open deformations. In fact, we 
$\eta$-expand
\beq
   \R (N^k - N'^k)\delta_k -  \R (T_\lambda -T_\lambda') \delta^\lambda = 
u^i \Big(- \tfrac12 \int_{L_0} \tilde \alpha_i \wedge \eta \lrcorner \beta^k  
\int_{L_0} \alpha_k + \tfrac12 \int_{L_0} \tilde \alpha_i \wedge \eta \lrcorner \alpha_\lambda  
\int_{L_0} \beta^\lambda \Big) + \ldots \ ,
\eeq
where we have used \eqref{u-chain} and \eqref{coord_expansion}. 
To compare this result, we also $\eta$-expand \eqref{fcor_rewrite} to find
\beq
   \R f_{\rm cor} - \R f_{\rm r} = 4 u^i \int_{L_0} \tilde \alpha_i \wedge \omega_\alpha  \int_{L_0} \eta \lrcorner \tilde \omega^\alpha + \ldots \ .
\eeq
This indicates that the result for $f_{\rm cor}$ cannot be complete. 
In particular, it is conceivable that a
contribution from the two-forms $\omega_a$ is missing which arises at higher order in the 
Kaluza-Klein reduction. This is similar to what was found in \cite{Jockers:2004yj,Grimm:2008dq}
for D7- and D5-branes on the type IIB side. 
It would be interesting to complete this computation to higher order and 
determine the fully corrected gauge coupling function.  For example, one loop corrections for the gauge-coupling function were
calculated for orbifold models in \cite{Honecker}.

\section{General deformations and the D- and F-term potential \label{general}}

In the previous section we considered D6-branes with a finite number of deformations 
arising from the expansion into harmonic forms on the brane world-volume. 
Using harmonic modes one infers that the scalar potential \eqref{DBI_potential} vanishes. A
non-vanishing potential precisely arises for deformations which violate the 
supersymmetry conditions that the three-cycle is special Lagrangian. In this section
we include such deformations into the discussion and analyze the $\cN=1$ encoding the  
geometry on the infinite field space. We discuss the K\"ahler potential and
show that the scalar potential \eqref{DBI_potential} indeed arises from a D-term, induced 
by a gauging, and a holomorphic superpotential. In order to do that we will keep the 
background geometry fixed and only consider the variations of the brane degrees of freedom.

\subsection{A local K\"ahler metric for general deformations of $L_0$}

In the general reduction performed in section \ref{gen-defD6} we already
included a whole tower of normal deformations of $L_0$ as well as the
whole tower of Kaluza-Klein modes in $F_{\rm D6}$ parameterizing variations
around a background connection $A_0$.
Together, these modes parameterize a neighborhood around $(L_0,A_0)$ in an
infinite dimensional field-space $\cV_o$.
We will focus on the neighborhood around a supersymmetric
$L_0$ and mainly be concerned with the local geometrical
structure of $\cV_o$. In order to do that we study the tangent space to $\cV_o$ at the special Lagrangian
$L_0$ with connection $A_0$. This tangent space is identified with
\beq \label{tangentspace}
  T_{(L_0,A_0)} \cV_o\ \cong\ TY|_{L_0} \cong NL_0 \oplus TL_0 \ .
\eeq
In this we can identify the $s_I$ introduced in \eqref{def-gen_theta} as basis of
sections of $NL_0$ and the $\tilde s_I^m = g^{mn}|_{L_0} (\hat \alpha_I)_n$ as sections
of $TL_0$. Note that in defining the tangent vector $\tilde s_I$ we have simply raised
the tangent index $m$ of the one-form $\hat \alpha_I$ introduced in \eqref{exp_AD6_gen} by the inverse of the induced metric $g_{mn}|_{L_0}$. This also means that we can identify
\beq
  T_{(L_0,A_0)} \cV_o\ \cong\ \Omega^1(L_0) \oplus \Omega^1(L_0)\ ,
\eeq
which  is naturally parameterized by the basis vectors $\theta_I$ and $\hat \alpha_I$ introduced in \eqref{def-gen_theta} and \eqref{exp_AD6_gen}.

Using the first identification in \eqref{tangentspace} the tangent space $T_{(L_0,A_0)} \cV_o$ admits a natural symplectic form
\beq \label{def-varphi}
  \varphi(X,Y) = \frac{1}{2} e^{-\phi} \int_{L_0} J(X,Y)|_{L_0} \vol_{L_0} \ .
\eeq
for $X,Y \in TY|_{L_0}$. It was shown in \cite{Hitchin:1999fh} that the two-form $\varphi$ 
on $\cV_o$ is actually closed. The tangent space 
\eqref{tangentspace} also admits a natural complex
structure $I$, which is the induced complex structure from the Calabi-Yau manifold $Y$.
At $L_0$ the complex structure $I$ identifies $TL_0$ with $NL_0$ such that
complex tangent vectors in $T_{(L_0,A_0)} \cV_o$ are given by
\beq
  \partial_{z^I} = \tfrac12 (s_I - i I s_I)\ , \qquad \partial_{\bar z^{\bar I}} = \tfrac12 (s_I + i I s_I)\ .
\eeq
Since this complex structure is formally integrable, the manifold $\cV_o$ is
K\"ahler, with K\"ahler form
\beq \label{general_Kaehler}
  \varphi(\partial_{z^I},\partial_{\bar z^J}) = \frac{i}{2} e^{-\phi} \int_{L_0} g(s_I,s_J) \vol_{L_0} = i \widehat \cG_{IJ}\ , \qquad  \varphi(\partial_{z^I},\partial_{z^J}) =   \varphi(\partial_{\bar z^I},\partial_{\bar z^J}) = 0 \ .
\eeq
Here we have used that $J(I s_I,s_J)=-g(s_I, s_J)$ and the fact that $L_0$ is Lagrangian
such that $J(s_I,s_J)=-J(I s_I,I s_J)=0$ for normal vectors $s_I$ to $L_0$.
This implies that $\widehat \cG_{IJ}$ is a K\"ahler metric, which is locally the
second derivative of a K\"ahler potential $K_o = K_o(z^I,\bar z^I)$.
Explicitly this means that 
\beq \label{cGIJ_z}
  \widehat \cG_{IJ} =  \partial_{z^I} \partial_{\bar z^J} K_{o} = \tfrac12 e^{-\phi} \int_{L_0} \theta_I \wedge * \theta_J\ ,
\eeq
with the forms $\theta_I$ as introduced in \eqref{def-gen_theta}. Note that the real part of the 
complex coordinates $z^I$ are the normal vectors $\eta^I$. This should 
be contrasted to the complex coordinates $\zeta^i$ which were the complexifications 
of the $u^i$ as discussed in section \ref{special_Lagr_mod}. In the appendix \ref{moment_maps} we
further analyze the symmetries of the symplectic form \eqref{def-varphi}. We argue that 
the first derivatives  of the K\"ahler potential $K_o$ are encoded by moment maps 
of these symmetries.

It is interesting to note that there is a natural generalization of the finite-dimensional 
analysis of section \ref{special_Lagr_mod} to the infinite dimensional deformation space. The key 
will be the use of the four-chain $\cC_4$ which interpolates between $L_0$ and $L_\eta$. 
Clearly, the natural generalization of the complex coordinates in \eqref{def-zetai_simple} is  
\beq \label{def-general_zeta_chain}
   \zeta^I = - i \int_{\cC_4} (J_c -\cF_{\rm D6}) \wedge \hat \beta^I \ ,
\eeq
where $\beta^I$ is the infinite basis of two-forms on $L_0$ which has been trivially extended to the 
chain $\cC_4$. We have also included the field strength $\cF_{\rm D6}$ on $\cC_4$ which is obtained from the 
gauge connection $\cA_{\rm D6}$ introduced in \eqref{cA-boundary}. A natural proposal for the K\"ahler potential $K_o$ is given 
by 
\beq \label{Ko_general}
  K_o(\zeta+ \bar \zeta) = - \frac12 \, \int_{\cC_4} J \wedge \hat \beta^I  \int_{\cC_4}\I (C\Omega) \wedge \hat \alpha_I \ .  
\eeq
This can be checked by performing an $\eta$-expansion around the supersymmetric 
cycle $L_0$. This yields the 
leading term
\bea \label{Koexpansion}
   K_o(\zeta+ \bar \zeta) &=& -\tfrac12 \int_{L_0} s_L \lrcorner J \wedge \hat \beta^I \int_{L_0} s_K \lrcorner \I (C\Omega)\wedge \hat \alpha_I \  \eta^L \eta^K  + \ldots  \\
                          &=& \tfrac14 e^{-\phi}\int_{L_0} \theta_L \wedge \hat \beta^I \int_{L_0} * \theta_K \wedge \hat \alpha_I \  \eta^L \eta^K  + \ldots \nn \\
                          &=& \tfrac12 \, \widehat \cG_{LK}\, \eta^L \eta^K + \ldots    \nn \\
                          &=& \tfrac{1}{8} \cG_{LK} (\zeta+\bar \zeta)^L \, (\zeta+\bar \zeta)^K  + \ldots\nn  
\eea
where here we mean by $\widehat \cG_{LK}, \cG_{LK}$ the leading order metrics independent of $\eta$. 
Here we have used \eqref{useful_id} on $L_0$ to rewrite the contraction $s_K \lrcorner \I (C\Omega)$ into the Hodge-star on $L_0$. 
Using \eqref{Koexpansion} one sees that \eqref{cGIJ_z} is satisfied. Let us stress that in general 
the evaluation of $K_o$ as a function of $\zeta^I + \bar \zeta^I$ is non-trivial due to the appearance 
of the chain $\cC_4$ in both integrals of  \eqref{Ko_general}. It would be very interesting the 
compute $K_o$ explicitly for specific orientifold examples, generalizing the superpotential computations 
of \cite{Sum1,Jockers,Alim:2009rf,Grimm:2009ef}.

\subsection{The superpotential and D-terms}

Having discussed the K\"ahler potential determining the kinetic terms, 
we will now examine the scalar potential 
in more detail. More precisely, we will work in a fixed background geometry by 
fixing K\"ahler and complex structure deformations and  
focus on the leading scalar potential $V_{\rm DBI}$ given in \eqref{DBI_potential}. We will show that $V_{\rm DBI}$ splits 
into an F-term and a D-term piece as 
\beq
V_{\rm DBI} = V_{F} + V_{D}\ ,
\eeq
with
\beq \label{D-termpot}
 V_D =  \frac{e^{3\phi}}{\cV^2} \int_{L_0} d^*\theta_\eta \wedge * d^*\theta_\eta
\eeq
and
\beq \label{VF_exp}
  V_{F} = \frac{e^{3\phi}}{\cV^2} \int_{L_0} d\theta_\eta \wedge * d\theta_\eta + (\tilde F -B_2 -d\theta_\eta^B) \wedge * (\tilde F -B_2 -d\theta_\eta^B)\ .
\eeq
We will show momentarily that $V_{F} = e^K  \cG^{IJ} \partial_{\zeta^I}W 
		\overline{\partial_{\zeta^J}W} $ can be obtained from a superpotential $W$ and the metric determined from $K_o$ 
using only the open string degrees of freedom.

To specify $W$ we aim to define a functional which picks out deformations
$\eta$ such that $L_\eta$ is a Lagrangian submanifold $J|_{L_{\eta}}=0$.
In section \ref{kin_action} we defined a chain $\cC_4$ with boundaries $L_\eta$ and $L_0$. 
Recall also that we extended the gauge field $A_{\rm D6}$ from $L_0$ to 
$\cC_4$ as in \eqref{cA-boundary}, such that the extension $\cF_{\rm D6} = d \cA_{\rm D6}$ 
satisfies  
\beq \label{cF-boundary}
  \cF_{\rm D6}|_{L_0} =  f_{\rm D6}\ ,\qquad \quad \cF_{\rm D6}|_{L_\eta} = f_{\rm D6}+ a^I d \hat \alpha_I\ .
\eeq
In the following we will again set again the D-brane flux $f_{\rm D6}$ to zero.
One next identifies the superpotential functional
\beq \label{general_W}
  W = \int_{\cC_4} (J_c - \cF_{\rm D6}) \wedge (J_c - \cF_{\rm D6}) 
\eeq
depending on the open string data as well as the complexified
K\"ahler form \eqref{expJB}. This is an extension of the
functional introduced in ref.~\cite{Thomas:2001ve}, 
since we have included the
B-field through the complex two-form $J_c$. Note that a superpotential 
of this form has been already discussed in \cite{Martucci:2006ij,Koerber:2006hh}.

Let us briefly study the holomorphicity properties of $W$. 
Clearly, $W$ is holomorphic with respect to variations of the
complexified K\"ahler form $J_c$ parameterized by the scalars
$t^a$ in \eqref{expJB}. However, note that one first has to express
$W$ as a function of the open fields $\zeta^I = u^I_c + i a^I$ introduced 
in \eqref{def-general_zeta_chain}.
To check that $W$ it is a holomorphic section in the $\zeta^I$ we show that 
$\partial_{\bar \zeta^I}W =(\partial_{u^I_c} + i \partial_{a^I})W = 0$.
The derivative with respect to Wilson lines is
\beq \label{super_a_derivative}
\partial_{a^I} W = 2 \int_{L_\eta}  (J_c - \cF_{\rm D6}) \wedge  \hat \alpha_I  =  2 \int_{L_0}  (J_c - \cF_{\rm D6}) \wedge  \hat \alpha_I  + 2 \int_{L^0} d (\eta \lrcorner J_c - a^J \hat \alpha_J )\wedge \hat \alpha_I +\ldots 
\eeq
To evaluate the derivative with respect to $u^I_c$ we expand the chain integral around the 
special Lagrangian cycle $L_0$ in terms of the deformations
\bea   \label{superpotgen_etaexp}
	W &=& 	2 \int_{L_0} \eta \lrcorner J_c \wedge (J_c- \tilde F) 
	+ \int_{L_0} \eta \lrcorner J_c \wedge \cL_\eta J_c  
	+ ... \\
&=&  2 \int_{L_0} (\eta \lrcorner J_c) \wedge (J_{c} - \cF_{\rm D6}) 
	+ \int_{L_0} \eta \lrcorner J_c \wedge d(\eta \lrcorner J_c + 2\, a^I \hat \alpha_I) + ... \nn
\eea
Recalling  $\eta \lrcorner J_c  = \theta_\eta^B + i\theta_\eta = i u_c^I \hat \alpha_I  + ... $ 
one sees that by comparing \eqref{super_a_derivative} with $\partial_{u_c^I} W$ obtained from \eqref{superpotgen_etaexp} that the 
superpotential is holomorphic in $\zeta^I=u^I_c + i a^I$.

It is now straightforward to determine the F-term potential using the expression \eqref{superpotgen_etaexp}.
The real part of the derivative of \eqref{superpotgen_etaexp} is given by  
\beq
	\R \, \partial_{\zeta^I} W
	= 2 \int_{L_0} d \theta_\eta \wedge \hat \alpha_I \ .
\eeq
Note that $d\theta_\eta$ is a 2-form in $L_0$ and therefore can be expanded in the infinite basis $*\hat \alpha_I$ as $d\theta_\eta = c^I * \hat \alpha_I$
The coefficients $c^I$ can be obtained by taking on both sides the 
wedge product with $\alpha_J$ and integrate on $L_0$. 
Inverting this relation for $c^I$ and taking the Hodge star one finds
\beq
* d\theta_\eta = \tfrac12 e^{-\phi} \hat \alpha_I \ \cG^{I J} \int_{L_0} \hat \alpha_J \wedge d\theta_\eta \ .
\eeq
We proceed analogously with the imaginary part $\I \, \partial_{\zeta^I} W$ obtained from \eqref{superpotgen_etaexp} and 
expand the two-form $(B-\tilde F + d\theta_\eta^B)$ in the $* \hat \alpha_I$ basis.
The F-term potential is thus given by 
\bea \label{ftermgen}
V_F&=& e^K\cG^{IJ} \partial_{\zeta^I}W 
		\overline{\partial_{\zeta^J}W}  \nn \\ 
&=&	\frac{e^{2D}}{2 \cV}   \int_{L_0} d\theta_\eta \wedge \hat \alpha_I \ 
		\cG^{IJ} \int_{L_0} \hat \alpha_J \wedge d\theta_\eta\nn \\
&&	+\frac{e^{2D}}{2 \cV}   \int_{L_0} (B-\tilde F+d\theta_\eta^B) \wedge \hat \alpha_I \ 
		\cG^{IJ} \int_{L_0} \hat \alpha_J \wedge (B-\tilde F+d\theta_\eta^B) 	\nn \\
&=&	 \frac{1}{\cV^2} e^{3\phi} \int_{L_0} d\theta_\eta \wedge * d\theta_\eta 
	+ (B - \tilde F + d\theta^B_\eta)\wedge * (B - \tilde F+d\theta^B_\eta) 
\eea
which agrees with the result \eqref{VF_exp} obtained from dimensional reduction, 
and reduces to the result of McLean \cite{McLean} in the limit of vanishing B field. 
As expected, the condition for vanishing of the potential and therefore 
to preserve supersymmetry is the closedness of $\theta_\eta$ and $\theta_\eta^B$, as 
well as the condition $(B-\tilde F)|_{L_0}=0$.

Finally, we also compute the D-term potential in \eqref{D-termpot} induced by the
gaugings of the scalars $\hat a^I$ in \eqref{exp_FD6_spbasis} and $(\xi^k,\tilde \xi_\lambda)$ in \eqref{gauge_xii}.
More precisely, these scalars are charged under the gauge transformations $A^I \rightarrow A^I + d\Lambda^I $
of the $U(1)$ vectors $A^I$ as
\beq
  \hat a^I \rightarrow \hat a^I - \Lambda^I\ , \qquad
  (\xi^k,\tilde \xi_\lambda) \rightarrow (\xi^k - \delta^k_I \Lambda^I,\tilde \xi_\lambda - \delta_{\lambda I} \Lambda^I)
\eeq
The potential arising from D-terms can be calculated by
\beq\label{Dfinite1}
V_D=\tfrac12 \R f^{AB} D_A D_B \ , \qquad
\partial_A D_I = K_{A \bar B} X_I^B \ ,
\eeq
where $X_I^B$ are the Killing symmetries appearing in the covariant derivative $D \xi^k = d\xi^k + X^k_I A^I$. 
Explicitly they take the form  
\beq
X^k_I =\int_{L_0} h_I \beta^k + \int_{\cC_4} dh_I \wedge \beta^k \ , \qquad \quad
	X_{I \lambda}
	= \int_{L_0} h_I \alpha_\lambda + \int_{\cC_4} dh_I \wedge \alpha_\lambda \ .
\eeq
The leading inverse gauge coupling function is simply
\beq
(\R  f_{\text{r}}^{-1})^{IJ} = \left(\int_{L_0}h_I h_J 2 \R (C\Omega)\right)^{-1}.
\eeq
Integrating \eqref{Dfinite1} we obtain the D-terms
\beq \label{D-terms}
   D_I = - 2 e^{2D} \left (\int_{L_0} h_I \I C \Omega  + \int_{\cC_4} dh_I \wedge \Im C\Omega \right) \ .
\eeq
We can expand the chain along an infinite set of brane deformations and obtain
\beq
 D_I=-2 e^{2D} \int_{L_0} h_I \I C \Omega -2 e^{2D} \int_{L_0} h_I d(\eta \lrcorner \I C \Omega) + \ldots\ ,
\eeq
where we have used that the functions $h_I$ are translated constantly along the chain.
Now we repeat a similar calculation as for the F-term, by expanding the three forms 
into $* h_I$ and noticing that on the $L_0$ cycle 
$\int h_J *h_I = e^\phi \int h_J h_I 2 \R (C\Omega) 
	= e^\phi \R f_{{\rm r} I J}$.
The potential is then,
\beq
  V_D = \frac{e^{3\phi}}{\cV^2} \int_{L_0} 4 \ \I C \Omega \wedge * \I C \Omega
 + 4 \ \I C \Omega \wedge  * d * \theta + d*\theta \wedge * d * \theta \ .
\eeq
From the condition $\I C\Omega|_{L_0} = 0$ only the last term survives, 
yielding the remaining term obtained from dimensional reduction. 
The vanishing of the D-term potential, which is necessary in a supersymmetric vacuum, happens when the two-form $*\theta_\eta$ is closed. 

\section{Mirror Symmetry with D-branes \label{mirror}}

In this final section we relate the Type IIA $\cN=1$ characteristic 
data found in the previous sections with the data for Type IIB orientifolds
with space-time filling D3-, D5- and D7-branes. In order to do that, we first review 
some basics of Type IIB orientifolds following \cite{Grimm:2004ua}. 
To define the orientifold set-up starting with Type IIB string theory 
compactified on a Calabi-Yau manifold $\tilde Y$, one acts with a discrete 
involutive symmetry ${\cal O}$ containing worldsheet parity $\Omega_p$. 
In Type IIB one still is left with two options of constructing such an involution. 
These correspond to the situations with O3/O7 or O5/O9 orientifold planes:
\beq\label{constrOB}
\begin{tabular}{lllll}
${\cal O}_1$ &=& $\Omega_p \sigma_B (-)^{F_L}\ ,\qquad$ &
$\sigma_B^* \Omega\ =\ - \Omega\ ,\qquad$  &O3/O7 ,\\[1ex]
${\cal O}_2$ &=& $\Omega_p \sigma_B$ \ ,
 &$ \sigma_B^* \Omega\ =\ \Omega\ ,$&   O5/O9 .
\end{tabular}
\eeq
Here $\sigma_B$ is a holomorphic 
(instead of antiholomorphic, as in the Type IIA case) 
involutive symmetry $\sigma_B^2=1$ of the Calabi-Yau 
target space, and ${F_L}$ is the space-time fermion number in the 
left-moving sector.
The subspace of fields which are invariant under the orientifold 
projection has to satisfy
\begin{equation} \label{fieldtransfB}
\begin{array}{lcl}
\\
\sigma^*_B  \phi &=& \   \phi\ , \\
\sigma^*_B    g &=& \  g\ , \\
\sigma^*_B    B_2 &=& -   B_2\ ,
\end{array}
\hspace{1cm}
\begin{array}{lcl}
\multicolumn{3}{c}{ \underline{O3/O7}} \\[2ex]
\sigma^*_B  C_0 &=& \ \  C_0\ , \\
\sigma^*_B    C_2 &=& -  C_2\ , \\
\sigma^*_B   \ C_4 &=& \ \  C_4\ ,
\end{array}
\hspace{1cm}
\begin{array}{lcl}
\multicolumn{3}{c}{ \underline{O5/O9}} \\[2ex]
\sigma^*_B  C_0 &=& -  C_0\ , \\
\sigma^*_B  C_2 &=& \ \  C_2\ , \\
\sigma^*_B  C_4 &=& -  C_4\ ,
\end{array}
\end{equation}
where the first column is identical for both involutions $\sigma_B$
in \eqref{constrOB}. The involution allows us to separate 
the cohomologies into even and odd eigenspaces 
$H^{(p,q)} = H^{(p,q)}_+ \oplus  H^{(p,q)}_-\ .$

Let us focus on the closed string sector for the moment. 
Locally the truncated moduli space of Type IIB orientifolds can then 
be written as a direct product 
\beq \label{modsp-B}
  \cM^{\rm K}_B \times \cM^{\rm Q}_B\ .
\eeq
Here $\cM^{\rm Q}_B$ is a K\"ahler manifold and 
spanned by the dilaton, the K\"ahler 
structure deformations, the NS-NS B-field 
and the R-R scalars. $\cM^{\rm K}_B$ is a special K\"ahler manifold 
spanned by the complex structure deformations 
of $\tilde Y$ respecting the constraints \eqref{constrOB}. 
In contrast, recall that in 
Type IIA $\cM^{\rm Q}_A$ is spanned by the dilaton, the complex structure 
deformations and the R-R scalars, 
while $\cM^{\rm K}_A$ is spanned by the K\"ahler deformations and the NS-NS B-field. 
The Type IIB effective theory also contains 
$h^{(2,1)}_+ (h^{(2,1)}_-)$ vector multiplets for 
orientifolds with $O3/O7 (O5/O9)$ planes, 
whereas in Type IIA one as $h_+^{(1,1)}$ vector multiplets.  
The number of multiplets from the closed string sector is shown in Table \ref{numberM}.\\
\begin{table}[h]
\begin{center}
\begin{tabular}{|l|c|c|c|} \hline
 \rule[-0.3cm]{0cm}{0.8cm}
multiplets& IIA$_Y$ \  $O6$ & IIB$_{\tilde Y}$ \  $O3/O7$ & IIB$_{\tilde Y}$ \  $O5/O9$ \\ \hline\hline
 \rule[-0.3cm]{0cm}{0.9cm}
 {vector multiplets} &   $h_+^{(1,1)}$ & $h_+^{(2,1)}$ & $h_-^{(2,1)}$ \\ \hline
\rule[-0.3cm]{0cm}{0.9cm}
 chiral multiplets in $\cM^{\rm K}$& $h_-^{(1,1)}$ & $h_-^{(2,1)}$ &
                      $h_+^{(2,1)} $   \\ \hline
 \rule[-0.3cm]{0cm}{0.9cm}
 chiral multiplets in $\cM^{\rm Q}$&$h^{(2,1)} + 1$&$h^{(1,1)} + 1$ &
                      $h^{(1,1)} + 1$   \\ \hline
\end{tabular}
\caption{Number of $N=1$ multiplets of orientifold compactifications.}\label{numberM}
\end{center}
\end{table}

Applying mirror symmetry to this $\cN=1$ set-up one expects 
that the $\cM^{\rm Q}_B$ space of type 
IIB should be identified with 
the $\cM^{\rm Q}_A$ moduli space 
of the mirror IIA, and similarly $\cM^{\rm K}_B$ 
with $\cM^{\rm K}_A$. 
Requiring $\tilde Y$ to be the mirror 
manifold of $Y$, the mirror map between
the moduli spaces implies that
for the different orientifold setups
\bea \label{matchchohm}
   O3/O7&: &\quad h^{(1,1)}_-(Y) = h^{(2,1)}_-(\tilde Y) \ , \qquad  h^{(1,1)}_+(Y) = h^{(2,1)}_+(\tilde Y) \ ,  \nn\\
   O5/O9&: &\quad h^{(1,1)}_-(Y) = h^{(2,1)}_+(\tilde Y)   \ , \qquad h^{(1,1)}_+(Y) = h^{(2,1)}_-(\tilde Y) \ ,
\eea
as well as $h^{(2,1)}(Y) = h^{(1,1)}(\tilde Y)$ for both set-ups. 
The mirror mapping for closed moduli is discussed in more detail in  
\cite{Grimm:2004ua}, and will be briefly recalled below.

In the following we want to extend the mirror identification 
to include the leading corrections due to the space-time filling 
D-branes. As we have seen, at leading order the 
moduli space $\cM^{\rm K}_A$ remains unchanged after 
the inclusion of open string moduli. This is also true for 
$\cM^{\rm K}_B$ on the Type IIB side. In section \ref{modulispace} 
we have shown that the open string moduli space of the D6-branes 
is fibered over the closed string moduli space $\cM^{\rm Q}_A$. 
The mirror equivalent of this statement has been established in 
\cite{Grana:2003ek,Jockers:2004yj,Grimm:2008dq} 
for $\cM^{\rm Q}_B$ and the moduli space of D3-, D5- or D7-branes. 
In the reminder of this section we 
will therefore focus on the discussion of the $\cM^{\rm Q}$
and establish the mirror map including the open degrees of 
freedom.

\subsection{Mirror of O3/O7 orientifolds}

The moduli space $\cM^{\rm Q}$
is obtained from the four-dimensional scalar parts of the fields $J, B_2,  C_2$, $C_4$. 
To make this more precise, we expand
\bea\label{expansionclosed}
   B_2 &=& b^k \, \omega_k\ ,\qquad C_2\ =\ c^k \, \omega_k\ ,
   \quad k=1,\ldots, h_-^{(1,1)}(\tilde Y)\ , 
\\
   J &=&  v^{\lambda} \, \omega_{\lambda}\ ,\qquad   C_4 =
   \rho_\lambda\ \tilde \omega^\lambda\ ,\quad \lambda = 1,\ldots,h_+^{(1,1)}(\tilde Y)\ .\nonumber
\eea
The complex coordinates and the K\"ahler potential 
which encode the local geometry of $\cM^{\rm Q}_B$ are \cite{Grimm:2004uq}
\bea \label{coordsO3/O7}
	\tau
&=&	C_0 + i e^{-\phi_B}\ ,  \qquad \quad G^k
=	c^k-\tau b^k \ , 
\\
	T_\lambda^{\prime B}
&=&	e^{-\phi_B}\tfrac{1}{2}\mathcal{K}_{\lambda \rho \sigma} v^{\rho} v^{\sigma} + i \rho_\lambda - i \tfrac12 \cK_{\lambda k l} b^k G^l
     	 \nn \ ,
\eea
and 
\beq \label{KaehlerO3/O7}
  K(\tau,G^k,T_\lambda^{\prime B}) =
-2 \ln \left[e^{-2\phi_B} \int_{\tilde Y} J \wedge J \wedge J \right] = \ln(e^{4 D_B}) \ .
\eeq
Here $D_B$ is the redefined four-dimensional dilaton. 
The K\"ahler potential has to be evaluated as a function of 
the moduli $\tau,G^k,T_\lambda'^B$ by solving \eqref{coordsO3/O7} for 
$v^a,\phi_B$ and inserting the result into \eqref{KaehlerO3/O7}.
The coefficients $\cK_{\lambda bc}$ are the intersection numbers of the basis 
$ \omega_{\lambda} $ of $H_+^{1,1}(\tilde Y)$ and
$ \omega_{a} $ of $H_-^{1,1}(\tilde Y)$, 
$\cK_{\lambda bc}=\int \omega_\lambda \wedge \omega_b \wedge \omega_c$.
Note that the above scalar fields can be also obtained from the expansion
\beq \label{moduliexpansionD3closed}
 - \R\, \Phi^{\rm ev} + i \sum_{n} e^{-B} \wedge C_{2n} 
= i \tau + i G^k \omega_k + T^{\prime B}_\lambda \tilde \omega^\lambda\ ,
\eeq
which has to be evaluated by matching the parts of different form degrees on both sides.
Here we have introduced the even form 
\beq \label{def-Phiev}
\Phi^{\rm ev} = e^{-\phi_B} e^{-B_2 + i J}
\eeq
following the notation of \cite{Benmachiche:2006df}.

Let us now recall the mirror map to the Type IIA coordinates without inclusion 
of the open string degrees of freedom. The $\cN=1$ coordinates $(N'^k,T'_\lambda)$
have been introduced in \eqref{def-N'T'}. Note that on a Calabi-Yau manifold 
we can use the rescaling invariance of $\Omega$ to fix one of the $X^I$ 
to be constant. At large complex structure there is a special real 
symplectic basis of $H^{3}(Y)$ which is distinguished 
by the logarithmic behavior of the solutions in the complex structure moduli 
of $Y$. In particular, this fixes a pair $(\alpha_0,\beta^0)$, by demanding that $X^0$, 
the fundamental period,
has no logarithmic singularity. One can use the rescaling of $\Omega$ to 
set the $\alpha_0$ period to a constant. 
Note that in the orientifold background $H^3(Y)$ splits
into $H_-^3$ and $H_+^3$. The component chosen to eliminate the 
rescaling property of $\Omega$ can be either in the positive or negative 
eigenspace of the orientifold projection. We will 
see momentarily these choices will correspond to 
different orientifold set-ups on the Type IIB side.

For the O3/O7 case 
we fix the component $X^0 \alpha_0$ in $H^3_+(Y)$. 
We define then the special coordinates $q$
and the scaling parameter $g_A$ as
\beq
	q^k = \frac{\R C X^k}{\R C X^0} \ , \qquad 
	q^\lambda=\frac{\I C X^\lambda}{\R C X^0} \ ,\qquad
	g_A^{-1} = 2 \R C X^0 \ .
\eeq
Recall that in the underlying $\cN=2$ theory, the periods of $\Omega$ are 
determined by a holomorphic pre-potential $\cF(X)$. 
Due to the homogeneity property of $\cF$ we can define a rescaled function $f$ as
\beq
	\cF(2 CX)=i (2 \R C X^0)^2 f(q^k,q^\lambda) \ 
\eeq
such that $C\Omega$ can be written as 
\begin{equation} \label{Omegaexplicit}
	2 C\Omega
	=g_A^{-1} \left [1 \alpha_0 + q^k \alpha_k + i q^\lambda \alpha_\lambda
	- f_\lambda \beta^\lambda - i (2 f - q^k f_k - q^\lambda f_\lambda) \beta^0 - i f_k \beta^k \right] \ ,
\end{equation}
where $(f_\lambda,f_k)$ are the derivatives of $f$ with respect to $(q^k,q^\lambda)$.
The coordinates $(N'^k,T'_\lambda)$ become in terms of these special coordinates
\begin{equation} \label{N'T'A}
	N'^0 = g_A^{-1} + i \xi^0 \, \qquad
	N'^k = g_A^{-1}q^k + i \xi^k \, \qquad
	T_\lambda^{\prime A} = g_A^{-1} f_\lambda + i \tilde \xi_\lambda \ .
\end{equation}

In order to provide complete match with the Type IIB side we 
need an explicit expression for $f_\lambda$ at the large complex 
structure limit of the Calabi-Yau manifold $Y$. The results will then 
be identified with the large volume results of Type IIB.
In this limit the $\cN=2$ pre-potential is given by
\beq
	\cF(X)=\frac{1}{6}\kappa_{I J K} \frac{X^I X^J X^K}{X^0} \ .
\eeq
Therefore, inserting the orientifold constraints and switching to
special coordinates we  find 
\beq
	f(q)=-\tfrac{1}{6} \kappa_{\lambda \mu \rho} q^\lambda q^\mu q^\rho
	+ \tfrac{1}{2} \kappa_{\lambda k l} q^\lambda q^k q^l \ ,
\eeq
such that one can readily evaluate the $T_\lambda^{\prime A}$ using \eqref{N'T'A}.
Now it is straightforward to relate the Type IIA coordinates with the ones 
from the Type IIB side
\beq \label{NandTD3}
 	(-i \tau, -iG^k)  	\leftrightarrow	(N'^0,N'^k) \qquad {\rm and} \qquad
	-T_\lambda^{\prime B} 	\leftrightarrow 	T_{\lambda}^{\prime A} \ ,
\eeq
with the matching of the cohomologies for the pair of mirror Calabi-Yau manifolds given in
Table \ref{mirror_cohomologyO3}.
In terms of the string moduli, the above relations translate into
\begin{align}\label{phi=g}
 	g_A^{-1}
&=	e^{-\phi_B} \ ,\qquad 
        &q^k &=\ - b^k\ ,  
        &q^\lambda &= v^\lambda\ ,
	  \\
	\xi_0 
&=	-C_0\ , \qquad
	&\xi^k &= -c^k+C_0 b^k\ , \quad 
        & \tilde \xi_\lambda  &=-\rho_\lambda 
	+ \frac12 \cK_{\lambda kl}c^k b^l
	- \frac{1}{2} C_0 \cK_{\lambda kl}b^k b^l\ . \nn
\end{align}

\begin{table}
\begin{center}
\begin{tabular}{|c|c|}\hline 
$H^3(Y)$					& $H^{\rm even} (\tilde Y)$			\\	\hline \hline
\rule[-.2cm]{0cm}{.6cm} $\alpha_0 \in H^3_+(Y)$ 		& $1$								\\ \hline
\rule[-.2cm]{0cm}{.6cm} $\alpha_k \in H^3_+(Y)$		& $\omega_k \in H^2_-(\tilde Y)$			\\ \hline
\rule[-.2cm]{0cm}{.6cm} $\alpha_\lambda \in H^3_-(Y)$	& $\omega_\lambda  \in H^2_+(\tilde Y)$		\\ \hline
\rule[-.2cm]{0cm}{.6cm} $\beta^k \in H^3_-(Y)$		& $\tilde \omega^k  \in H^4_-(\tilde Y)$		\\ \hline
\rule[-.2cm]{0cm}{.6cm} $\beta^\lambda \in H^3_+(Y)$	& $\tilde \omega^\lambda \in H^4_+(\tilde Y)$	\\ \hline
\rule[-.2cm]{0cm}{.6cm} $\beta^0 \in H^3_-(Y) $	& $\cV^{-1} \text{vol}_{\tilde Y}$					\\ \hline
\end{tabular}
\end{center}
\caption{The mirror mapping from the basis of $H^3(Y)$ to the
basis of even cohomologies of the mirror Calabi-Yau $\tilde Y$ in
$O3/O7$ orientifold setups.} \label{mirror_cohomologyO3}
\end{table}


\subsubsection*{Inclusion of D3 brane moduli}

In the discussion of mirror symmetry with D-branes we first consider 
the setup with spacetime filling D3 branes. 
The $\cN=1$ characteristic data were analyzed in \cite{Grana:2003ek}.
The brane is a point in the internal space $\tilde Y$, such that the
brane deformations $\eta$ are described by six scalar fields $\phi^I$ 
corresponding to the possible movements in $\tilde Y$. These fields naturally combine into 
complex fields $\phi^i$, $\phi^{\bar \jmath}$ with $i, \bar \jmath =1,2,3$
if one uses the inherited complex structure of the Calabi-Yau manifold.
Clearly, there are no Wilson line moduli for D3-branes since there is no internal 
one-cycle on the brane.
It turns out that, up to second order in the fields, 
only the coordinates $T_\lambda^{\prime B}$ are corrected by 
the open moduli \cite{Grana:2003ek}
\beq \label{coordsD3open}
	 \R\, T_\lambda^B = \R\, T_\lambda^{\prime B}
     	+i (\omega_\lambda)_{i \bar \jmath}(\phi_0)\, \phi^i \phi^{\bar \jmath}  \ ,
\eeq
where the two-form $(\omega_\lambda)_{i \bar \jmath}$ has to be evaluated at the 
point $\phi_0$ around which the D3-brane fluctuates. More generally, it was 
argued in ref.~\cite{DeWolfe:2002nn} that the D3-brane correction to $T_\alpha$ can 
be expressed through the K\"ahler potential $K_{\tilde Y}$ for the Calabi-Yau metric as
\beq \label{coordsD3openKY}
    \R\, T_\lambda^B = \R\, T_\lambda^{\prime B} - \partial_{v^\lambda} K_{\tilde Y}(\phi_0 + \phi)\ ,
\eeq 
where $v^\lambda$ are the K\"ahler moduli introduced in \eqref{expansionclosed}.
To obtain \eqref{coordsD3open} one expands $K_{\tilde Y}$ around the point $\phi_0$ as 
\beq \label{KY_expansion}
    K_{\tilde Y}(\phi_0 + \phi) = K_{\tilde Y}^0 + 2 \R \big[ (K_{\tilde Y})_i^0 \phi^i\big]   
     +\R \big[ (K_{\tilde Y})_{i j }^0  \phi^i \phi^j ] +  (K_{\tilde Y})_{i \bar \jmath }^0  \phi^i \bar \phi^j + \ldots \ ,  
\eeq
where $K_{\tilde Y}^0$, and $ (K_{\tilde Y})_i^0, (K_{\tilde Y})_{i j }^0, (K_{\tilde Y})_{i \bar \jmath }^0$ are the K\"ahler potential and its 
$\phi^i$-derivatives evaluated at $\phi_0$. Since the coefficients are constant, the first three terms in \eqref{KY_expansion} can be absorbed 
by a holomorphic redefinition into a new $T_\lambda^B$. Clearly, this does not change the complex structure on 
the $\cN=1$ moduli space. Using $(K_{\tilde Y})_{i \bar \jmath }^0 = - i J_{i \bar \jmath }^0 = - i v^\lambda (\omega_\lambda)_{i \bar \jmath }(\phi_0)$ one then 
recovers \eqref{coordsD3open}.

Let us now turn to the discussion of mirror symmetry. We aim to match the corrected 
coordinates $T_\lambda^B$ as well as the un-corrected $G^k$ and $\tau$ 
with the Type IIA side. This implies that we must have up to quadratic order in 
the brane moduli that 
\bea \label{matchopenD3}
   -2 \partial_{V^\lambda} (e^{2D_A} K_o) &=& \partial_{v^\lambda} K_{\tilde Y}(\phi_0 + \phi) 
                                         \,  \cong \, - i (\omega_\lambda)_{i \bar j} \phi^i \phi^{\bar j} \\
	\partial_{V_0} (e^{2D_A} K_o) 
	&=& \partial_{V_k} (e^{2D_A} K_o)
= 0 \ , \nn
\eea
where the $\cong$ indicates that one has to apply the transformation which 
identifies \eqref{coordsD3openKY} and \eqref{coordsD3open}. Using the 
fact that $V^\lambda = - e^{2D_{B}} e^{-\phi_B} v^\lambda$, as inferred from \eqref{Omegaexplicit},
the identification \eqref{matchopenD3} implies 
\beq
   K_o(\phi,\bar \phi) = \tfrac12 e^{-\phi_B} K_{\tilde Y}\ .
\eeq
The number of open moduli must coincide, so the number of brane deformations 
on the Type IIB must equal the number of brane and Wilson line moduli 
on the Type IIA side. 
Since this number is given by the number of non-trivial one-cycles in $L_0$, 
we must have $b^1(L_0)=3$. 
However, recall that the open moduli space in Type IIA has 
shift symmetries, $\I \zeta^i \rightarrow \I \zeta^i + c^i$, for constants $c^i$.
These are not manifested in the Type IIB side for a general $K_{\tilde Y}$, since 
the Calabi-Yau metric has no continuous symmetries. As we recall below, this 
can be attributed to the fact that instanton contributions break these symmetries
and are not included in this leading order identification.

Before commenting on the corrections to the mirror construction let us make contact to the 
chain integral form of the K\"ahler potential as given in \eqref{Ko_general}. For 
a D3-brane we simply have to introduce a one-chain $\cC_1$ which starts at $\phi_0$ 
and ends at the point in $\tilde Y$ to which the D3-brane has moved. We also introduce a 
basis of complex normal vectors $s_i$ to the point $\phi_0$ and dual $(1,0)$-forms $s^j_{(1)}$ such that 
\beq
   s_i \lrcorner s^j_{(1)} = \delta_i^j\ .
\eeq
Note that the index $i,j$ are counting here the number of such normal vectors. In case we only include the massless 
modes, one has $i,j=1,\ldots,3$. The complex structure of $s_i$ and $s^i_{(1)}$ is induced by the 
complex structure of $\tilde Y$, and hence depends on the complex structure moduli. In 
fact one can use the no-where vanishing $(3,0)$-form $\Omega$ on $\tilde Y$ and introduce 
a bi-vector $s^j$ 
such that $s^j_{(1)} = \bar s^{j} \lrcorner \Omega$.
To propose a form for $K_o$ one trivially extended $s_i, \bar s^i$ to the chain $\cC_1$ and writes 
\beq
    K_o = \tfrac{i}{4} e^{-\phi_B } \int_{\cC_1} s_i \lrcorner J  \int_{\cC_1} \bar s^{i} \lrcorner \Omega + c.c.\ .
\eeq
This form of $K_o$ is very suggestive and yields upon expanding the chain integral the desired leading order expression \eqref{coordsD3open}.
Moreover, we will see in the following that a generalization of this $K_o$ also arises for D7-brane, and one 
can generally write in O3/O7 orientifolds for the deformations of a $D (p+3)$-brane 
\beq \label{Kdef_suggest}
  K^{\rm def}_o = \tfrac{i}{4} \int_{\cC_{p+1}} s_I \lrcorner  \I \Phi^{\rm ev}  \int_{\cC_{p+1}} \bar s^{I} \cdot \Omega + c.c.\ .
\eeq 
where $\Phi^{\rm ev}$ has been introduced in \eqref{def-Phiev}, and $\cC_{p+1}$ is a  $(p+1)$-chain which ends on 
the internal parts of the D-branes and its reference cycle. 
Moreover, $s_I$ is an appropriate basis of complex normal vectors and $s^J$ are their 
duals as we discuss below. 

Before giving a more careful treatment of the other D-brane configurations let us 
first comment on a more intuitive understanding of mirror symmetry which we 
will apply below. 
It was argued  by Strominger, Yau and Zaslow \cite{Strominger:1996} that the 
Calabi-Yau manifold $\tilde Y$ can be viewed as a three-torus fibration 
with singular fibers. This manifold can be endowed with a semi-flat metric.
In a local patch avoiding possible singular points the metric of the Calabi-Yau 
manifold can be written as
\beq \label{semi-flat}
 ds^2=g_{a b}(\tilde u) d\tilde u^a d\tilde u^b + 2 g_{i a}(\tilde u) d\tilde a^i d\tilde u^a + g_{i j}(\tilde u) d\tilde a^i d\tilde a^j \ ,
 \qquad i, a=1,2,3 \ ,
\eeq
where $\tilde a^i$ are the coordinates on the $T^3$ fiber and $\tilde u^a$ of the base. Since the 
coefficient functions in \eqref{semi-flat} are independent of $\tilde a^i$ the shift symmetry is 
now manifest. In fact, introducing complex coordinates as in the Type IIA setting
a K\"ahler metric in \eqref{semi-flat} can be obtained 
from a K\"ahler potential $K_{\tilde Y}(\tilde u)$ which is independent of $a^i$.  
The argument for the existence of such a $T^3$-fibration with a metric of the form 
\eqref{semi-flat} away from singularities proceeds precisely via mirror symmetry 
of a pointlike D-brane on $\tilde Y$ which is mapped to a D-brane which wraps 
a three-torus \cite{Strominger:1996}.
Having found a $T^3$-fibration in the Type IIB set-up one can equally use T-duality 
along all $T^3$-directions to analyze the setting.
Since T-duality exchanges Neumann and Dirichlet boundary conditions, 
it exchanges the dimensionality of the brane for each wrapped cycle that is T-dualized. 
Starting with a D3-brane on such a fibered Calabi-Yau manifold, 
T-duality on the fiber will turn the brane into a D6-brane wrapping 
the $T^3$-fiber. The D6-brane then has $b^{1}(L_0)=3$ deformation moduli 
in the direction of the base, and there are also $b^{1}(L_0)=3$ Wilson line moduli will be along the torus.

In the following it will be more important that we can use the SYZ-picture also for D7- and D5-branes
present in a Type IIB reduction. Clearly, both types of branes will map to D6-branes under mirror symmetry. 
Away from the singular fibers one can obtain a clearer picture of the wrappings of the 
D6-branes as indicated in Table \ref{mirrortable}.

\begin{table}[h]
\begin{center}
\begin{tabular}{|c || c | c | | c | c | | c | c||} \hline
		& D6		& D3	& D6		& D7		& D6		& D5 	\\ \hline
		& $\times$	& 	& 		&$\times$ 	&  		&$\times$  \\ \cline{2-7}
T$^3$	& $\times$	& 	& 		&$\times$ 	&$\times$ 	&    		\\ \cline{2-7}
		& $\times$	& 	&$\times$	&  		&$\times$ 	&    		\\ \hline
		&  		&  	& 	 	& 		&  		& 		\\ \cline{2-7}
Base		&  		&  	&$\times$ 	&$\times$	&  		&  		\\ \cline{2-7}
		&  		& 	&$\times$ 	&$\times$	&$\times$ 	&$\times$	\\ \hline
\end{tabular}
\caption{It is summarized how mirror symmetry acts on different  
brane configurations. The table shows the six dimensions of the Calabi-Yau manifold, 
split into base and fiber. $\times$ indicate the directions wrapped by each 
brane. Mirror symmetry acts as T-duality on all directions of the $T^3$-fiber. 
It exchanges Dirichlet and Neumann boundary conditions, while it does not act 
on the base. Different wrappings of a D6-brane
correspond to different branes in the Type IIB side.}\label{mirrortable}
\end{center}
\end{table}

\subsubsection*{Inclusion of D7 brane moduli}

Let us now discuss mirror symmetry for the D7-brane case. The effective 
action for a pair of moving D7-branes was computed in \cite{Jockers:2004yj}. 
In this setup, the brane wraps a four-cycle $S^{(1)}$ while its orientifold image wraps a non-intersecting 
$S^{(2)}$. One can view the whole configuration as a single D7-brane wrapping a divisor 
$S_+ = S^{(1)} + S^{(2)}$. Brane deformations and Wilson line moduli can be expanded in 
terms of 
\bea \label{expansionD7}
 \chi&=&\chi^A s_A + \bar{\chi}^{\bar A} \bar s_{\bar A},\quad \quad A=1,\ldots,h^{(2,0)}_- \left(S_+\right),\\
 a&=& a^I \gamma_I+ \bar a^{\bar I} \bar \gamma_{\bar I} \, \, \, \quad \quad \quad I=1,\ldots,h_-^{(0,1)} (S_+)\ ,\nn
\eea
where $s_A$ and $\gamma_I$ are complex normal vectors to $S^{(1)}$ and $(0,1)$-forms on $S^{(1)}$, respectively. 
The complex type of $s_A$ and $\gamma_I$ is induced by the complex structure of $\tilde Y$. Moreover, 
one can use the holomorphic $(3,0)$-form $\Omega$ on $\tilde Y$ to map the $s_A$ 
to $(2,0)$-forms $\mathcal{S}_A = s_A \lrcorner \Omega$ on $S^{(1)}$.
The four-dimensional fields are thus the $h^{(2,0)}_- + h^{(1,0)}_-$ 
complex scalars $\chi^A$ and $a^I$, respectively.

Including the open string degrees of freedom, the chiral coordinates $(\tau,G_a,T_\lambda^{\prime B})$ are shifted to \cite{Jockers:2004yj}
\bea \label{coordsD7}
	S&=&\tau+\mathcal{L}_{A\bar B}\chi^A \bar{\chi}^{\bar B}  ,  \qquad \quad G^k\ =\ c^k-\tau b^k \ ,  \\
	T_\lambda^{B}
&=&	\tfrac{1}{2} e^{- \phi_B} \mathcal{K}_{\lambda \rho \sigma} v^{\rho} v^{\sigma} + i \rho_\lambda - i \tfrac12 \cK_{\lambda k l} b^k G^l
	+i\mathcal{C}_{\lambda {I\bar J}} a^I \bar a^{\bar J} \nn \ .
\eea
The coupling functions $\mathcal{L}_{A\bar B}$ and $\mathcal{C}_{\lambda \, I\bar J} $ for the basis of 
brane deformations and Wilson line moduli on the four-cycle are given by
\beq
 \mathcal{L}_{A\bar B} = \frac{\int_{S_{+}} \mathcal{S}_A \wedge \bar \cS_B}{\int_{\tilde Y} \Omega \wedge \bar \Omega}\ ,\qquad \quad
 \mathcal{C}_{\lambda \, I\bar J} = \int_{S_{+}} \omega_\lambda \wedge \gamma_I \wedge \bar \gamma_{\bar J} \ .
\eeq
Since the closed moduli are the same, we proceed in the same way 
as we did for the closed and the D3-brane cases, identifying the 
coordinates as \eqref{NandTD3}. Analogously to the D3-brane case, 
we expand up to second order in the open moduli and match 
both theories by
\beq \label{matchopenD7}
	\partial_{V^\lambda} (e^{2D_A} G_{i j}) u^i u^j \cong  i  \cC_{\lambda I \bar J} a^I \bar a^{\bar J} \ ,\qquad
\partial_{V_0} (e^{2D_A} G_{i j}) u^i u^j  \cong  i \cL_{A \bar B} \chi^A \bar {\chi}^{\bar B} \ , \qquad
  \partial_{V_k} (e^{2D_A} G_{i j}) u^i u^j \cong  0 \ ,
\eeq 
where we have indicated that as in the D3-brane case one will need to make the shift symmetry 
manifest before finding complete match. Crucially one has to split the Type IIA coordinates into 
two sets $\zeta^I$ and $\zeta^A$ and identify 
\beq
  \zeta^I \cong a^I \ , \qquad \zeta^A \cong \chi^A\ .
\eeq  
One notes that Wilson line moduli and brane deformations do 
not mix on the Type IIB side which seems to be in contrast to 
the general form on the Type IIA side.
We will argue later how this splitting can be understood 
from the SYZ-picture of mirror symmetry.

As already suggested in \eqref{Kdef_suggest} one expects that the open corrections to the 
$\cN=1$ coordinates can again be given in terms of chain integrals. Let us first give 
the expression for $K_o$ which encodes upon differentiation with respect to $V^\lambda,V_0,V_k$ the 
the corrections in $T_\lambda, N^0, N^k$. Explicitly, we propose
\beq \label{KoD7}
    K_o = \tfrac{i}{4} \int_{\cC_5} s_A \lrcorner  \I\, \Phi^{\rm ev}  \int_{\cC_5} \bar s^{A} \wedge \Omega  
         + \tfrac{i}{4} \int_{\cC_5} \cF_{\rm D7} \wedge \gamma_I \wedge \I \Phi^{\rm ev}  \int_{\cC_5} \cF_{\rm D7} \wedge \bar \gamma^I +c.c.\ ,
\eeq
where $\Phi^{\rm ev}$ is given in \eqref{def-Phiev}.
Here we have used a five-chain $\cC_5$ ending on the D7-brane and a reference four-cycles $S^0_+$.
Note that similar to the D6-brane case we have to introduce a dual basis $s_A$ and $s^{A}$. To do that 
we use the fact that no-where vanishing $(3,0)$-form $\Omega$ provides an identification
\beq
   \Omega: \quad NS_+  \, \rightarrow   TS_+^* \wedge TS_+^* \ ,
\eeq
of normal vectors with two-forms of $S_+$.
Hence, 
in the Type IIB setting we adopt this basis to the complex structure by demanding that $s_A$ is a complex 
normal vector in $H^0_+(NS_+)$ and $s^A$ is a $(2,0)$-form in $H^{(2,0)}_-(S_+)$ on $S^0_+$. Similarly, $\gamma_I$ is a $(0,1)$-form as introduced 
above and $\gamma^J$ is a $(1,2)$-form in $H^{(1,2)}_-(S^0_+)$. These forms are defined to be dual and hence satisfy 
\beq
   \int_{S^0_+} \bar s^{A} \wedge (s_B \lrcorner \Omega)=  \delta_B^A\ ,\qquad  \qquad \int_{S^0_+} \gamma_I \wedge \bar \gamma^J = \delta_I^J\ .
\eeq
As in the D6-brane case we have to extend these forms to the chain. It is interesting to note that the expression 
\eqref{KoD7} indeed reproduces the leading order corrections after differentiating with respect to  $V^\lambda,V_0,V_k$.

\subsection{Mirror symmetry for O5-orientifolds and D5-branes}

Let us now discuss the second Type IIB set-up which is obtained 
by an involution with O5-planes as fix-point set. 
The bulk $\cN=1$ coordinates of the moduli space $\cM^{\rm Q}$ are given as functions 
of the zero-modes in the expansion
\bea\label{expansionclosedO5}
  J &=&  v^{k}\, \omega_{k}\ ,\qquad C_2\ =\ \tilde C_2 + c^k\, \omega_k\ ,
\quad k=1,\ldots, h_+^{(1,1)}(\tilde Y)\ , \\
 B_2 &=& b^\lambda \, \omega_\lambda\ ,\qquad    C_4 =
 \rho_\lambda \ \tilde \omega^\lambda\ ,\quad \quad \quad\ \ \lambda = 1,\ldots,h_-^{(1,1)}(\tilde Y)\ .\nonumber
\eea
Note the difference that we have used forms of different $\sigma$-parity 
in the expansion for the R-R-fields, $C_2$ and $C_4$ as required for the second  
orientifold projection in \eqref{fieldtransfB}. 
While $C_0$ has been projected out $C_2$  now contains a four-dimensional 
two-form $\tilde C_2(x)$ which together with the dilaton $\phi_B$
form the bosonic content of a linear multiplet. However, $\tilde C_2$ can 
be dualized to a scalar field $h$ and form with $\phi_B$ a chiral multiplet.
The $\cN=1$ coordinates which span $\cM^{\rm Q}$ are thus the
$h^{(1,1)}+1$ complex fields
\bea\label{def-tA}
  t'^k& =& e^{-\phi_B} v^k - i c^k\ , \qquad \quad 
  P_\lambda = \cK_{\lambda \rho k} b^\rho t^k + i\rho_\lambda\ , \\
  S &=& e^{-\phi_B} \cV + i h - \tfrac{i}{2} \rho_{\lambda} b^{\lambda}
- \tfrac{1}{2} P_{\lambda} b^{\lambda} 
, \nn
\eea
Formally the K\"ahler potential is the same as in the O3/O7-case given in \eqref{KaehlerO3/O7}.
However, it now has to be evaluated as a function of the coordinates $t'^k,P_\lambda$ and $S$
by using there explicit form \eqref{def-tA}. 
Similar to \eqref{moduliexpansionD3closed} we can write
\beq \label{moduliexpansionD5closed}
 - \I\, \Phi^{\rm ev} + i \sum_{n} e^{B_2} \wedge C_{2n} 
= - t'^k \omega_k + P_\lambda \tilde \omega^\lambda + S {\rm vol}_{\tilde Y} .
\eeq

Let us turn to the discussion of the mirror Type IIA side to this 
construction. As explained above the second set-up with O5-planes 
is obtained by choosing the three-form $\alpha_0$ for the fundamental 
period $X^0$ to lie in the negative eigenspace $H^{3}_-(\tilde Y)$.
Again we will perform a rescaling of $\Omega$ setting the 
coefficient of $\alpha_0$ to be constant. 
The special coordinates are then given by 
\beq
	g_A^{-1} = 2 \I C X^0\ ,\qquad q^k = \frac{\R C X^k}{\I C X^0} \ , \qquad 
	q^\lambda=\frac{\I C X^\lambda}{\I C X^0}\ .
\eeq
Now the rescaled prepotential $f$ is given by $\cF(2 CX)=- i (2 \I C X^0)^2 f(q^k,q^\lambda) $. 
This allows us to rewrite $C\Omega$ in the rescaled coordinates as
\begin{equation}
	2 C\Omega
	=g_A^{-1} \left [q^k \alpha_k + i \alpha_0 + i q^\lambda \alpha_\lambda
	+ f_\lambda \beta^\lambda - (- 2 f + q^k f_k + q^\lambda f_\lambda) \beta^0 + i f_k \beta^k \right] \ .
\end{equation}
Moreover, we can use the special coordinates to write $(N'^k,T'^A_\lambda,T'^A_0)$ as
\begin{equation}
	N'^k = g_A^{-1}q^k + i \xi^k \, \qquad
	T_0^{\prime A} = g_A^{-1} (-2f+q^\lambda f_\lambda+ q^k f_k)+i\tilde \xi_0 \qquad
	T_\lambda^{\prime A} = - g_A^{-1} f_\lambda + i \tilde \xi_\lambda \ .
\end{equation}
With $f$ in the large complex structure limit
\beq
	f(q)=\tfrac{1}{6} \kappa_{k l m}  q^k q^l q^m
	-\tfrac{1}{2} \kappa_{l \mu \rho} q^l q^\mu q^\rho
	 \ .
\eeq
this allows us to write
\beq
T_0^{\prime A} = g_A^{-1} \left(\tfrac{1}{6} \kappa_{k l m}q^k q^l q^m 
	- \tfrac12 \kappa_{\mu \lambda k} q^\mu q^\lambda q^k\right) + i \tilde \xi_0 \ , \qquad
T_\lambda^{\prime A} = g_A^{-1} \kappa_{\lambda \mu k} q^\mu q^k+ i \tilde \xi_\lambda \ .
\eeq
The mirror mapping is then realized by
\beq \label{NandTO5}
 	t'^k  	\leftrightarrow	N'^k \qquad {\rm and} \qquad
	(S,P_{\lambda})  \leftrightarrow (T_0^{\prime A},T_\lambda^{\prime A})\ .
\eeq
In terms of the Kaluza-Klein modes this amounts to the identification of the closed moduli
\begin{align}
 &g_A^{-1} = e^{- \phi_B}\ , \qquad &q^k& = v^k\ , \qquad &q^\lambda& = b^\lambda\ ,\qquad  \\
  &\tilde \xi_0 =  h - \rho_\lambda b^\lambda\ + \tfrac{1}{2} \cK_{l\lambda \kappa} c^l b^\lambda b^\kappa \ ,  &\xi^k& = -c^k\ , \qquad
  &\tilde \xi_\lambda& = \rho_\lambda\ - \cK_{\lambda \kappa l} c^l b^\kappa\ .
  \nn
\end{align}
The identification of the basis elements on the Type IIA and Type IIB side is given in Table \ref{mirror_cohomology)5}.

\begin{table}[h]
\begin{center}
\begin{tabular}{|c|c|}\hline 
\rule[-.2cm]{0cm}{.6cm} $H^3(Y)$					& $H^{\rm even} (\tilde Y)$			\\	\hline \hline
\rule[-.2cm]{0cm}{.6cm} $\alpha_0 \in H^3_-(Y)$ 		& $1$								\\ \hline
\rule[-.2cm]{0cm}{.6cm} $\alpha_k \in H^3_+(Y)$		& $\omega_k \in H^2_+(\tilde Y)$			\\ \hline
\rule[-.2cm]{0cm}{.6cm} $\alpha_\lambda \in H^3_-(Y)$	& $\omega_\lambda  \in H^2_-(\tilde Y)$		\\ \hline
\rule[-.2cm]{0cm}{.6cm} $\beta^k \in H^3_-(Y)$		& $\tilde \omega^k  \in H^4_+(\tilde Y)$		\\ \hline
\rule[-.2cm]{0cm}{.6cm} $\beta^\lambda \in H^3_+(Y)$	& $\tilde \omega^\lambda \in H^4_-(\tilde Y)$	\\ \hline
\rule[-.2cm]{0cm}{.6cm} $\beta^0 \in H^3_+(Y)$		& $\cV^{-1} vol_{\tilde Y}$					\\ \hline
\end{tabular}
\end{center}
\caption{The mirror mapping from the basis of $H^3(Y)$ to the
basis of even cohomologies of the mirror Calabi-Yau $\tilde Y$ in
$O5/O9$ orientifold setups.} \label{mirror_cohomology)5}
\end{table}

\subsubsection*{Inclusion of D5 brane moduli}

We now consider a pair of D5-branes on curves $\Sigma^{(1)}$ and 
$\Sigma^{(2)}$ which are interchanged under the orientifold involution.
We call the positive union of $\Sigma^{(1)}$ and $\Sigma^{(2)}$ by $\Sigma_+ = \Sigma^{(1)}+\Sigma^{(2)}$
Again we view this as a single D5-brane on the quotient space.  
The open moduli for a single D5-brane, corresponding to 
complex brane deformations $\chi^A,\, A= 1,\ldots \text{dim}\, H^{0}_-(N\Sigma_+)$ and Wilson line moduli 
$a^I,\, I=1,\ldots , h^{(0,1)}_-(\Sigma_+)$, correct the $\cN=1$ coordinates according to \cite{Grimm:2008dq}
\bea\label{def-tA_2}
  t^k& =& t'^k +\cL ^k _{A \bar B} \chi^A \bar \chi^{\bar B}
, \nn\\
  P_\lambda &=& \cK_{\lambda \rho k} b^\rho t'^k + i\rho_\lambda\ , \\
  S &=& e^{-\phi_B} \cV + i h - \tfrac{i}{2} \rho_{\lambda} b^{\lambda}
- \tfrac{1}{2} P_{\lambda} b^{\lambda}
+ \cC_{I \bar J} a^I \bar a^{\bar J} \ . \nn
\eea
Here we have introduce the couplings 
\beq
  \cL^k_{A \bar B} = - i \int_{\Sigma_+} s_A \lrcorner \bar s_B \lrcorner \tilde \omega^k\ , \qquad \quad
\cC_{I \bar J} = i \int_{\Sigma_+} \gamma_I \wedge \bar \gamma_{\bar J}
\eeq
The K\"ahler potential now has to be evaluated as a function of $t^k,P_\lambda,S$ as well as 
the open coordinates $\chi^A$ and $a^I$.

In order to discuss mirror symmetry to the D6-brane set-up we again compare the 
form of the $\cN=1$ coordinates. Expanding to second order in the open corrections we find
\beq \label{matchopenD5}
	- \partial_{V_k} (e^{2D_A} G_{i j}) u^i u^j \cong \cL ^k _{A \bar B} \chi^A \bar \chi^{\bar B}, \quad
	- \partial_{V_0} (e^{2D_A} G_{i j}) u^i u^j \cong  \cC _{I \bar J} a^I \bar a^{\bar J} \ , \quad - \partial_{V^\lambda} (e^{2D_A} G_{i j}) u^i u^j \cong 0 \ .
\eeq
More interestingly, we can also directly compare the open K\"ahler potential $K_o$. To do that, we give a chain integral 
expression for the D5-brane case. We introduce a the three-chain $\cC_3$ ending on a reference cycle $\Sigma_+^0$ and 
the two-cycle to which the brane has moved. The open K\"ahler potential then takes the form 
\beq \label{KoD5}
    K_o = - \tfrac{i}{4}\int_{\cC_3} s_A \lrcorner  \R\, \Phi^{\rm ev}  \int_{\cC_3} \bar s^{A} \cdot \Omega  
         - \tfrac{i}{4} \int_{\cC_3} \cF_{\rm D5} \wedge \gamma_I \wedge \R\, \Phi^{\rm ev}  \int_{\cC_3} \cF_{\rm D5} \wedge \bar \gamma^I + c.c. \ ,
\eeq
where $\Phi^{\rm ev}$ is given in \eqref{def-Phiev}. Note that this expression has a similar structure as \eqref{KoD7}. However, due to 
the lower dimensionality of the chain the four-form part of $\R\, \Phi^{\rm ev}$ is picked up in the first term of \eqref{KoD5}, while the 
zero-form part of $\R\, \Phi^{\rm ev}$ contributes in the second term of \eqref{KoD5}. In the case of a D5-brane the $(3,0)$-form 
$\Omega$ on $\tilde Y$ provides a map
\beq
   \Omega: \quad N\Sigma_+ \otimes N\Sigma_+ \ \rightarrow \ T\Sigma_+^*\ ,
\eeq
taking two normal vectors to a one-form on $\Sigma_+$.
This allows us to introduce a basis $s^A$ of $H^{0}(T\Sigma^0_+ \otimes \overline{N\Sigma}^0_+)$ which is 
dual to the normal vectors $s_A$. Hence, the $\cdot$ in \eqref{KoD5} indicates that the vector part of $s^A$ is 
inserted, while the form part of $s^A$ is wedged with $\Omega$.  We also introduce complex one-forms $\gamma^J$ on $\Sigma_+^0$ which are dual to the $(0,1)$-forms 
$\gamma_I$ used in the expansion determining the complex Wilson line scalars $a^I$. Explicitly, the $s^A,\gamma^I$ have to satisfy 
on the reference $\Sigma_+^0$ that
\beq
  \int_{\Sigma^0_+} s_A \lrcorner \bar s^B \cdot \Omega = \delta_A^B \ ,\qquad \quad  \int_{\Sigma^0_+} \gamma_I \wedge \bar \gamma^J = \delta_I^J \ ,
\eeq
As in the D6-brane case the basis forms and vectors have to be extended trivially to the chain $\cC_3$ to evaluate the 
open K\"ahler potential \eqref{KoD5}. One can now check that the expansion \eqref{KoD5} leads upon differentiation with 
respect to $V_k,V^0,V^\lambda$ the leading order corrections in \eqref{def-tA_2}.

\subsection{General remarks on the structure of the couplings}

In this subsection we address the question if there is a simple way to understand the mappings
of \eqref{matchopenD5},  \eqref{matchopenD3} and  \eqref{matchopenD7} using the SYZ-picture of mirror symmetry. 
For example for D5-branes the $(\partial_{V_k} (e^{2D_A} G_{i j}),\partial_{V_0} (e^{2D_A} G_{i j}))$
correct the coordinates $t^k$ and $S$ by brane deformations and 
Wilson line moduli as demanded by the mirror identification \eqref{matchopenD5}. In contrast, the coordinates $P_\lambda$ do not receive any 
contributions from open moduli and hence $\partial_{V^\lambda} (e^{2D_A} G_{i j})$ has 
to vanish in the D6-brane set-up mirror dual to a D5-brane. To analyze this question in the SYZ-picture, first let us look at the gauge coupling functions. In the 
limit of vanishing open string moduli they are given by the analogous 
to the D6-brane gauge coupling function $f_{\rm D6} = N^k \int_L \alpha_k - T_\lambda \int_L \beta^\lambda$, 
\beq\label{gaugecouplingsIIB}
f_{\rm D3}=\tau \ , \qquad f_{\rm D5}=t^\Sigma \int_{\Sigma_+} \omega_\Sigma  \ , 
\qquad f_{\rm D7}=T_S \int_{S_+} \tilde \omega^S \ ,
\eeq
where $\Sigma_+$($S_+$) is the curve(divisor) wrapped by the D5(D7)-brane, 
and they are obtained from a basis of homology by
\bea
&\left[\Sigma_+\right]=n^k \left[\Sigma_k\right] \ , \qquad \Sigma_k \in H_2^+(Y) \ & \text{and}\\
&\left[S_+\right]=n_\lambda [S^\lambda ] \ , \qquad S^\lambda \in H_4^+(Y) \ .& \nn
\eea
Therefore the forms appearing in \eqref{gaugecouplingsIIB} are, 
in terms of the cohomology basis, $\omega_\Sigma=n^k \omega_k$ and 
$\tilde \omega^S=n_\lambda \tilde \omega^\lambda$.

From the four internal dimensions
the D7-brane wraps, locally two of them are along the $T^3$-fiber and 
the other two on the base, as seen from table \ref{mirrortable}. The 
mirror D6-brane, on the other hand, wraps one dimension on $T^3$-fiber and two dimensions 
on the base. It is also inferred from the gauge coupling function of the D7-brane
\eqref{gaugecouplingsIIB} that $\tilde \omega^\lambda$ sits on the brane,
therefore having two ``legs'' on the 3-Torus and two on the base.
We define thus the notation $\tilde \omega^\lambda : (bbtt)$, 
where $b$ and $t$ correspond to base and torus components.
Table \ref{mirror_cohomologyO3} shows that $\tilde \omega^\lambda$ 
on the Type IIB side is mapped on the Type IIA side to $\beta^\lambda$. 
Therefore, from table \ref{mirrortable}, 
since $\beta^\lambda$ must sit on the mirror D6-brane, 
it should satisfy $\beta^\lambda : (bbt)$. $\beta^\lambda$ must be dual 
to $\alpha_\lambda$ on the Calabi-Yau manifold Y, thus $\alpha_\lambda : (btt)$.
A similar analysis can be done for the D5 and D3-Branes, from where 
we obtain $\alpha_k : (btt)$, $\beta^k : (bbt)$, $\beta^0:(bbb)$ and $\alpha_0:(ttt)$.


One can now analyze the open moduli corrections to the 
$\cN=1$ chiral coordinates from the metric derivatives $\partial_{V_0} \widehat \cG_{ij}$,
$\partial_{V_k} \widehat \cG_{ij}$ and $\partial_{V^\lambda} \widehat \cG_{ij}$.
As a simple example we consider the D3-brane case. 
We can rewrite the corrections 
in terms of the normal deformations $\eta^i$
\beq \label{metricmirror0}
    \R (N'^0-N^0) = \partial_{V_0} (e^{2 D_A} \widehat \cG_{ij}) \eta^i \eta^j 
	= \frac12 \int_{L_0} \hat \alpha_k \wedge \eta \lrcorner \beta^0 \int_{L_0} \hat \beta^k \wedge \eta \lrcorner J \ .
\eeq
Since the brane wraps the three-torus,
both integrands in \eqref{metricmirror0} must be of the form $(ttt)$. 
The normal directions of this D6-brane are all on the base, so $\eta \lrcorner \beta^0 : (bb)$, 
making the first integral vanish. 
Therefore there is no correction to $N'^0 = i\tau$ coming from 
$\partial_{V_0} \widehat \cG_{ij}$, as was already seen in \eqref{matchopenD3}.
By repeating the analysis to $\partial_{V^\lambda} \widehat \cG_{ij}$ 
and $\partial_{V_k} \widehat \cG_{ij}$ one shows that 
only the latter can be non-vanishing, and analysing in the same fashion 
the corrections for the D5 and the D7 cases we obtain the same 
relations as \eqref{matchopenD5} and  \eqref{matchopenD7}.

One can realize then that brane deformations with normal 
direction $\eta$ along one cycle of the 3-torus on the Type IIA side 
are mapped to Wilson line moduli along the T-dual cycle 
on the Type IIB side, while brane deformations along the 
base are mapped to brane deformations on the Type IIB side, also along the base.
In the opposite direction, brane deformations on the Type IIB side 
along the 3-torus are mapped to Wilson line moduli 
along the dual cycle on the Type IIA side.

\section{Conclusions}

In this paper we have derived the four-dimensional $\cN=1$ effective action 
of IIA and IIB Calabi-Yau orientifolds including single space-time filling 
D$p$-branes by performing a Kaluza-Klein reduction. In particular, we 
derived the $\cN=1$ characteristic data of the open-closed system
for space-time filling D6-branes in an Type IIA Calabi-Yau orientifold. In the  
determination of the K\"ahler potential $K$ we showed that the complex $\cN=1$ 
open coordinates appear in $K$ only through a redefinition of the closed coordinates. 
$K$ itself can be viewed as a function of real three- and two-forms.  
In the presence of D6-branes these forms have localized corrections with 
the open coordinates. In addition to the kinetic terms of 
the scalars we have also determined the holomorphic 
gauge coupling function of the brane and bulk $U(1)$ gauge 
fields including possible mixed terms.  

We have discussed the $\cN=1$ characteristic data of the orientifold 
compactifications both for a finite as well as for the infinite dimensional 
case. In a fixed background Calabi-Yau geometry a D6-brane on a special 
Lagrangian cycle $L_0$ has $b^1(L_0)$. The scalar potential vanishes for 
these deformations. Considering a general normal deformation of $L_0$ 
the superpotential \eqref{general_W} and D-terms \eqref{D-terms} are induced. These have been determined 
explicitly and were shown to be given in terms of chain integrals over 
a four-chain $\cC_4$ ending on the reference cycle $L_0$ and the deformed cycle 
$L_\eta$. We also argued that the corrections to the closed string coordinates 
can be formulated as chain integrals. In particular, we introduced a 
K\"ahler potential $K_o$ which depends on both open and closed deformations and 
encodes the corrections to the $\cN=1$ closed coordinates. $K_o$ as given in \eqref{Ko_general} 
contains two chain integrals involving both the K\"ahler form $J$ 
as well as the holomorphic three-form $\I C\Omega$.
When restricting to a finite dimensional deformation space $K_o$ was shown to restrict
to the K\"ahler potential introduced by Hitchin on the moduli space of 
special Lagrangian submanifolds with $U(1)$ connection.

In the last part of the paper we related our Type IIA results to 
the $\cN=1$ data for Type IIB orientifold compactifications
with D3-, D5-, or D7-branes by using mirror symmetry. The 
SYZ proposal to view the internal manifold as $T^3$ fibration, with 
possibly resolved singular fibers, allowed us the match of the 
$\cN=1$ data for branes and orientifold planes of different dimensionalities 
with the D6/O6 set-up. The mirror map has been evaluated in special limits
of the closed and open moduli space. 
It will be interesting to extend this analysis to the interior of the 
open-closed moduli space. The general chain integral expressions for 
the $\cN=1$ coordinates, K\"ahler potential and gauge coupling function 
might allow to compute quantum corrections using geometric methods on 
one side of the mirror correspondence and applying the mirror map.

There are various further directions in which our results can be extended. It is well-known 
that D6-branes in Type IIA string theory are obtained from specific geometries, so-called 
Taub-NUT spaces, in an M-theory. More precisely, one expects that the D6/O6 compactifications 
considered in this work naturally lift to a compactification of M-theory on 
a $G_2$ manifold. The $\cN=1$ data found in this paper will naturally embed into 
the $\cN=1$ data of non-singular $G_2$ reduction found in \cite{Harvey:1999as,Gukov:1999gr,Beasley:2002db}. One expects 
that similar issues as for D7/O7 compactifications embedded into F-theory arise \cite{Grimm:2010ks}. 
Also for the D6/O6 compactifications it will be interesting to understand the origin of the 
flux independent St\"uckelberg gaugings as it was found in \cite{Grimm:2010ez} for F-theory set-ups.
Moreover, it will be interesting to generalize the set-up to intersecting D6-branes 
including the possibility of fundamental matter. This will modify the $\cN=1$ data 
in both orientifold and M-theory compactifications \cite{Blumenhagen:2006ci,Lust:2004cx}.

In flux compactifications the backreaction is often so strong that the 
compactification manifold cannot be a Calabi-Yau manifold. This implies that
one has to compute the effective action by looking at variations around a 
new non-Calabi-Yau solution. These often can be described using generalized 
geometry as discussed for $\cN=1$ vacua, for example in \cite{Grana:2005sn,Benmachiche:2006df,Koerber:2007xk}, 
and the review \cite{Koerber:2010bx}. It would 
be interesting to extend our results to such a generalized setting. Note that 
formally this is rather straightforward by replacing $e^{J_c}$ and $\R(C\Omega)$ 
by general pure spinors in all our expressions. However, it will be desirable 
to show if one still can explicitly compute the $\cN=1$ data by finding non-trivial 
example threefolds which are described by generalized geometric methods and cannot 
be analyzed in either symplectic or complex geometry.

\vskip 3cm

{\noindent  {\Large \bf Acknowledgments}}
 \vskip 0.5cm

We gratefully acknowledge helpful discussions with Ralph Blumenhagen, Raphael Flauger, 
Denis Klevers, Albrecht Klemm, Dieter L\"ust, Stephan Stieberger, Gonzalo Torroba, and  Marco Zagermann. 
We like to thank Timo Weigand and Max Kerstan for coordinating submission of their related work \cite{Kerstan:2011dy}.
TG is thankful 
to the KITP, Santa Barbara 
and the Bethe Center for Theoretical Physics where part of this work was done.
The work of DL was supported by 
European Union 7th network program ``Unification in the LHC era" (PITN-GA-2009-237920)
and the Bonn-Cologne Graduate School of Physics and Astronomy (BCGS).

\vskip 1cm

\noindent {\bf \LARGE Appendices   }

\begin{appendix}

\section{Derivation of the K\"ahler metric \label{derivation_K}}

Let us now discuss the derivation of the K\"ahler metric and
compare the result with the effective action for the
D6-brane found by dimensional reduction, \eqref{eqn:DBIlight} and \eqref{eqn:CSlight}.
Firstly, we note that the metrics for $\R M^K$ and the pure $\xi^K$ terms match
the result found from the reduction of the closed string action, since
$\tilde K^{KL}= (G_{kl},G^{\lambda \kappa},G^\lambda_k)$, as 
described in \cite{Grimm:2004ua}.
We need then to check the terms involving open string moduli $\zeta^i$.
From the reduction of the action the metrics $\cG_{ij}$ and $\widehat \cG_{ij}$ are
\beq \label{cG_lambda}
   \widehat \cG_{ij} = \mu_{ki} \, \lambda_{j}^k \ ,\qquad \cG_{ij} = \mu_{ik}\, (\lambda^{-1})_j^k\ ,
\eeq
where, recalling equations \eqref{theta-periods} and \eqref{def-theta},
\beq \nn
  e^{-\phi} \theta_i = \lambda^j_i \tilde \alpha_j\ \ , \qquad \theta_i = s_i \lrcorner J|_{L_0} \ ,
\eeq
\beq \nn
\tfrac12  e^{-\phi} *\theta_i = \mu_{j i}\, \tilde \beta^j\ \ , 
\qquad *\theta_i = - 2 e^{\phi} s_i \lrcorner \I (C \Omega)|_{L_0} \ .
\eeq
The coefficients $\mu_{ij}$ and $\lambda_i^j$ are calculated to be
\beq
e^{2D}\mu_{ij}=\frac12 \int_L \tilde \alpha_i \wedge s_j \lrcorner (V^\kappa \alpha_\kappa+ V_k \beta^k)\ ,
\qquad
\lambda_i^j = \int_L \tilde \beta^j \wedge s_i \lrcorner J \ ,
\eeq
also making use of the relations $\int \tilde \alpha_i \wedge \tilde \beta^j = \delta_i^j$.
To leading order, the $V$ derivatives of $\mu_{i j}$ are
\beq
   \frac{\partial}{\partial V^\lambda}(e^{2D} \mu_{ij})
   =
   \frac12 \int_L \tilde \alpha_i \wedge s_j \lrcorner  \alpha_\lambda
    \ ,  \qquad
    \frac{\partial}{\partial V_k}(e^{2D} \mu_{ij})
   =
   \frac12 \int_L \tilde \alpha_i \wedge s_j \lrcorner  \beta^k
    \ ,
\eeq
On the other hand, $\lambda_i^j$ is independent of $(V^\lambda,V_k)$, 
at least for leading order complex structure deformations. This implies using \eqref{correctedK}, 
\eqref{def-zeta_withB} and \eqref{cG_lambda} that
\beq
	  \tilde K_{\zeta^i \bar \zeta^j} 
	= \frac{\partial(e^{2D}\cG_{ij})}{\partial V_K} V^K 
	= e^{2D} \cG_{ij}\ ,
\eeq
which is in accord with the result \eqref{eqn:DBIlight} found from dimensional reduction. 
The derivatives of the metric with respect to $(V^\lambda,V_k)$ are given explicitly 
by (for first order deformations)
\beq 
	\partial_{V^\lambda} (e^{2D}\cG_{ij}) 
	=  \frac12 \int_{L} \tilde \alpha_i \wedge s_l \lrcorner \alpha_\lambda 
		(\int_{L} \tilde \beta^j \wedge s_l \lrcorner J)^{-1} \ ,
	\quad
	\partial_{V_k} (e^{2D}\cG_{ij})
	= \frac12 \int_{L} \tilde \alpha_i \wedge s_l \lrcorner \beta^k 
		(\int_{L} \tilde \beta^j \wedge s_l \lrcorner J)^{-1} \ .
\eeq
The derivatives of the metric $\widehat \cG_{ij}$ are, in turn,
\beq 
	\partial_{V^\lambda} (e^{2D}\widehat \cG_{ij}) 
	= \frac12 \int_{L} \tilde \alpha_l \wedge s_i \lrcorner \alpha_\lambda 
		\int_{L} \tilde \beta^l \wedge s_j \lrcorner J \ ,
	\qquad
	\partial_{V_k} (e^{2D}\widehat \cG_{ij}) 
	=  \frac12 \int_{L} \tilde \alpha_l \wedge s_i \lrcorner \beta^k 
		\int_{L} \tilde \beta^l \wedge s_j \lrcorner J \ .
\eeq
To also check
the mixing terms of the Wilson lines $a^i$ with the scalars $\zeta^K$ we expand
\beq
\frac{\partial K_o}{\partial \zeta^i}= \tfrac12 \mu_{ij}|_{\rm fix}\, \eta^j +\ldots \ ,
\eeq
to lowest order in the
$\eta^i$. This yields the lowest order expression for $\tilde K^K_{\zeta^i}$ evaluated
to be
\beq
  \tilde K^k_{\zeta^i} = \hat \cI^k_i \ , \qquad \tilde K^{\lambda}_{\zeta^i} = \hat \cI_{i\lambda}\ ,
\eeq
where were used equations \eqref{def-cI_exact} and \eqref{def-cI}
\beq
  \hat \cI^k_i = \int_L \tilde \alpha_i \wedge \eta \lrcorner  \beta^k + \ldots \ ,\qquad
  \hat \cI_{i \lambda} = \int_L \tilde \alpha_i \wedge \eta \lrcorner  \alpha_\lambda + \ldots  \ .
\eeq

\section{Supergravity with several linear multiplets} \label{linm}

In this appendix we want to show, in a step by step way, 
how does the dualization from linear to chiral multiplets work, 
following \cite{Grimm:2004ua}. We want to relate the effective action in terms of 
linear multiplets $(V_K, C^2_K)$, obtained by generalizing a result in \cite{Binetruy:2000zx},
\bea\label{kineticlinear}                
\cL &=& - \tilde K_{\zeta^i\bar \zeta^j}\, d\zeta^i \wedge * d \bar \zeta^{j}
  + \tfrac{1}{4} \tilde K_{V_K V_L}\,
  dV_K \wedge * dV_L \\
  && + \tilde K_{ V_K V_L}\, dC_K^2 \wedge * dC_L^2
     -  i \, dC_K^2 \wedge \nn
\big(\tilde K_{V_K \zeta^i}\,d\zeta^i -\tilde K_{V_K \bar \zeta^i}\,d\bar \zeta^i\big)
\ ,
\eea
with the one with chiral multiplets, \eqref{general_lin},
\bea
 \cL^{\rm kin} &=& - (\tilde K_{\zeta^i \bar \zeta^j} 
+ \tilde K_{\zeta^i}^K \tilde K_{KL} \tilde K_{\bar \zeta^i}^L)\ d\zeta^i \wedge * d\bar \zeta^j \nn
    \\
&&  + \tilde K_{K L}\left( d\R M^I \wedge * \R M^J + d\xi^K \wedge * d\xi^J \right) 
- 2 \, \tilde K_{K L} \tilde K^L_{ \zeta^i} \left(d\R M^I \wedge * du^j +  d\xi^I \wedge * d a^j\right) \nn \ .
\eea
In \eqref{kineticlinear}  
$\tilde K(V,\zeta,\bar \zeta)$ is a function of the scalars $V_K$ 
and the chiral multiplets $\zeta^i$. The function $\tilde K$ encodes 
the dynamics of the fields, and we would like to relate it to the 
K\"ahler potential from \eqref{general_lin}. The standard procedure 
is to eliminate the fields $C^2_K$ in favor of its duals $\xi^K$ 
by introducing an appropriate term to the action
\beq
\cL \to \cL + \delta \cL\ , \qquad			
  \delta \cL\ =\
  - 2 \xi^K\, dC_K^3\ =\ - 2 C^3_K \wedge d \xi^K\ ,\
\eeq
where $\xi^K (x)$ is a Lagrange multiplier. By solving the equations 
of motion for $\xi^K$ one finds $dC_K^3=0\,$ such that locally 
$C^3_K=dC^2_K$, giving $\delta \cL=0\,$ as expected.
One can use the equations of motion of $C_K^3$,
\beq
  *C^3_K = \tilde K^{V_K V_L}\Big(d \xi^L + \tfrac{i}2 					 
  \big(\tilde K_{V_L \zeta^i}\,d\zeta^i
 -\tilde K_{V_L \bar \zeta^i}\,d\bar \zeta^i \big)\Big)
\eeq
to eliminate it from \eqref{kineticlinear},
\bea \label{eff_act1}
\cL &=& - \tilde K_{\zeta^i \bar \zeta^j}\, d\zeta^i \wedge * d \bar \zeta^j		 
  + \tfrac{1}{4} \tilde  K_{V_K V_L}\,
  dV_K \wedge * dV_L  \\
  && +  \tilde K^{V_K V_L} \Big(d\tilde \xi^L -
  \I \big(\tilde K_{V_L \zeta^j}\,d\zeta^j\big)\Big)\wedge *
  \Big(d\tilde \xi^L -
  \I \big(\tilde K_{V_L \zeta^i}\,d\zeta^i\big)\Big) \ .\nn
\eea
For our particular case, we can further simplify this equation. 
Comparing \eqref{eqn:CS}
with the Chern-Simons action \eqref{KK_RR}, one can notice 
that the field $C^2$ couples, to first order, 
with the imaginary part of $\zeta^i$, namely $a^i$.
We can assume that $\tilde K$ is a function only of 
$V_L$ and the real part of $\zeta^i$, $\R \zeta^i=u^i$. 
We will see shortly that this assumption agrees with 
our results (indications that $\tilde K$ depends 
only on $\R \zeta^i$ can be inferred from section \ref{modulispace}, 
as in equation \eqref{open_metric}). The effective Lagrangian 
\eqref{eff_act1} thus simplifies to
\bea \label{linaction}
\cL &=& - \tfrac{1}{4}\tilde K_{u^i u^j}\, d\zeta^i \wedge * d \bar \zeta^j
  + \tfrac{1}{4} \tilde K_{V_K V_L}\, 							
  dV_K \wedge * dV_L \\
  && +  \tilde K^{V_K V_L}
  \Big(d\tilde \xi^K - \tfrac{1}{2}\tilde K_{V_K u^i}\, d\, \I \zeta^i\Big)\wedge *
  \Big(d\tilde \xi^L - \tfrac{1}{2}\tilde K_{V_L u^j}\, d\, \I \zeta^j\Big) \ .\nn
\eea
We would like to relate this $N=1$ Lagrangian to the standard Lagrangian of 
chiral multiplets $\Phi=(M^I,\zeta^i)$
\bea \label{lagsugrachiral0}
\cL &=& - K_{\Phi \bar \Phi}\ d\Phi \wedge * d\bar \Phi  \\
&=&      - K_{\zeta^i \bar \zeta^j}\ d\zeta^i \wedge * d\bar \zeta^j
         - K_{M^I \bar M^J}\left( d\R M^I \wedge * \R M^J + d\xi^K \wedge * d\xi^J \right) \nn \\		 
     &&  - 2 K_{M^I \bar \zeta^j}\ \left(d\R M^I \wedge * du^j +  d\xi^I \wedge * d a^j\right) \nn.
\eea
and relate the K\"ahler metrics $K_{\Phi \bar \Phi}$ with derivatives 
of the function $\tilde K$, as in equation \eqref{general_lin}. This is obtained
by performing a Legendre transform with respect to the fields $M^K$,
\beq \label{LegKP}
 K(M,\zeta) = \tilde K(V, \zeta + \bar \zeta)  + (M^K +\bar M^K) V_K					 
\eeq
where $V_K(\zeta,M)$ is written as a function of the complex fields $\zeta^i$ and implicitly of new field $M^K$, defined as
\beq\label{defT}
M^K = - \tfrac{1}2 \tilde K_{V_K} + i \xi^K.
\eeq
One can see $(M^K + \bar M^K)$ as the conjugate coordinate to $V_K$.
To see that equations \eqref{lagsugrachiral0} and \eqref{linaction} are indeed related 
by this Legendre transformation, one has to calculate
the derivatives of $K$ in terms of the derivatives of $\tilde K$. 
One starts by differentiating \eqref{defT},
\bea \label{derL}
  \frac{\partial V_K}{\partial M^L} &=& - \tilde K^{V_K V_L}\ ,\\
  \frac{\partial V_K}{\partial \zeta^j} &=&
\frac{1}{2}\frac{\partial V_K}{\partial M^L} \frac{\partial M^L}{\partial u^j}				 
= \tfrac{1}{2} \tilde K^{V_K V_L} \tilde K_{V_L u^j}\ \nn .
\eea
Using these expressions one easily calculates the first derivatives of the K\"ahler
potential \eqref{LegKP} as
\beq \label{Kder}
  K_{M^K} = V_K\ , \qquad K_{\zeta^i} = \tfrac{1}{2} \tilde K_{u^i}\ .					 
\eeq
Applying the equations \eqref{derL} once more when differentiating \eqref{Kder}
one finds the K\"ahler metrics
\bea \label{Km1}
  K_{M^K \bar M^L} &=& - \tilde K^{V_K V_L}\ , \quad
  K_{M^K \bar \zeta^i}\ =\tfrac{1}{2} \tilde K^{V_K V_L} \tilde K_{V_L u^i}\ , \nn \\ 			
  K_{\zeta^i \bar \zeta^j} &=& \tfrac{1}{4} \tilde K_{u^i u^j} + \tfrac{1}{4}
                     \tilde K_{ u^i V_K }\, \tilde K^{V_K V_L}\, \tilde K_{V_L u^j}\ ,
\eea
with inverses
\bea \label{invKm1}
  K^{M^K \bar M^L} &=& - \tilde K_{V_K V_L}
                    + \tilde K_{ u^i V_K }\, \tilde K^{u^i u^j} \, \tilde K_{V_L u^j}\ , 		
                    \nn \\
  K^{M^K \bar \zeta^j} & = & 2 \tilde K^{u^i u^j}\, \tilde K_{ u^i V_K }\ , \quad
  K^{\zeta^i \bar \zeta^j} \ = \ 4 \tilde K^{u^i u^j}\ .
\eea
Finally, one checks that $K(T,N)$ is indeed the K\"ahler potential for the
Lagrangian \eqref{linaction}. This is done by inserting in the definition
of $T_\kappa$ and the K\"ahler metrics obtained above into \eqref{lagsugrachiral0}, yielding back \eqref{linaction}.
\section{Mixing of brane and bulk $U(1)$ vectors} \label{ap_mix}
In this Appendix we analyze the 4D effective action for all the massless 
spacetime vector fields that appear after dimensional reduction.
They are the $A^\alpha$ and $A_\alpha$ components coming 
from the combination of RR and $B_2$ bulk fields \eqref{KK_RR}, 
and $A$, the massless vector component of the U(1) field $A_{\text{D6}}$ 
on the brane, \eqref{exp_AD6_gen}. 
The duality relation between $C_3$ and $C_5$ implies a electric-magnetic 
duality between $A^\alpha$ and $A_\alpha$. To avoid the overcounting of 
degrees of freedom, we consider both fields, but each weighted by a factor of one half, 
as in \cite{Jockers:2004yj}. This procedure gives the action
\bea \label{vectoraction}
S^{(4)}_{\text{vec}} &=& -\int \tfrac{1}{2} \R f_{\text{r}\, }\, F\wedge * F
        + \tfrac{1}{2} \I f_{\text{r}\, }\, F\wedge F \\
&&	+ \tfrac{1}{4} ( \I \cN_{\alpha \beta}
                       + \R \cN_{\alpha \gamma} \I \cN^{\gamma \delta} \R \cN_{\delta \beta})
                        dA^\alpha \wedge * dA^\beta \nn \\
     && + \tfrac{1}{4} \I \cN^{\alpha \beta} dA_\alpha \wedge * dA_\beta
        - \tfrac{1}{2} \R \cN_{\alpha \gamma} \I \cN^{\gamma \beta} dA_\beta \wedge * dA^\alpha
        - \Delta_\alpha dA^\alpha \wedge F
        - \tilde \cJ^\alpha dA_\alpha \wedge F \ , \nn
\eea
where $F=dA$, $\Delta_\alpha= (a^j \Delta_{j \alpha} + \Gamma_{\alpha} )$ and
\beq
	\I \cN_{\alpha \beta} = - \int_Y \omega_\alpha \wedge * \omega_\beta \qquad
	\I \cN^{\alpha \beta} = (\I \cN_{\alpha \beta})^{-1} = - \int_Y \tilde \omega^\alpha \wedge * \tilde \omega^\beta \qquad
	\R \cN_{\alpha \beta} = - b^a \cK_{a \alpha \beta} \ .
\eeq
Recalling the duality relation \eqref{cA-def} for the $\cA$ fields
\beq
	\left. e^B d\cA \right|_6 = \left. - *_{10} \left(e^B d\cA \right) \right|_4 \ ,
\eeq
we obtain, for $A^\alpha$ and $A_\alpha$,
\beq 
       d(A_\alpha \tilde \omega^\alpha) + dA^\beta b^a \omega_a \wedge \omega_\beta=
       -*dA^\gamma *_6 \omega_\gamma \ .
\eeq
We take the wedge product of the above expression with $\omega_\alpha$ and integrate to obtain
the duality relation
\beq \label{dualrelation}
 d A_\alpha= \I \cN_{\alpha \beta} * dA^\beta+ \R \cN_{\alpha \beta} dA^\beta \ .
\eeq
From the variation of action \eqref{vectoraction}, we obtain the equations of motion 
for $A_\alpha$ and $A^\alpha$,
\bea \label{eomvec1}
   &\tfrac{1}{2} ( \I \cN_{\alpha \beta}
                       + \R \cN_{\alpha \gamma} \I \cN^{\gamma \delta} \R \cN_{\delta \beta}) \ d * dA^\beta
 	 - \tfrac{1}{2}  \R \cN_{\alpha \gamma} \I \cN^{\gamma \beta} \ d*dA_\beta
         - \Delta_\alpha d F = 0 \ ,&  
	\\
	&\tfrac{1}{2} \I \cN^{\alpha \beta} \ d * dA_\alpha
	- \tfrac{1}{2}  \R \cN_{\alpha \gamma} \I \cN^{\gamma \beta} \ d * dA^\alpha
	- \tilde \cJ^\beta d F = 0 \ . & \nn
\eea
However, if one takes the exterior derivative of equation \eqref{dualrelation} and compare 
with \eqref{eomvec1}, one notes that the equations are not compatible. That is,
the equations of motion and the duality constraints cannot be simultaneously 
satisfied.
In order to make the duality relation consistent, 
one should modify the field strengths as
\beq \label{redefduals}
	dA^\alpha \rightarrow G^\alpha = dA^\alpha - 2 \tilde \cJ^\alpha F \ ,
	\qquad dA_\alpha \rightarrow G_\alpha = dA_\alpha + 2 \Delta_{\alpha} F \ ,
\eeq
as well as the duality relation \eqref{dualrelation} by the same redefinition. 
This modified action becomes then
\bea \label{vectoraction2}
S^{(4)}_{\text{vec}} &\rightarrow& 
	- \int \tfrac{1}{4} ( \I \cN_{\alpha \beta}
                       + \R \cN_{\alpha \gamma} \I \cN^{\gamma \delta} \R \cN_{\delta \beta})
                        G^\alpha \wedge * G^\beta
	- \tfrac{1}{2} \R \cN_{\alpha \gamma} \I \cN^{\gamma \beta} G_\beta \wedge * G^\alpha 
	\\
	&& + \tfrac{1}{4} \I \cN^{\alpha \beta} G_\alpha \wedge * G_\beta 
	+ \tfrac{1}{2} \R f_{\text{r}\, }\, F\wedge * F
        + \tfrac{1}{2} \I f_{\text{r}\, }\, F\wedge F
       	- \Delta_\alpha G^\alpha \wedge F 
	- \tilde \cJ^\alpha G_\alpha \wedge F \       . \nn
\eea
The equations coming from this action are
\bea \label{eqdual}
	& dG^\alpha = - 2 \tilde \cJ^\alpha dF \ , \qquad
	dG_\alpha = 2 \Delta_{\alpha} dF \ , \qquad
	G_\alpha = \I \cN_{\alpha \beta} * G^\beta + \R \cN_{\alpha \beta} \ G^\beta  \ , & \\
	&\tfrac{1}{2} ( \I \cN_{\alpha \beta}
                       + \R \cN_{\alpha \gamma} \I \cN^{\gamma \delta} \R \cN_{\delta \beta}) \ d * G^\beta
 	- \tfrac{1}{2}  \R \cN_{\alpha \gamma} \I \cN^{\gamma \beta} \ d*G_\beta
        - \Delta_\alpha d F = 0   \ , & \nn
	\\
	&\tfrac{1}{2} \I \cN^{\alpha \beta} \ d * G_\alpha
	- \tfrac{1}{2}  \R \cN_{\alpha \gamma} \I \cN^{\gamma \beta} \ d * G^\alpha
	- \tilde \cJ^\beta d F = 0 \ .& \nn 
\eea
The first two equations follow directly from \eqref{redefduals}, the 
third is the imposed duality condition, and the two remaining are the 
equations of motion for $A^\alpha$ and $A_\alpha$. One can check that 
they are now consistent, by starting with the equation of motion for one
of the fields and obtaining the equation for the dual field after imposing 
the duality conditions.

As was mentioned, the duality condition implies that the degrees of 
freedom for the fields are not independent. To eliminate 
the dependence of $A_\alpha$ in favor of its dual, 
we now treat the field strength $G_\alpha$ as an 
independent field, and add to the action the term
\beq
 \delta S=- \tfrac12 dA^\alpha \wedge (G_\alpha - 2 \Delta_{\alpha} F) 
+ \lambda (dG_\alpha - 2\Delta_{\alpha} dF) \ ,
\eeq
where $\lambda$ is an auxiliary field acting as a Lagrange multiplier. 
The equations for this modified action are the same as \eqref{eqdual}, 
but now they all come from variations on the fields $A^\alpha$, 
$G_\alpha$ and $\lambda$. Having the equations for $G_\alpha$, 
we now substitute them back into the action, and obtain
\bea
	S^{(4)}_{\text{vec}} &=& - \int \tfrac{1}{2} \R f_{\text{r}\, }\, F\wedge * F
    	+ \tfrac12 \I f_{\text{r}\, }\, F\wedge F \\
&& 	+ \tfrac{1}{2} dA^\alpha \wedge
       	(\I \cN_{\alpha \beta} * G^\beta +\R \cN_{\alpha \beta} G^\beta - 2 \Delta_\alpha F ) \nn \\
&&	- \Delta_{\alpha} (dA^\alpha - 2 \tilde \cJ^\alpha F ) \wedge F
    	- (\I \cN_{\alpha \beta} *G^\beta 
		+ \R \cN_{\alpha \beta} G^\beta) \tilde \cJ^\alpha \wedge F \nn   \\
&=&	-\int \tfrac{1}{2} (\R f_{\text{r}\, } 
	+ 4 \I \cN_{\alpha \beta} \tilde \cJ^\alpha \tilde \cJ^\beta) \, F \wedge * F
	+ \tfrac12 (\I f_{\text{r}} + 4 \Delta_{\alpha} \tilde \cJ^\alpha 
	+ 4 \R \cN_{\alpha \beta} \tilde \cJ^\beta \tilde \cJ^\alpha) F \wedge F \nn \\
&&	-2 \I \cN_{\alpha \beta} \tilde \cJ^\beta dA^\alpha \wedge * F \nn
	-2 (\Delta_{\alpha} + \tilde \cJ^\beta \R \cN_{\alpha \beta} ) dA^\alpha \wedge F \\
&&	+\tfrac12 \I \cN_{\alpha \beta} dA^\alpha \wedge * dA^\beta
	+\tfrac12 \R \cN_{\alpha \beta} dA^\alpha \wedge dA^\beta \ ,\nn
\eea
from where we can extract a corrected gauge coupling function $f_{\text{cor}}$ 
for the brane U(1) gauge fields,
\beq
  \R f_{\text{cor}} = \R f_{\text{r}\, } 
	+ 4 \I \cN_{\alpha \beta} \tilde \cJ^\alpha \tilde \cJ^\beta \ , \quad
  \I f_{\text{cor}} = \I f_{\text{r}} + 4 \Delta_{\alpha} \tilde \cJ^\alpha 
	+ 4 \R \cN_{\alpha \beta} \tilde \cJ^\beta \tilde \cJ^\alpha \ ,
\eeq
a gauge coupling function $f_\alpha$ for the mixing between brane and bulk gauge bosons,
\beq
   \R f_\alpha = -4 \I \cN_{\alpha \beta} \tilde \cJ^\beta \ , \quad
    \I f_\alpha = -4 (\Delta_{\alpha} + \tilde \cJ^\beta \R \cN_{\alpha \beta}) \ ,
\eeq
and the gauge coupling function for the vector field $A^\alpha$ from the bulk \eqref{RRcoupling},
\beq    
f_{\alpha \beta} = -i \bar \cN_{\alpha \beta} \ .
\eeq

\section{Symmetries and moment maps} \label{moment_maps}

In this appendix we will have a closer look at the symmetries of the symplectic
form \eqref{def-varphi}. This is crucial for the determination of the first
derivatives of the K\"ahler potential $K_o$.

Before entering the detailed study of our set-up, let us quickly recall some general facts about moment
maps. Denote by $G$ a Lie group preserving the symplectic form $\varphi$ on a manifold $\cV_o$, and
by $\mathfrak{g}$ the Lie algebra of $G$.
There is a map identifying an element $\xi \in \mathfrak{g}$ with a vector field $X(\xi)$,
and by the invariance of $\varphi$ under $G$ and the fact that $d\varphi=0$ one has
\beq \label{invariance_of_varphi}
  \cL_{X(\xi)} \varphi = d(X(\xi)\lrcorner \varphi) = 0 \ .
\eeq
The moment map is a function $\mu: \cV_0 \rightarrow \mathfrak{g}^*$, where
$ \mathfrak{g}^*$ is the dual to $\mathfrak{g}$ under some paring $\langle \cdot,\cdot \rangle$,
which satisfies
\beq \label{def-moment_map}
  d \langle \mu,\xi \rangle = X(\xi) \lrcorner \varphi\ .
\eeq

In our example there are two Lie group actions, which will help us to study $K_o$.
As before we will be mostly interested in a local analysis around $L_0$. Therefore it
suffices to specify the associated Lie algebras $\mathfrak{g}_1$ and $\mathfrak{g}_2$
\beq
  \mathfrak{g}_1 = \Omega^2_{\rm ex}(L_0)\ , \qquad \mathfrak{g}_2 = \Omega^1_{\rm ex}(L_0)\ ,
\eeq
where $\Omega^i_{\rm ex}(L_0)$ are the exact $i$-forms on $L_0$.
In order to check that these indeed preserve the symplectic form $\varphi$ given
in \eqref{def-varphi} we have to specify the maps from $\mathfrak{g}_i$ to tangent
vectors $T_{(L_0,A_0)}\cV_o$. So given an exact two-form $\xi$ in $\mathfrak{g}_1$
and an exact one-form $\tilde \xi$ in $\mathfrak{g}_2$ one defines a tangent vector
$\tau(\xi)$ and a normal vector $\eta(\tilde \xi)$ by demanding that
\beq \label{xi_tildexi}
 (\tau(\xi) \lrcorner \Omega_1)|_{L_0} = \xi \ , \qquad (\eta(\tilde \xi) \lrcorner J)|_{L_0} = \tilde \xi\ .
\eeq
It turns out to be useful to identify also the tangent vectors $\tau(\xi)$ with normal
vectors to $L_0$ using the complex structure $I$ on $Y$. One first notes that the
normal bundle to $L_0$ admits the split
\beq \label{normal_split}
   NL_0 = (NL_0)^{\rm harm} \oplus (NL_0)^{\rm ex} \oplus (NL_0)^{\rm cex}\ .
\eeq
This split is performed in such a way that, e.g.~$X \in (NL_0)^{\rm harm}$ yields
a harmonic one-form $(X \lrcorner J)|_{L_0}$. Similarly, one defines $(NL_0)^{\rm ex}$ and
$(NL_0)^{\rm cex}$ corresponding to exact and co-exact one-forms. By Hodge-decomposition of
one-forms each normal vector has a unique decomposition under \eqref{normal_split}. Returning
to the two Lie algebras, one has
\beq
  \eta(\tilde \xi) \in (NL_0)^{\rm ex}\ ,\qquad  I\tau(\xi) \in (NL_0)^{\rm cex}\ ,
\eeq
for $\tilde \xi \in \mathfrak{g}_2$ and $\xi \in \mathfrak{g}_1$. The latter follows from the
fact that $(I \tau(\xi)\lrcorner J)|_{L_0} = 
- 2 e^\phi *(\tau(\xi)\lrcorner \Omega_1)|_{L_0}   
= -2 e^\phi *\xi$, using \eqref{useful_id} and \eqref{xi_tildexi}. 
Since $\xi$ is exact one concludes that $(I \tau(\xi)\lrcorner J)|_{L_0}$
is co-exact, i.e.~is annihilated by $d^*$.

Next we need to check the invariance \eqref{invariance_of_varphi}
of the symplectic form $\varphi$. We do that by first noting that on
a special Lagrangian $L_0$ the form $\varphi$ can be written as
\beq \label{new_varphi_form}
 \varphi(X,Y) =  \int_{L_0} (Y \lrcorner J)|_{L_0} \wedge (X\lrcorner \Omega_1)|_{L_0}
              - (X \lrcorner J)|_{L_0} \wedge (Y\lrcorner \Omega_1)|_{L_0} \ ,
\eeq
which is deduced by inserting $X$ and $Y$ into $J \wedge \Omega_1 =0$.
We have to check that $d(\tau(\xi) \lrcorner \varphi) =0 $ for all $\xi \in \mathfrak{g}_1$, and similarly
for the action of $\mathfrak{g}_2$. This is straightforward when using $\varphi$ in the form \eqref{new_varphi_form}.
For $\tau(\xi) \lrcorner \varphi$ only the term in the expression \eqref{new_varphi_form}
containing $(\tau(\xi) \lrcorner \Omega_1)|_{L_0}$ is
non-zero since $(\tau(\xi) \lrcorner J)|_{L_0}$ vanishes on the Lagrangian cycle $L_0$. Together
with the fact that $L_0$ is compact this yields the desired invariance under the action of $\mathfrak{g}_1$.
Similarly on checks invariance under the action of $\mathfrak{g}_2$ using the term in \eqref{new_varphi_form}
containing $(\eta(\tilde \xi) \lrcorner J)|_{L_0}$ and the 
fact that $(\eta(\tilde \xi)\lrcorner \Omega_1)|_{L_0}=0$. The invariance ensures the
existence of two moment maps $\mu_1$ and $\mu_2$ for the respective Lie algebras.

In a next step we determine the dual Lie algebras $\mathfrak{g}_1^*$ and $\mathfrak{g}_2^*$ with respect
to the pairing $\langle \alpha , \beta \rangle = \int_{L_0} \alpha \wedge \beta$. This implies that
$\mathfrak{g}_i^*$ is the space of non-closed $i$-forms, i.e.
\beq \label{def-frakg*}
  \mathfrak{g}^*_1 = \frac{\Omega^1(L_0)}{\Omega^1_{\rm cl}(L_0)}\ , \qquad
  \mathfrak{g}_2^* = \frac{\Omega^2(L_0)}{\Omega^2_{\rm cl}(L_0)}\ ,
\eeq
where $\Omega^i_{\rm cl}(L_0)$ are closed $i$-forms. Finally, one determines the
moment maps $\mu_i$ of $\mathfrak{g}_i$ which obey \eqref{def-moment_map} by
direct calculation
\beq
  \mu_1 = [\hat \mu_1] \ ,\qquad \mu_2 = [\hat \mu_2]\ ,
\eeq
where $\hat \mu_i$  are the non-closed $i$-forms which have been introduced in \eqref{def-muhat},
and the brackets $[\cdot]$ mean that these maps are only defined up to
closed forms as required in \eqref{def-frakg*}. This is checked by evaluating
\bea \label{determine_moment}
   \frac{d}{d{z^I}} \int_{L_0} \hat \mu_1 \wedge \xi &=& \int_{L_0} (s_I \lrcorner J)|_{L_0} \wedge \xi =
                                      \varphi(\tau(\xi),\partial_{z^I}) \ ,\\
   \frac{d}{d{z^I}}  \int_{L_0} \hat \mu_2 \wedge \tilde \xi &
=& \int_{L_0} (s_I \lrcorner \Omega_2)|_{L_0} \wedge \tilde \xi 
= \varphi(\eta(\xi),\partial_{z^I})\ . \nn
\eea
For the first equalities we have evaluated the derivative in the direction $\partial_{z^I}$ by
taking the Lie derivative of the expression under the integral with respect to $\partial_{z^I}$.
In evaluating the second equalities we used the form \eqref{new_varphi_form} of $\varphi$
and the fact that $(\partial_{z^I} \lrcorner \Omega_1)|_{L_0} = i (\partial_{z^I} \lrcorner \Omega_2)|_{L_0} $.
A rather compact way to rewrite the moment maps is by using the chain $\cC_4$
introduced in \eqref{def-cC4}. Since $J$ and $\Omega_2$ vanish on $L_0$
one thus has
\beq
  \langle \mu_1 , \xi \rangle = \cI (J , \xi)\ ,\qquad  
\langle \mu_2 , \tilde \xi \rangle = \cI (\Omega_2 , \tilde \xi)\ ,
\eeq
where $\cI$ is the chain integral introduced in \eqref{def-cI_gen}.

We have just found an explicit characterization of the symmetries of $\cV_o$
around $L_0$. In the following want to make contact to the parameterization used
in section \ref{D6reduction} for the Kaluza-Klein 
modes of the D6-brane.
Let us recall that using a Hodge-decomposition with respect to the induced metric the mode expansion
for $A_{\rm D6}$ reads
\beq
  A_{\rm D6} = a^i\, \tilde \alpha_i + a^I_{\rm ex}\, d h_I + a^{J}_{\rm cex}\, d^* \gamma_J\ .
\eeq
Note that the Hodge-star metric orthogonally splits under this decomposition as in \eqref{eqn:DBI}.
We also want to split the normal vectors $s_I$ appearing in \eqref{def-gen_theta}.
As in the decomposition \eqref{normal_split} one picks a basis of
normal vectors  $s_i,s_I^{\rm ex} ,s_J^{\rm cex}$ and corresponding
one-forms $\theta_I = (s_I \lrcorner J)|_{L_0}$
such that $\theta_i = \tilde \alpha_i$, $\theta_I^{\rm ex} = d h_I$ and $\theta_J^{\rm cex} = d^* \gamma_J$.
This implies that the metrics \eqref{metric_a} and \eqref{eqn:metrics} are identical in this basis. Moreover, from
\eqref{general_Kaehler} we inferred that
\beq
  \partial_{z^I} \partial_{\bar z^J} K_o = \frac12 e^{-\phi} \int_{L_0} \theta_I \wedge * \theta_J\ .
\eeq
which can be adapted to the Hodge-decomposition of the one-forms $\theta_I$ under
an appropriate split of complex coordinates $z^I = (z^i,z^I_{\rm ex},z^J_{\rm cex})$.
In this leading order analysis we thus find that $\I z^i = a^i$, $\I z^I_{\rm ex} = a^I_{\rm ex}$
and $\I z^J_{\rm cex} = a^J_{\rm cex}$.

One can proceed and determine the first derivatives of the K\"ahler potential
in the coordinates $z^I$ by using the moment map analysis of the previous
section. In fact, one infers from \eqref{determine_moment} that
\beq
  \partial_{z^I_{\rm ex}} K_o = \int_{L_0} \mu_2 \wedge \theta^{\rm ex}_I \ , \qquad
  \partial_{z^J_{\rm cex}} K_o = \int_{L_0} *\mu_1 \wedge \theta^{\rm cex}_J \ .
\eeq
It is straightforward to evaluate $\partial_{z^I_{\rm ex}} K_o$ and $\partial_{z^J_{\rm cex}} K_o $
at leading order in the deformations using \eqref{normal_exp_JOm}. They take the simple  form
\beq
  \partial_{z^I} K_o = \int_{L_0} \theta_{\eta} \wedge * \theta_I \ .
\eeq

\end{appendix}

\end{document}